\documentclass[
 aps,%
 11pt,%
 final,%
 notitlepage,%
 oneside,%
 onecolumn,%
 nobibnotes,%
 nofootinbib,%
 superscriptaddress,%
 noshowpacs,%
 centertags]%
 {revtex4}

 \usepackage{graphicx}

\newcommand{\aaps}{A\&AS\ }
\newcommand{\aap}{A\&A\ }
\newcommand{\mnras}{MNRAS\ }
\newcommand{\aj}{AJ\ }
\newcommand{\apjs}{ApJS\ }
\newcommand{\pasp}{PASP\ }

\begin{document}


\newcommand{\kms}{km/s}
\newcommand{\CVnIDist}{\mbox{$D=4.28$~Mpc}}                    
\newcommand{\CVnIVsys}{\mbox{$V_{\rm LG}=295\pm15$ \kms}}      
\newcommand{\CVnIVmed}{\mbox{$V_{\rm LG}=287$ \kms}}           
\newcommand{\CVnIVsigma}{\mbox{$\sigma=51$ \kms}}              
\newcommand{\CVnIRharm}{\mbox{$R_h=340$ kpc}}
\newcommand{\CVnIRmean}{\mbox{$\langle R\rangle=455$ kpc}}
\newcommand{\CVnIRta}{\mbox{$R_0=1.04\pm0.15$ Mpc}}
\newcommand{\CVnILBtot}{\mbox{$L_B=1.61\times10^{10}$ $L_{\odot}$}}
\newcommand{\CVnIMvir}{\mbox{$M_{\rm vir}=1.93\times10^{12}$ $M_{\odot}$}}     
\newcommand{\CVnIMproj}{\mbox{$M_{p}=2.56\times10^{12}$ $M_{\odot}$}}      
\newcommand{\CVnIMta}{\mbox{$M_{R_0}=2.38\times10^{12}$ $M_{\odot}$}}
\newcommand{\CVnIMLvir}{\mbox{$(M/L)_{\rm vir}=120$ $(M/L)_{\odot}$}}          
\newcommand{\CVnIMLproj}{\mbox{$(M/L)_{p}=159$ $(M/L)_{\odot}$}}           
\newcommand{\CVnIMLta}{\mbox{$(M/L)_{R_0}=148$ $(M/L)_{\odot}$}}
\newcommand{\CVnITcr}{\mbox{$T_{\rm cr}=R_h/\sigma=6.5$ Gyr} }                 










\title{Distances to Dwarf Galaxies of the Canes Venatici I Cloud
}

\author{D.~I.~Makarov}
\email{dim@sao.ru}
\affiliation{Special Astrophysical Observatory, Russian Academy of Sciences, Nizhnii Arkhyz, 369167 Russia }

\author{L.~N.~Makarova}
\email{lidia@sao.ru}
\affiliation{Special Astrophysical Observatory, Russian Academy of Sciences, Nizhnii Arkhyz, 369167 Russia }

\author{R.~I.~Uklein}
\email{uklein@sao.ru}
 \affiliation{Special Astrophysical Observatory, Russian Academy of Sciences, Nizhnii Arkhyz, 369167 Russia }

\begin{abstract}
We determined the spatial structure of the scattered concentration
of galaxies in the Canes Venatici constellation. We redefined the
distances for 30~galaxies of this region using the deep images
from the Hubble Space Telescope archive  with the WFPC2 and ACS
cameras. We carried out a high-precision stellar photometry of the
resolved stars in these galaxies, and determined the photometric
distances by the tip of the red giant branch (TRGB) using an
advanced technique and modern calibrations. High accuracy of the
results allows us to distinguish the zone of chaotic motions
around the center of the system. A group of galaxies around M\,94
is characterized by the median velocity \CVnIVmed, distance
\CVnIDist, internal velocity dispersion \CVnIVsigma{} and total
luminosity \CVnILBtot{}. The projection mass of the system amounts
to \CVnIMproj, which corresponds to the mass--luminosity ratio of
\CVnIMLproj. The estimate of the mass--luminosity ratio is
significantly higher than the typical ratio $M/L_B\sim30$  for the
nearby groups of galaxies. The CVn\,I cloud of galaxies contains
4--5 times less luminous matter compared with the well-known
nearby groups, like the Local Group, M\,81 and Centaurus\,A. The
central galaxy   M\,94 is at least  $1{\rm ^m}$ fainter than any
other central galaxy of these groups. However, the concentration
of galaxies in the Canes Venatici  may have a comparable  total
mass.
\end{abstract}

\maketitle

\section{INTRODUCTION}

\begin{figure*}
\includegraphics[width=0.64\textwidth,clip]{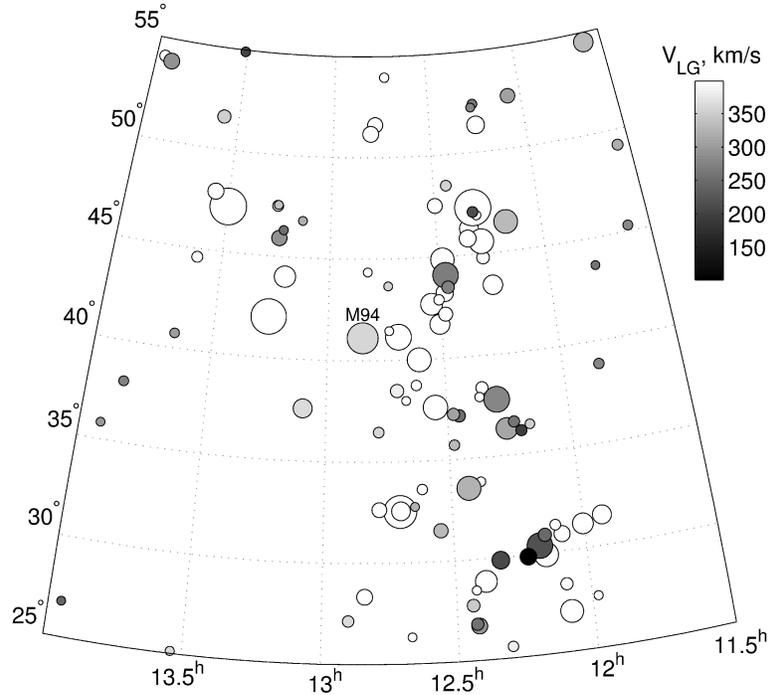}
\caption{The distribution of galaxies in the Canes Venatici
constellation on the celestial sphere. The sizes of the circles
are inversely proportional to the absolute magnitudes of objects.
The line-of-sight radial velocities of the CVn\,I cloud galaxies
with $V_{\rm LG}<400$~\kms{} are shown by the shades of gray,
while the white circles correspond to the galaxies of the distant
background.} \label{fig:sky:Makarov_n}
\end{figure*}

\begin{figure}
\includegraphics[width=0.65\textwidth]{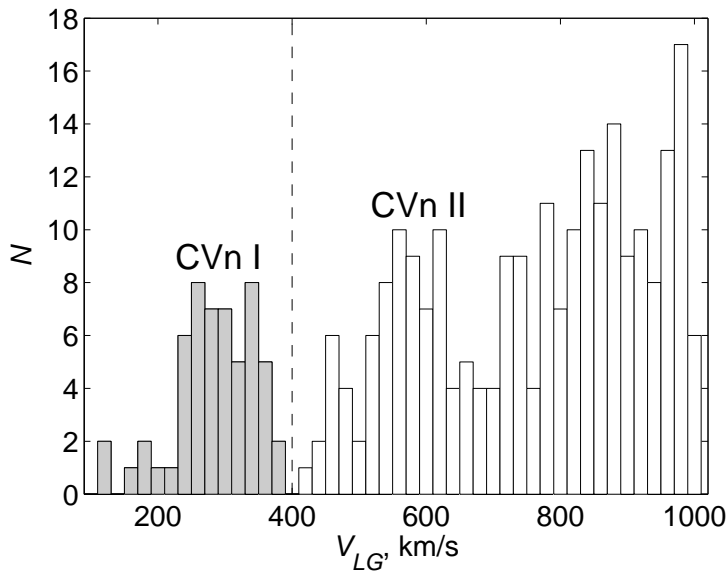}
\caption{The line-of-sight velocity distribution of galaxies in
the direction of Canes Venatici
constellation.}\label{fig:vlg:Makarov_n}
\end{figure}

The distribution of nearby galaxies of the Local Volume in the sky
reveals a significant concentration of objects in a small region
located in the Canes Venatici constellation
(Fig.~\ref{fig:sky:Makarov_n}). This feature has been noted by
Karachentsev (1966) and de Vaucouleurs
(1975)~\cite{karachentsev1966:Makarov_n,
vaucouleurs1975:Makarov_n}. This complex is mostly populated by
dwarf galaxies of late morphological types. Two peaks are
distinguished in the distribution of the line-of-sight velocities
of galaxies in the sky region \mbox{$\alpha = (11.5, 14.0)$,}
\mbox{$\delta = (+20^{\circ}, +60^{\circ})$}
(Fig.~\ref{fig:vlg:Makarov_n}). The first peak located around
\mbox{$V_{\rm LG}=300$ \kms{}} corresponds to the CVn\,I cloud,
while the CVn\,II concentration has an average velocity of about
560~\kms.   The Canes Venatici I cloud clearly differs from the
other nearby galaxy clusters, such as the Local Group, M\,81 or
the groups in the Centaurus by the absence of a clearly prominent
gravitational center and looks diffuse. The concentration of
galaxies in the Canes Venatici constellation has repeatedly
attracted the attention of researchers. In the series of
papers~\cite{makarova+1997:Makarov_n,tikhonov+1998:Makarov_n,makarova+1998:Makarov_n,karachentsev+1998:Makarov_n}
the structure of the complex was studied by the photometry of the
brightest blue stars in these galaxies. The use of the Hubble
Space Telescope (HST) has significantly improved the distance
accuracy and allowed to study the kinematics of the CVn\,I cloud
of galaxies~\cite{karachentsev+2003:Makarov_n}. A   blind survey
of the sky in neutral hydrogen~\cite{HIsurvey2009:Makarov_n} was
recently conducted in the region of nearby groups of galaxies in
the Canes Venatici in order to study the functions of the
H\,I-masses in dwarf galaxies. Only one object from this survey
has no optical identification. In addition, Kaisin and
Karachentsev~\cite{Hasurvey2008:Makarov_n} have investigated the
current star formation of galaxies in the Canes Venatici according
to the H$\alpha$ survey data. The authors did not find any
significant correlations between the star formation rates (SFR) in
the galaxies and their neighborhood.

A rapid progress in the deep observations of the Canes Venatici
galaxies, performed at the HST/ACS and HST/WFPC2, as well as a
significant improvement of distance measurement  by the tip of the
red giant branch (TRGB) method allow us to refine the structure of
this unusual concentration of galaxies. In this paper, we
determined the distances for~30~galaxies of the CVn\,I cloud,
using the optimized TRGB method~\cite{TRGB1:Makarov_n} and new
zero point calibration~\cite{TRGB2:Makarov_n}.

\section{STELLAR PHOTOMETRY OF GALAXIES IN THE CANES VENATICI I CLOUD}

The sample of galaxies from the Canes Venatici complex is
presented in Table~\ref{t:observ:Makarov_n}. Direct images of the
galaxies obtained with the ACS/HST and WFPC2/HST were taken from
the archive of the Hubble Space Telescope. The images of all the
galaxies were obtained in the  F606W and F814W filters, with the
exception of NGC\,4214, which was observed in F555W and F814W.
Standard initial reduction of images is done ``on the fly.'' Thus,
the user gets the images with subtracted dark frames, flat
field-corrected, and accounted for the presence of ``bad'' rows
and some ``hot''/``cold'' pixels. The images of the studied
galaxies are demonstrated in Fig.~\ref{fig:ima1:Makarov_n}.

\begin{table*}
\caption{Parameters of the HST observations
for 30~galaxies of the Canes Venatici I cloud}
\label{t:observ:Makarov_n}
\medskip
\begin{tabular}{l|@{~~}c@{~~}|@{~~}c@{~~}|@{~~}r@{~~}|@{~}c@{~}|@{~~}r@{/}l@{~}}
\hline
~~~~Name    & RA (J2000) Dec    &~~Camera &Proposal& Filters    & \multicolumn{2}{c}{$T_{\rm exp}$, s} \\
\hline
UGC\,6541 & 113328.9$+$491418 & WFPC2    &  8601 & F814W/F606W &  600&600  \\
NGC\,3738 & 113548.6$+$543122 & ACS/WFC  & 12546 & F814W/F606W &  450&450  \\
NGC\,3741 & 113606.0$+$451708 & WFPC2    &  8601 & F814W/F606W &  600&600  \\
UGC\,6817 & 115052.9$+$385251 & WFPC2    &  8601 & F814W/F606W &  600&600  \\
NGC\,4068 & 120401.9$+$523519 & ACS/WFC  &  9771 & F814W/F606W &  900&1200 \\
NGC\,4163 & 121209.2$+$361010 & ACS/WFC  &  9771 & F814W/F606W &  900&1200 \\
UGCA\,276 & 121458.1$+$361306 & WFPC2    &  8601 & F814W/F606W &  600&600  \\
NGC\,4214 & 121539.2$+$361939 & WFPC2    &  6569 & F814W/F555W & 1300&1300 \\
UGC\,7298 & 121630.1$+$521340 & WFPC2    &  8601 & F814W/F606W &  600&600  \\
NGC\,4244 & 121729.5$+$374826 & ACS/WFC  & 10523 & F814W/F606W &  735&735  \\
UGC\,7559 & 122705.0$+$370836 & WFPC2    &  8601 & F814W/F606W &  600&600  \\
UGC\,7577 & 122741.7$+$432939 & WFPC2    &  8601 & F814W/F606W &  600&600  \\
NGC\,4449 & 122811.0$+$440535 & WFPC2    &  5971 & F814W/F606W & 1400&1400 \\
UGC\,7605 & 122838.7$+$354304 & WFPC2    &  8601 & F814W/F606W &  600&600  \\
IC\,3687  & 124215.1$+$383010 & WFPC2    &  8601 & F814W/F606W &  600&600  \\
KK\,166   & 124913.1$+$353646 & WFPC2    &  8601 & F814W/F606W &  600&600  \\
M\,94     & 125053.0$+$410712 & ACS/WFC  & 10523 & F814W/F606W &  730&730  \\
IC\,4182  & 130549.6$+$373618 & WFPC2    &  8584 & F814W/F606W & 2600&2600 \\
UGC\,8215 & 130803.6$+$464941 & ACS/WFC  &  9771 & F814W/F606W &  900&1200 \\
UGC\,8308 & 131322.7$+$461913 & WFPC2    &  8601 & F814W/F606W &  600&600  \\
UGC\,8320 & 131428.2$+$455511 & WFPC2    &  8601 & F814W/F606W &  600&600  \\
UGC\,8331 & 131529.8$+$472959 & ACS/WFC  & 10905 & F814W/F606W & 1148&938  \\
NGC\,5204 & 132936.5$+$582510 & WFPC2    &  8601 & F814W/F606W &  600&600  \\
UGC\,8508 & 133044.4$+$545441 & WFPC2    &  8601 & F814W/F606W &  600&600  \\
UGC\,8638 & 133919.6$+$244631 & ACS/WFC  &  9771 & F814W/F606W &  900&1200 \\
UGC\,8651 & 133953.8$+$404421 & ACS/WFC  & 10210 & F814W/F606W & 1209&1016 \\
UGC\,8760 & 135051.2$+$380116 & ACS/WFC  & 10210 & F814W/F606W & 1189&998  \\
UGC\,8833 & 135448.5$+$355016 & ACS/WFC  & 10210 & F814W/F606W & 1189&998  \\
KK\,230   & 140710.4$+$350340 & ACS/WFC  &  9771 & F814W/F606W &  900&1200 \\
UGC\,9128 & 141556.5$+$230320 & ACS/WFC  & 10210 & F814W/F606W & 1174&985  \\
\hline
\end{tabular}
\vspace{10mm}
\end{table*}

\begin{figure*}
\includegraphics[width=0.26\textwidth]{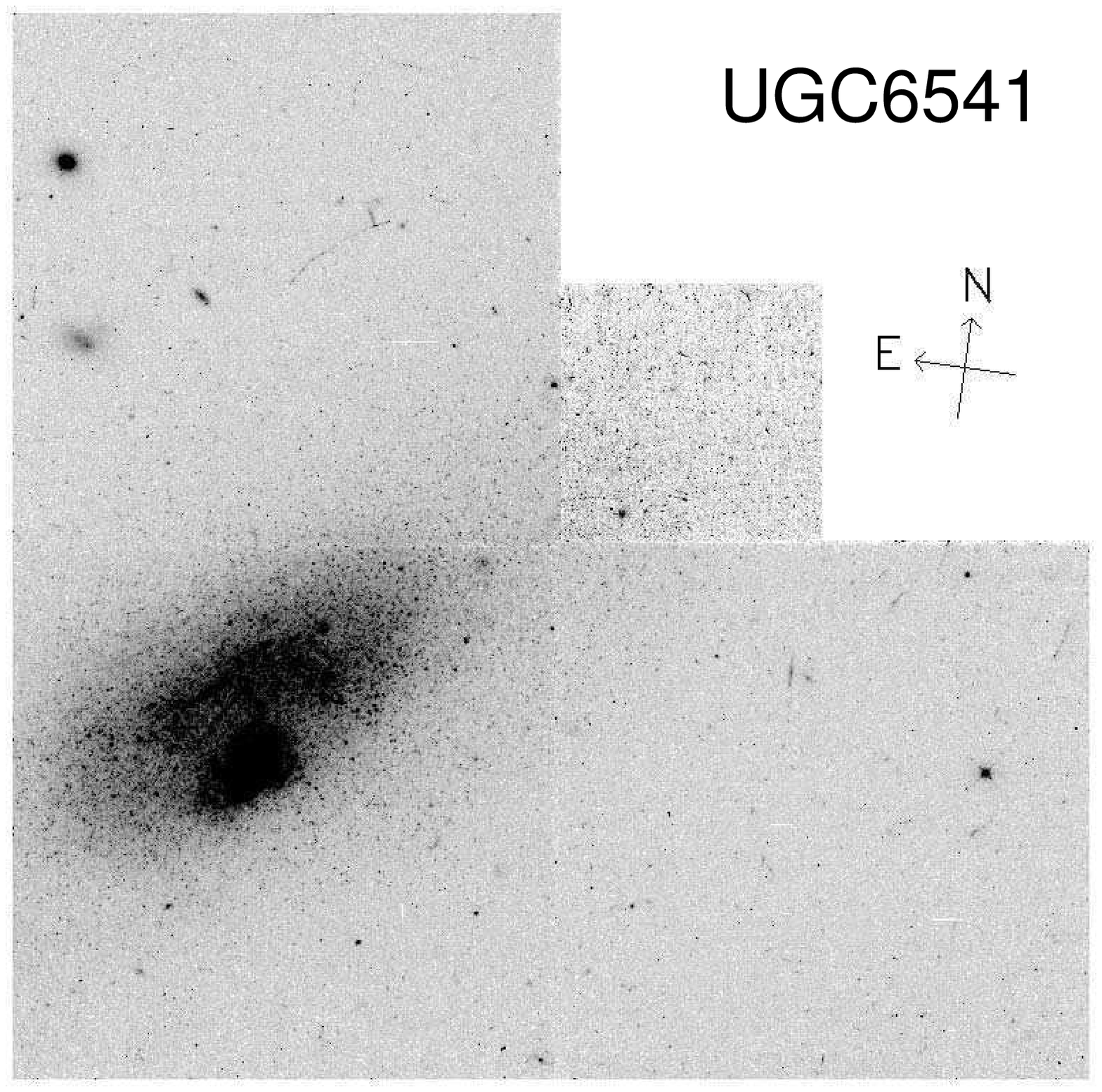}
\includegraphics[width=0.26\textwidth]{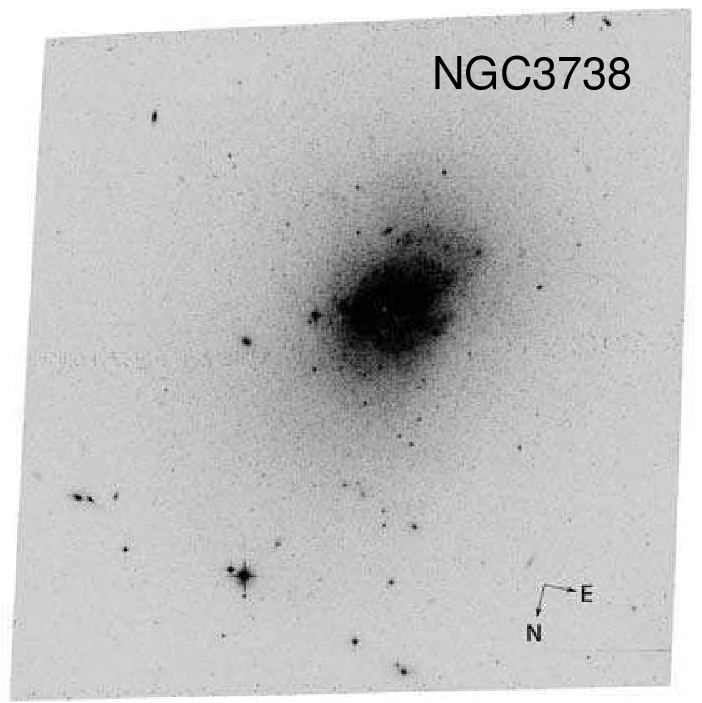}
\includegraphics[width=0.26\textwidth]{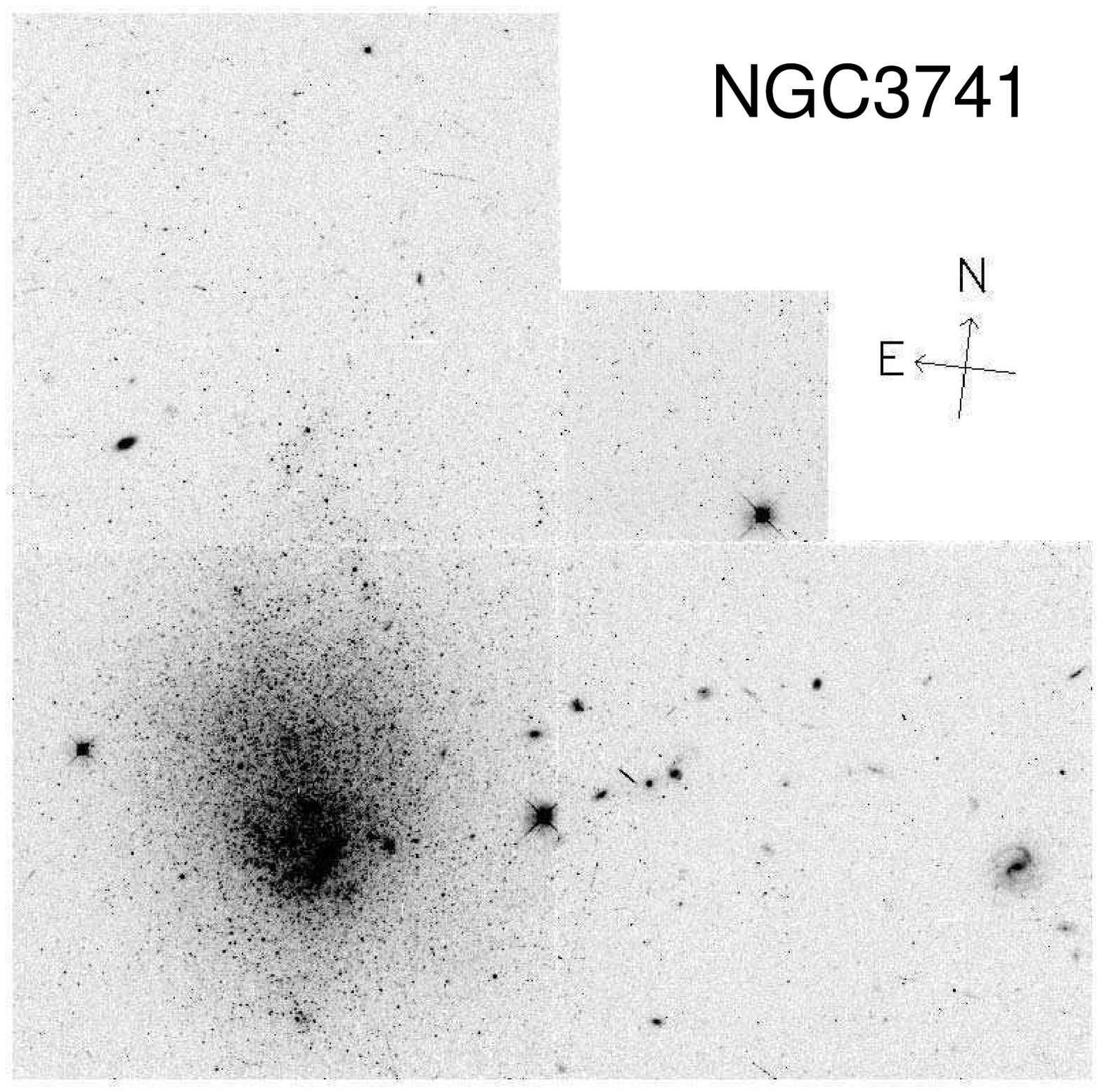}
\includegraphics[width=0.26\textwidth, bb = 74 310 518 752,clip]{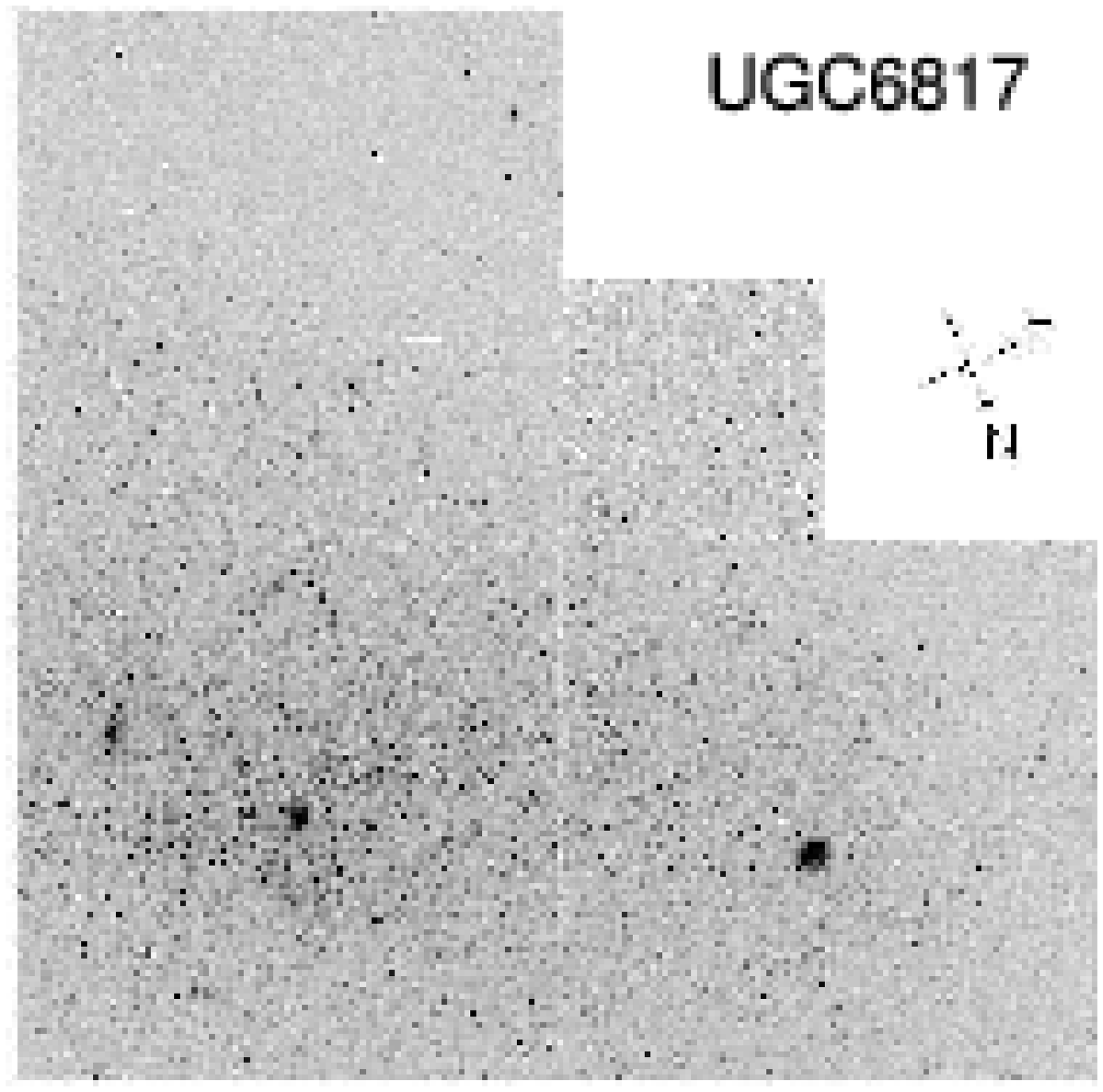}
\includegraphics[width=0.26\textwidth]{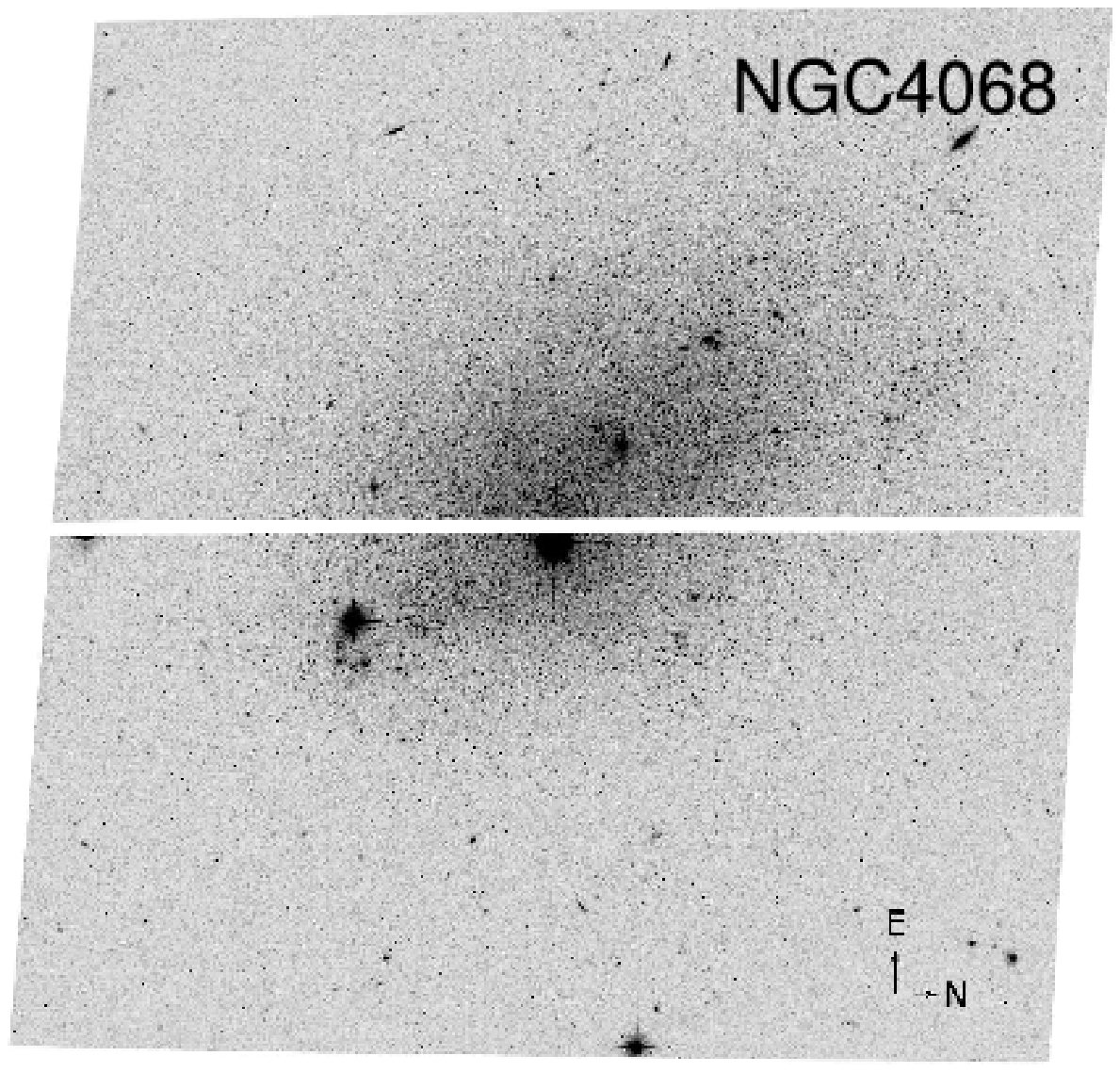}
\includegraphics[width=0.26\textwidth]{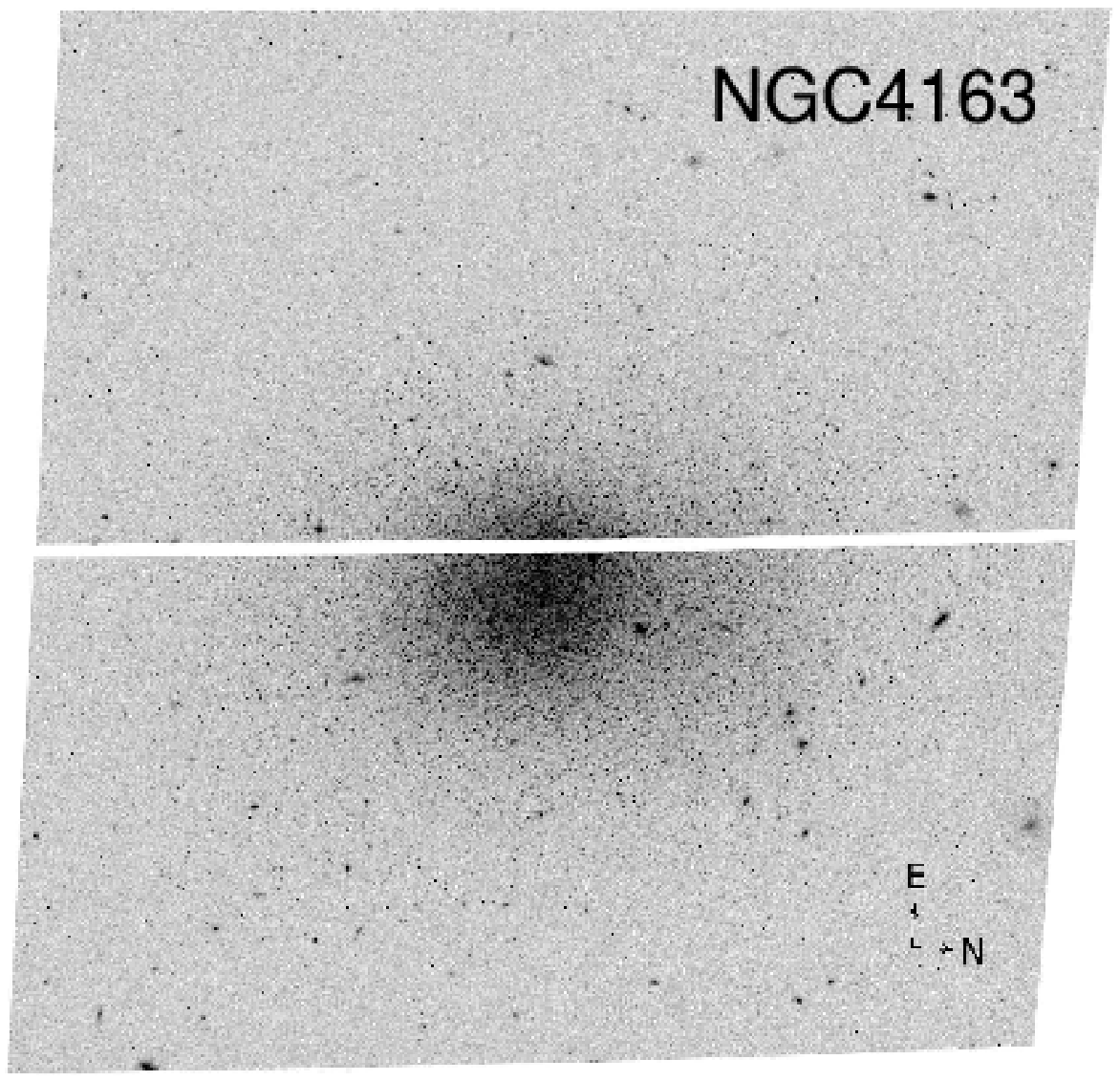}
\includegraphics[width=0.26\textwidth, bb = 74 310 518 752,clip]{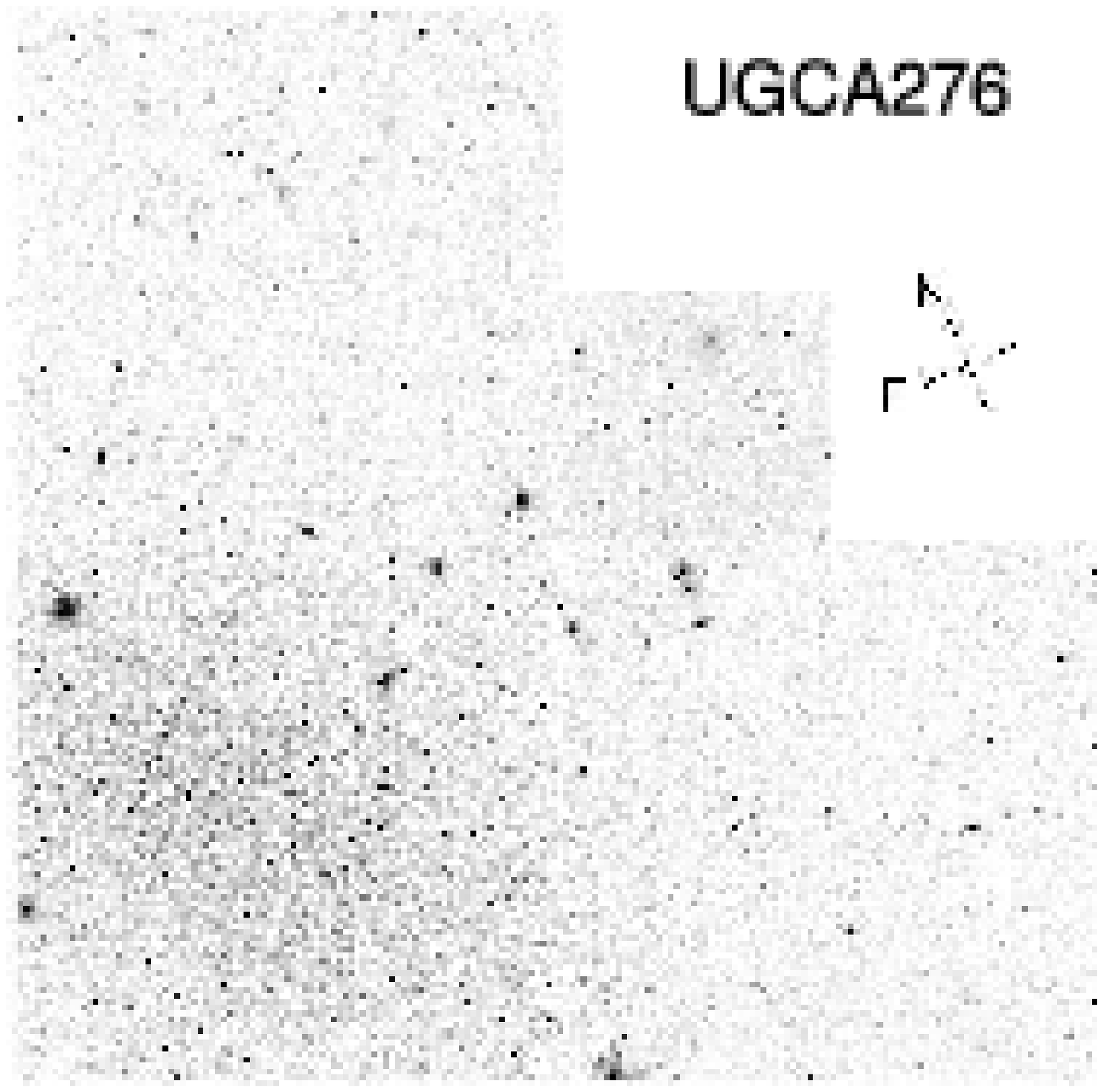}
\includegraphics[width=0.26\textwidth, bb = 74 310 518 752,clip]{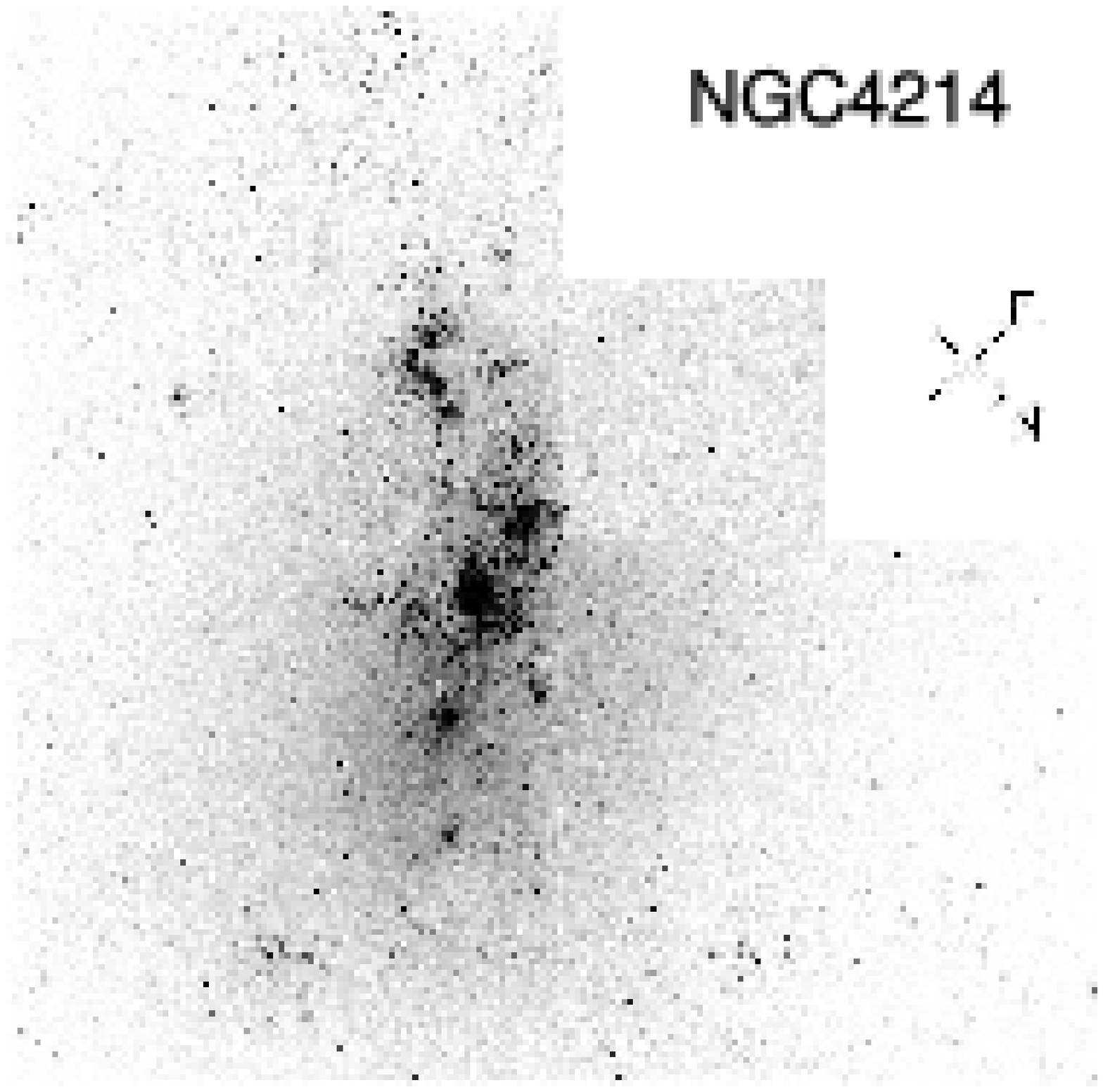}
\includegraphics[width=0.26\textwidth, bb = 74 310 518 752,clip]{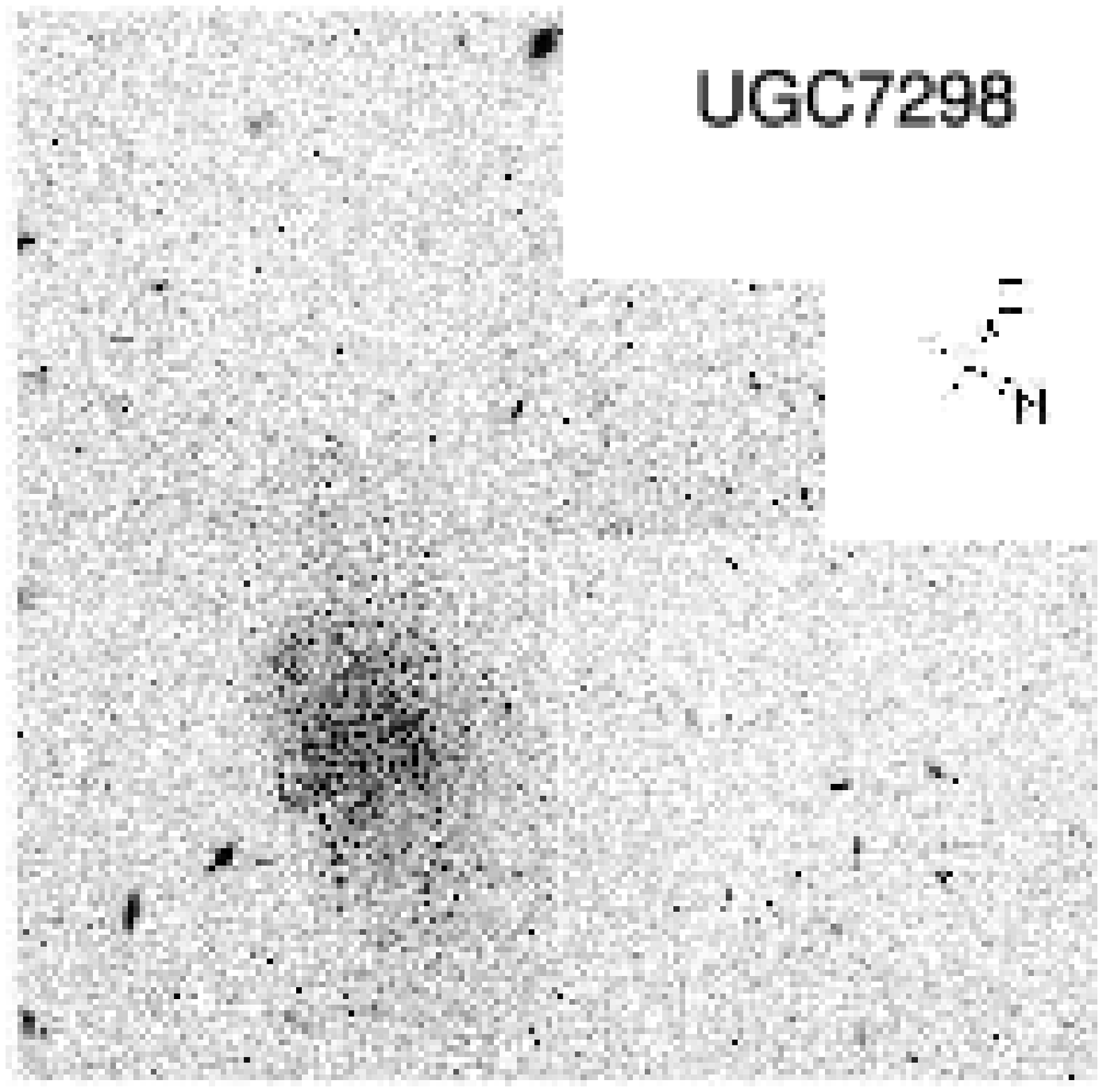}
\includegraphics[width=0.26\textwidth]{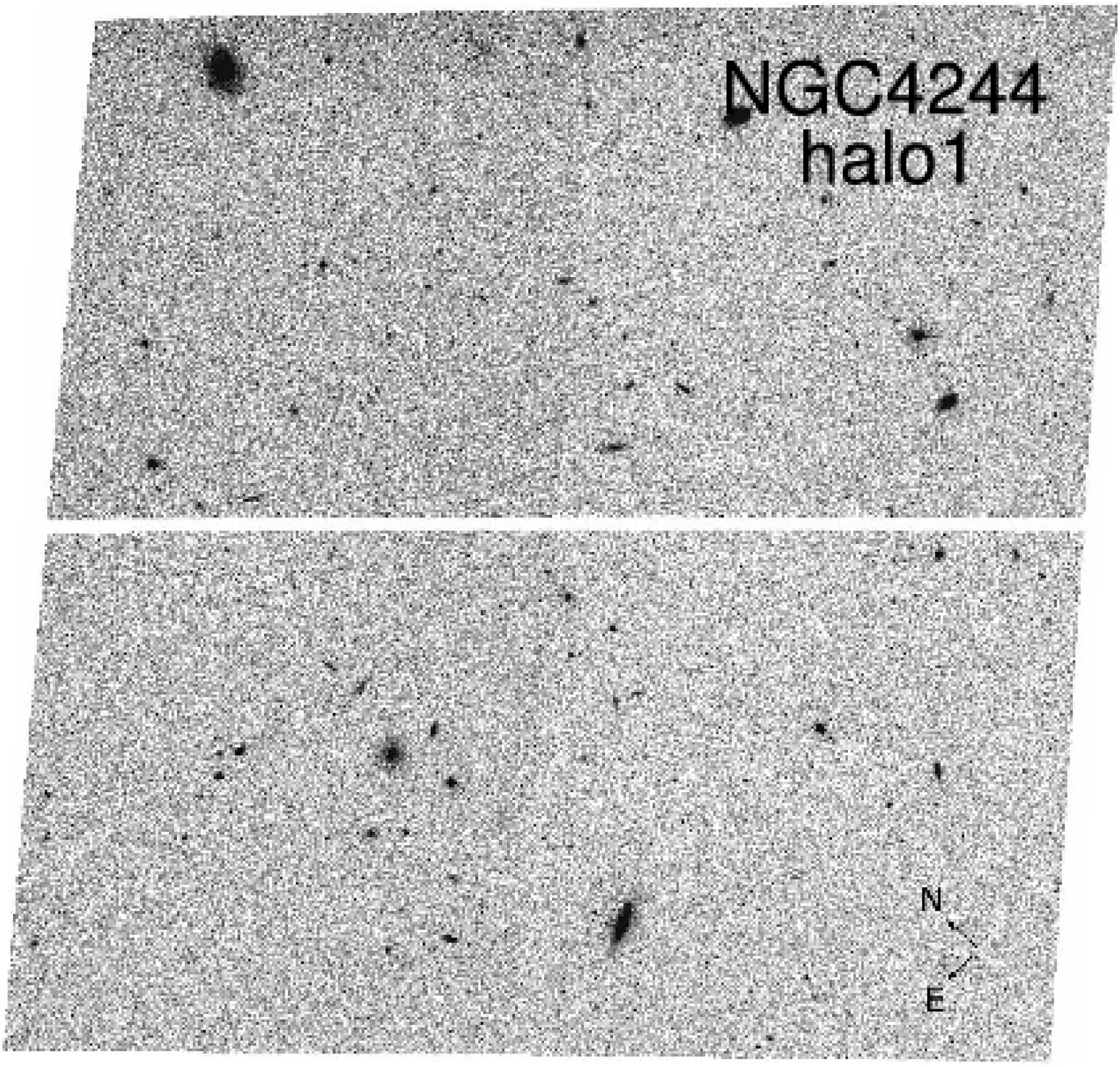}
\includegraphics[width=0.26\textwidth, bb = 74 310 518 752,clip]{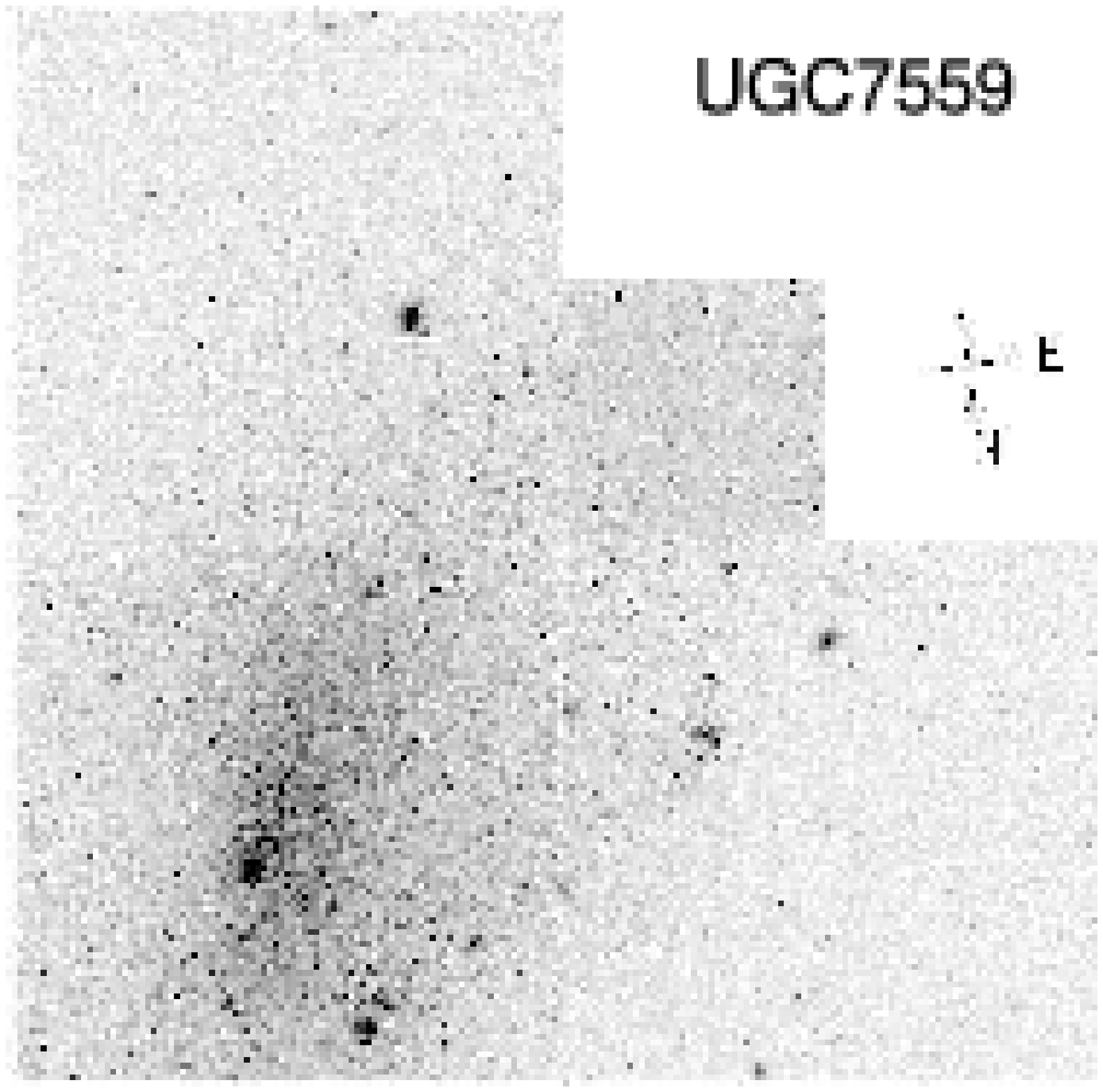}
\includegraphics[width=0.26\textwidth, bb = 74 310 518 752,clip]{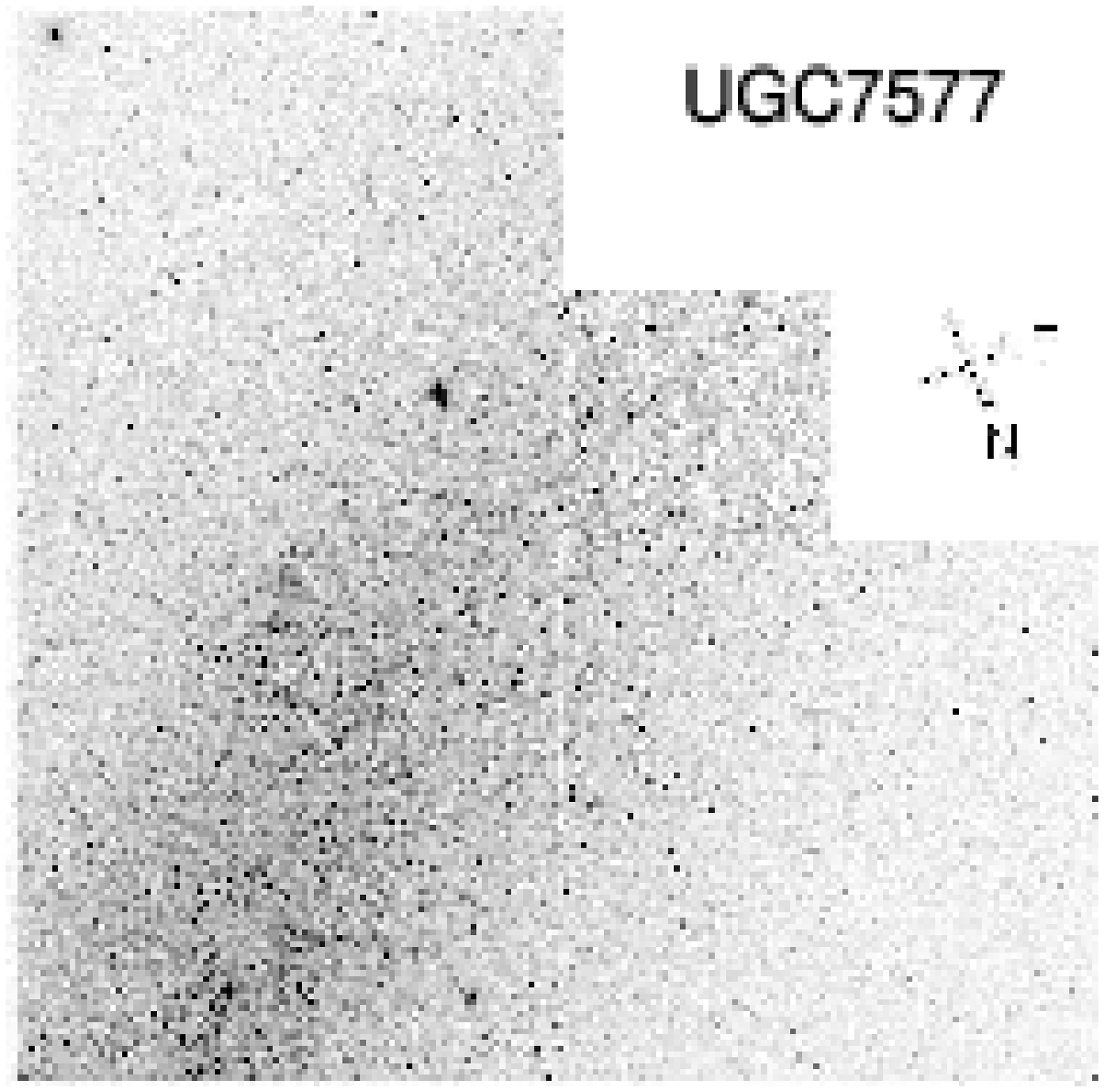}
\includegraphics[width=0.26\textwidth, bb = 74 310 518 752,clip]{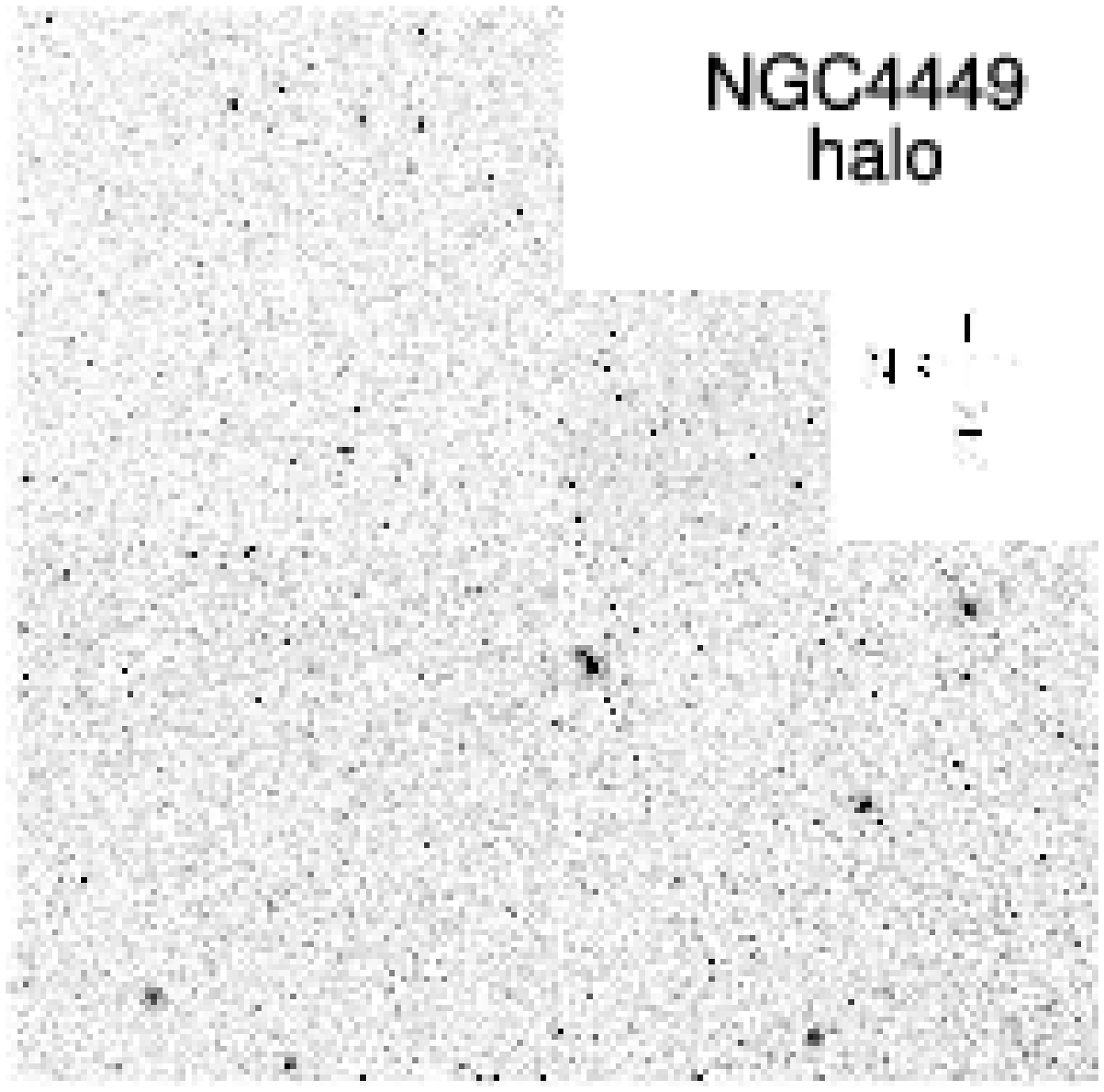}
\includegraphics[width=0.26\textwidth, bb = 74 310 518 752,clip]{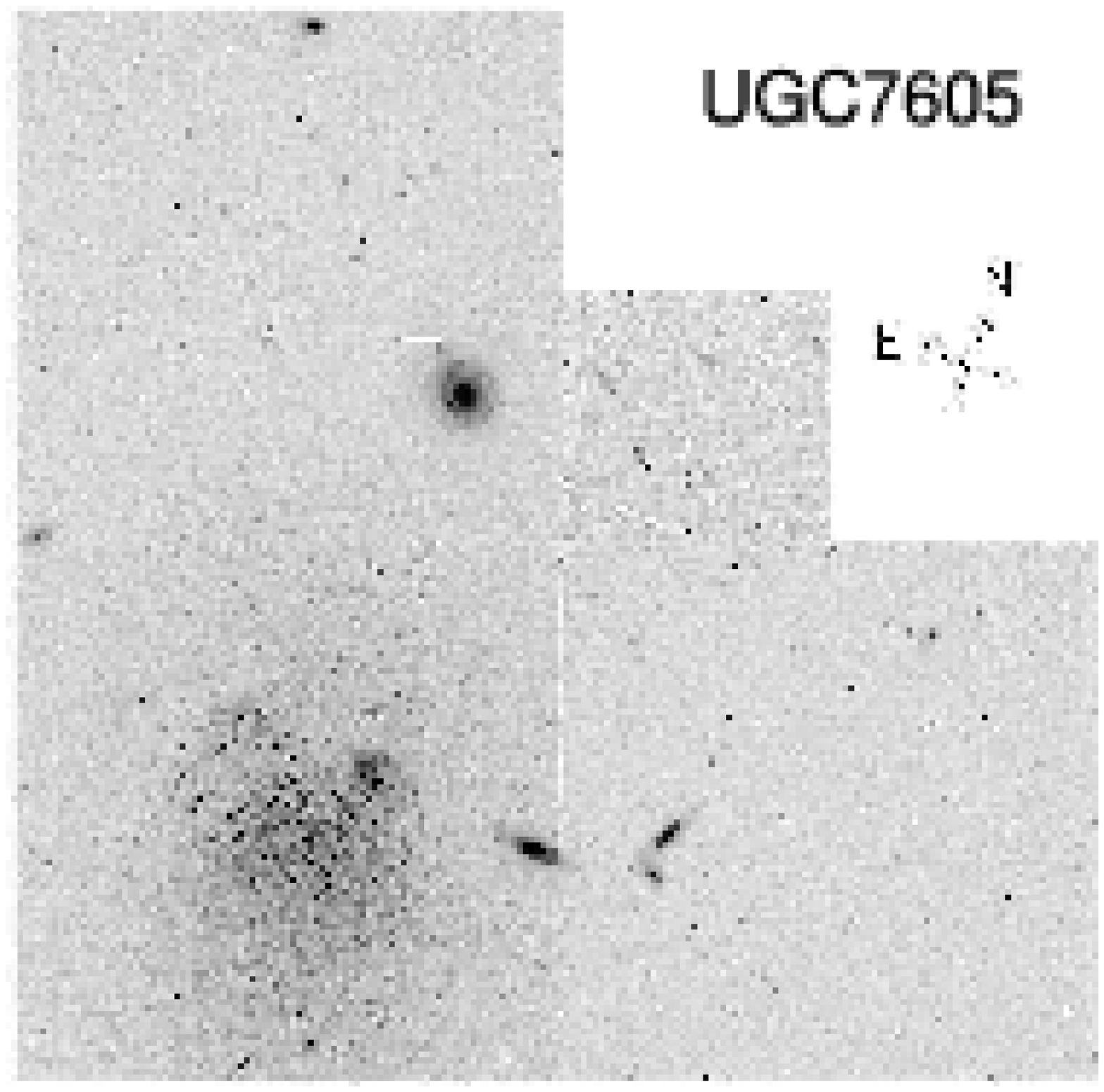}
\includegraphics[width=0.26\textwidth, bb = 74 310 518 752,clip]{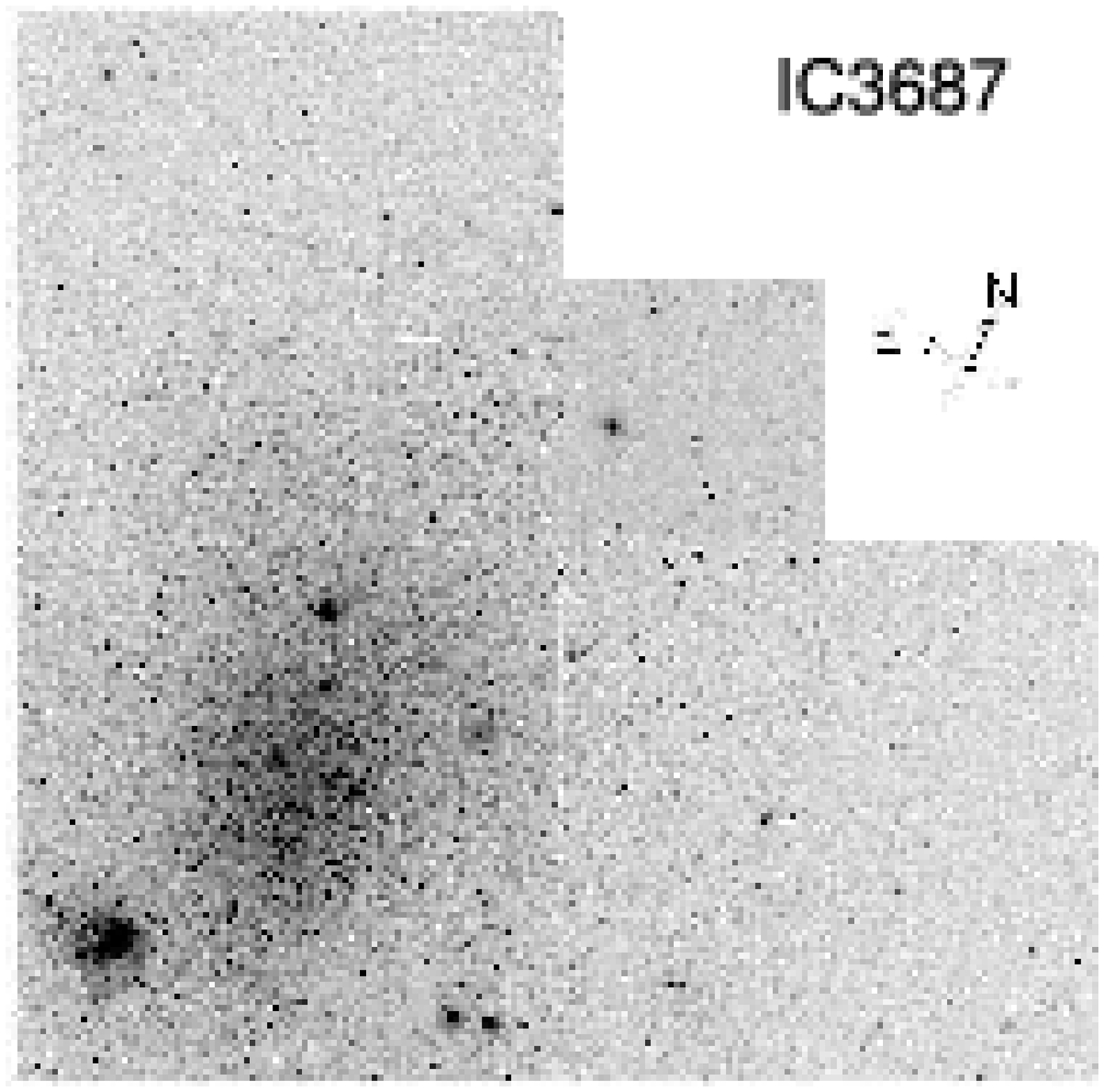}
\caption{The images of the studied galaxies in the Canes Venatici
I cloud, obtained with the HST WFPC2 or ACS.}
\label{fig:ima1:Makarov_n}
\end{figure*}

\addtocounter{figure}{-1}
\begin{figure*}
\includegraphics[width=0.26\textwidth, bb = 74 310 518 752,clip]{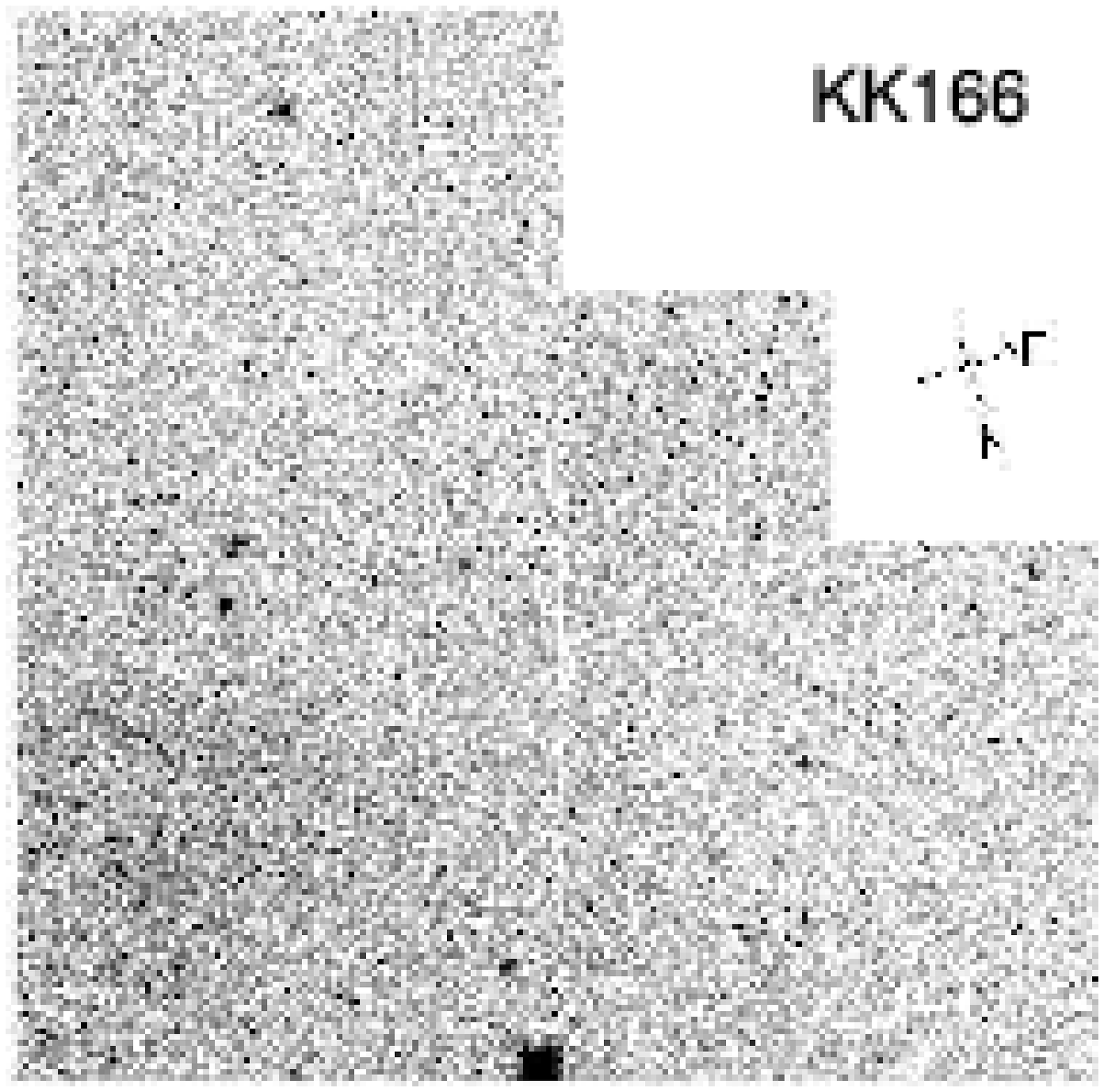}
\includegraphics[width=0.26\textwidth]{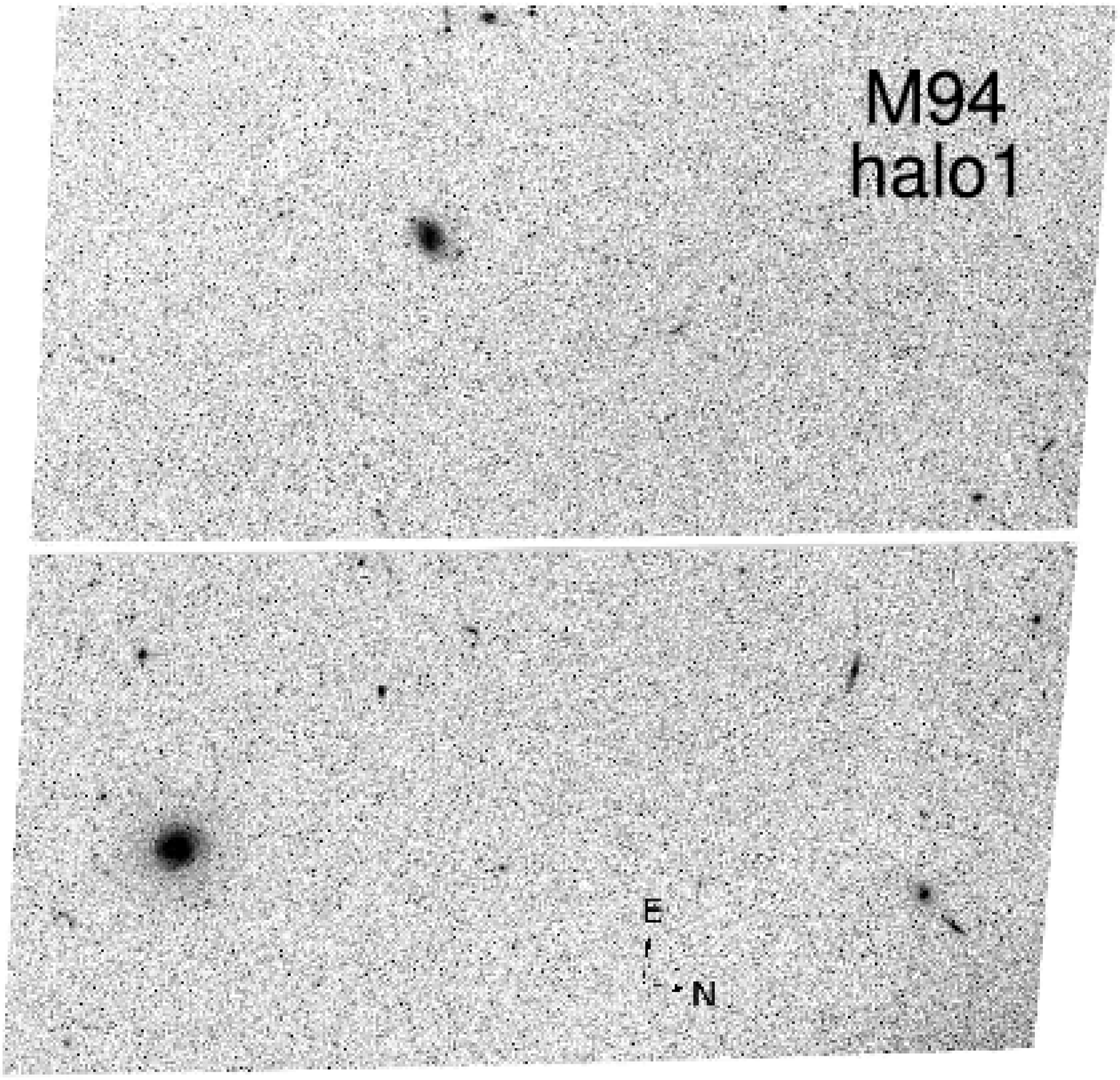}
\includegraphics[width=0.26\textwidth, bb = 74 310 518 752,clip]{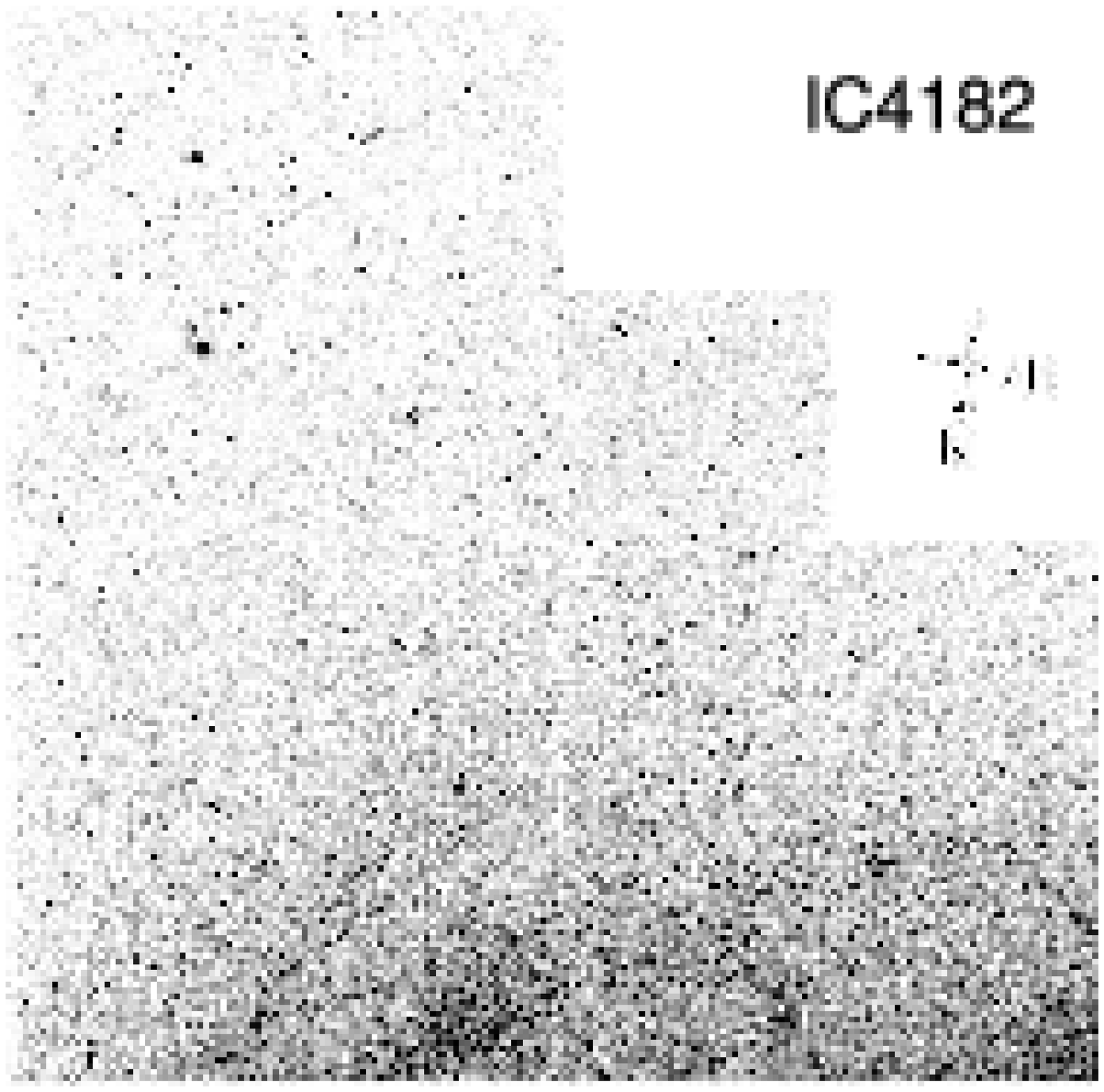}
\includegraphics[width=0.26\textwidth]{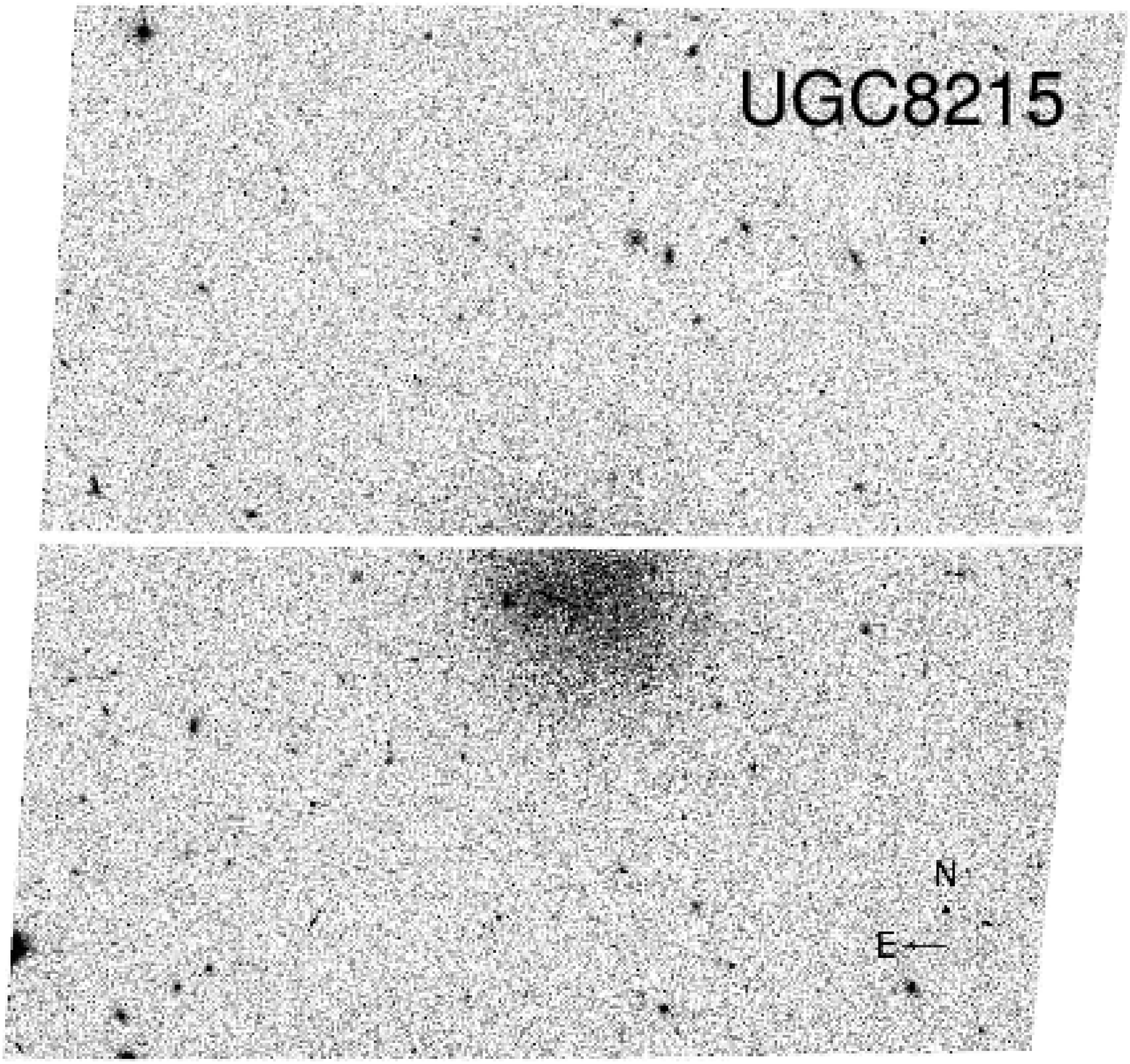}
\includegraphics[width=0.26\textwidth]{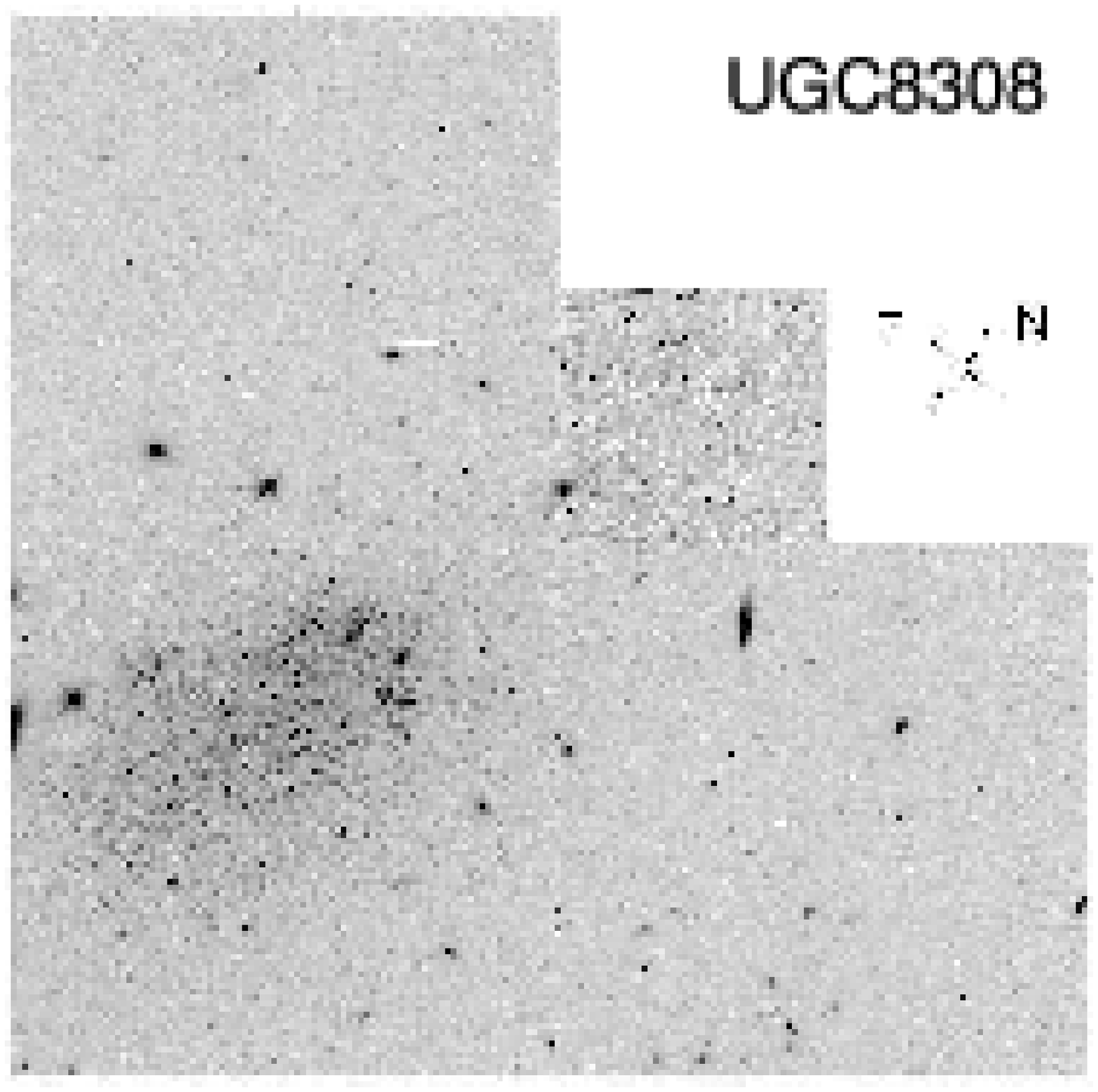}
\includegraphics[width=0.26\textwidth, bb = 74 310 518 752,clip]{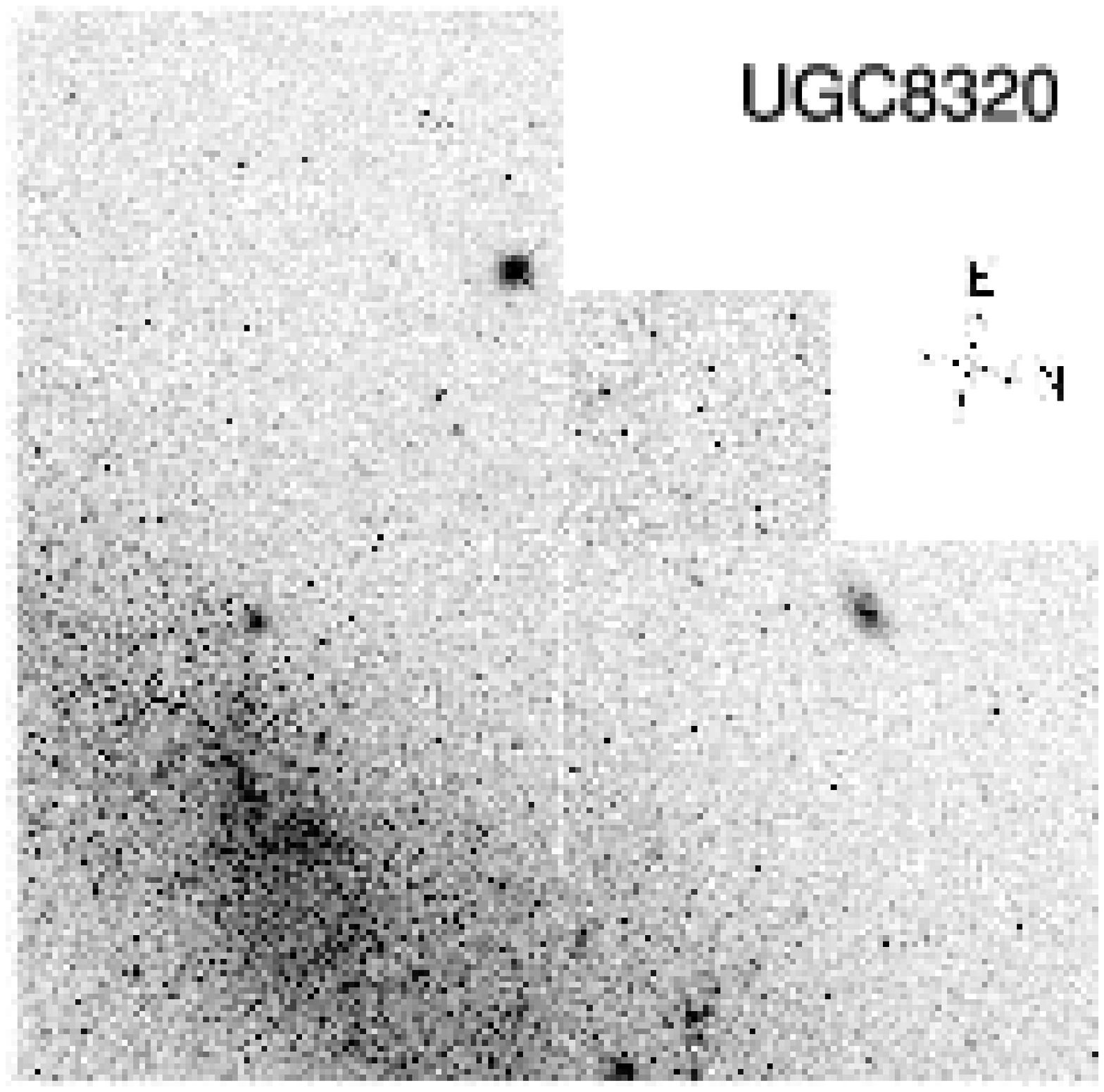}
\includegraphics[width=0.26\textwidth]{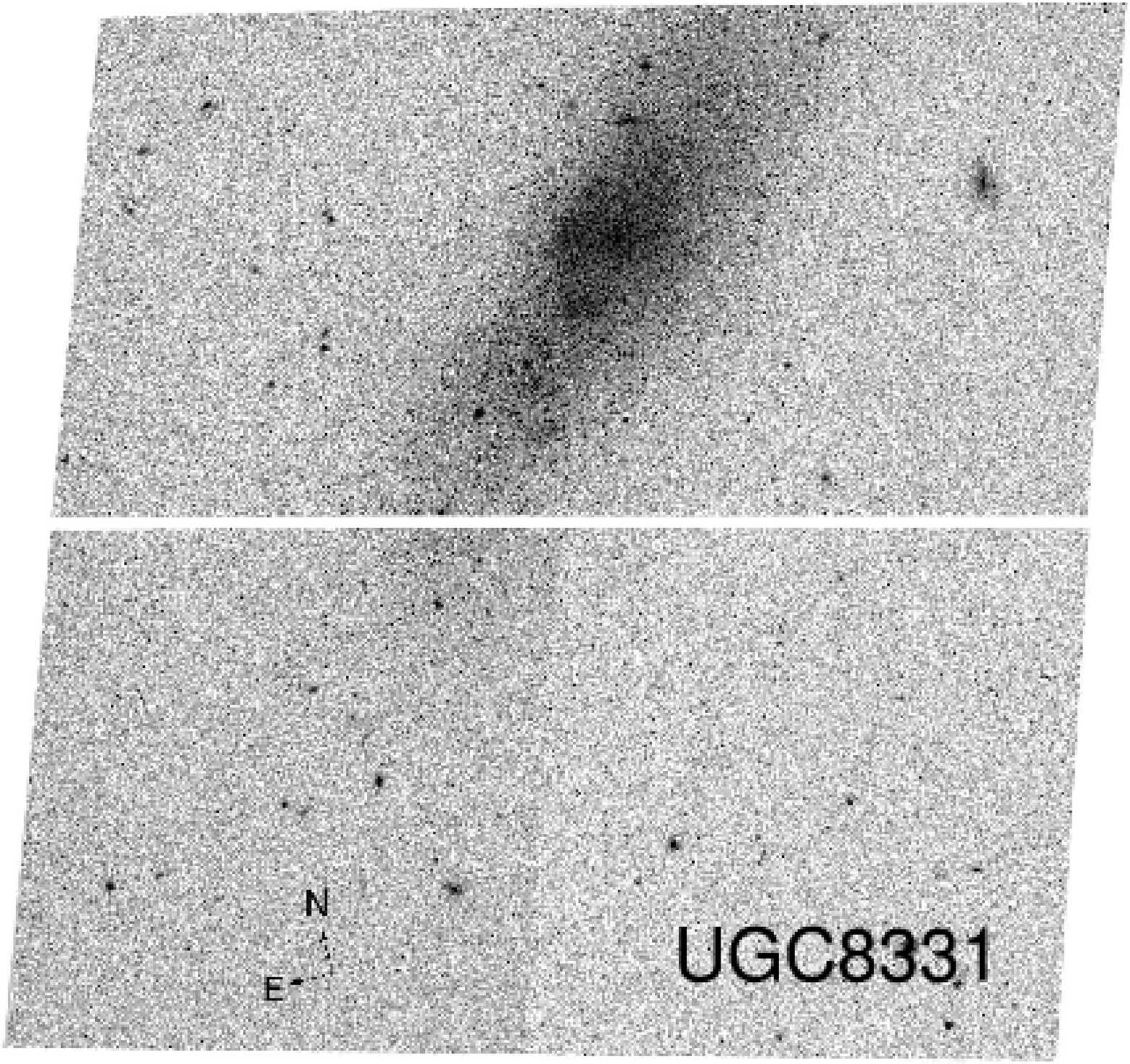}
\includegraphics[width=0.26\textwidth, bb = 74 310 518 752,clip]{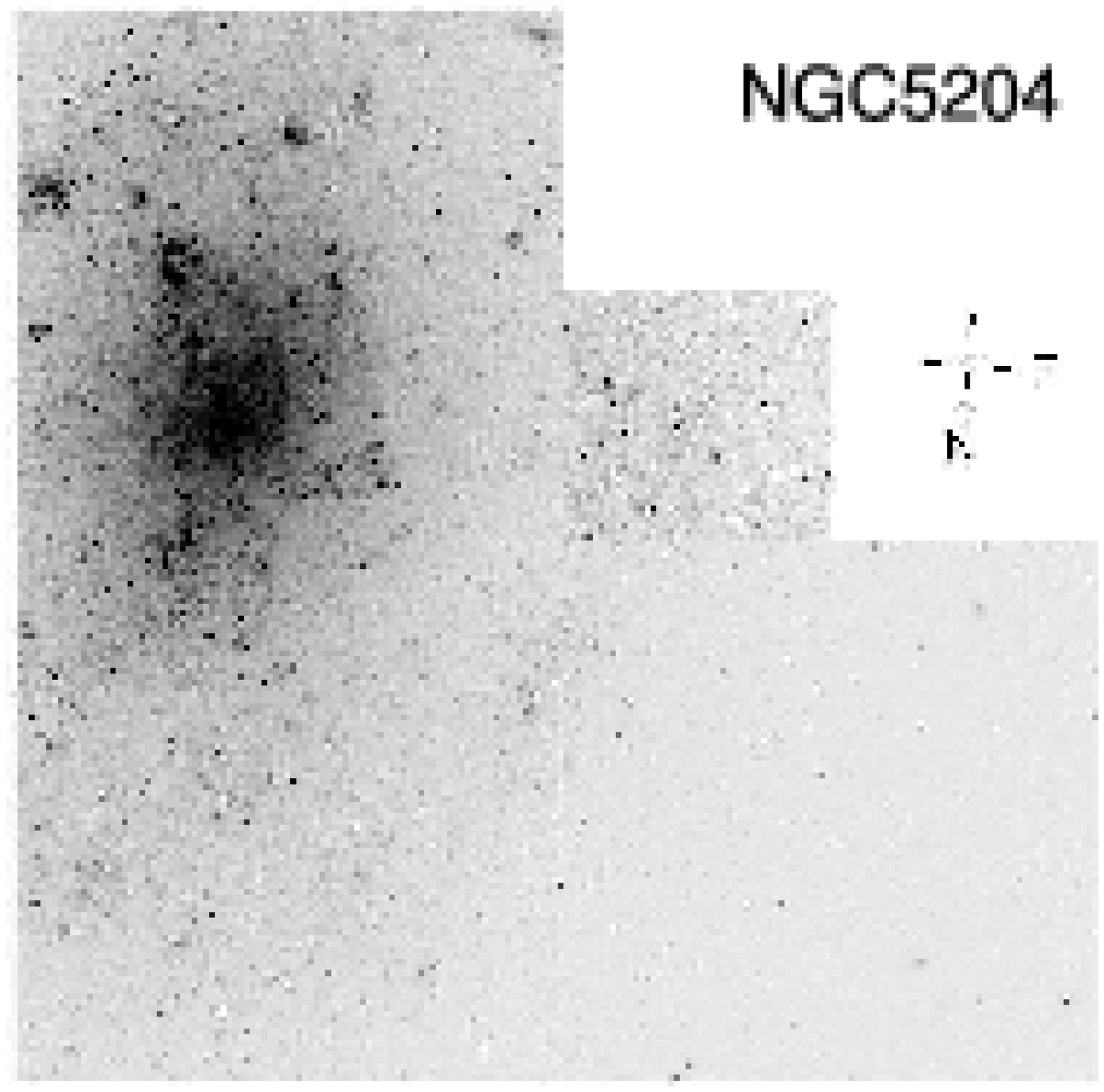}
\includegraphics[width=0.26\textwidth, bb = 74 310 518 752,clip]{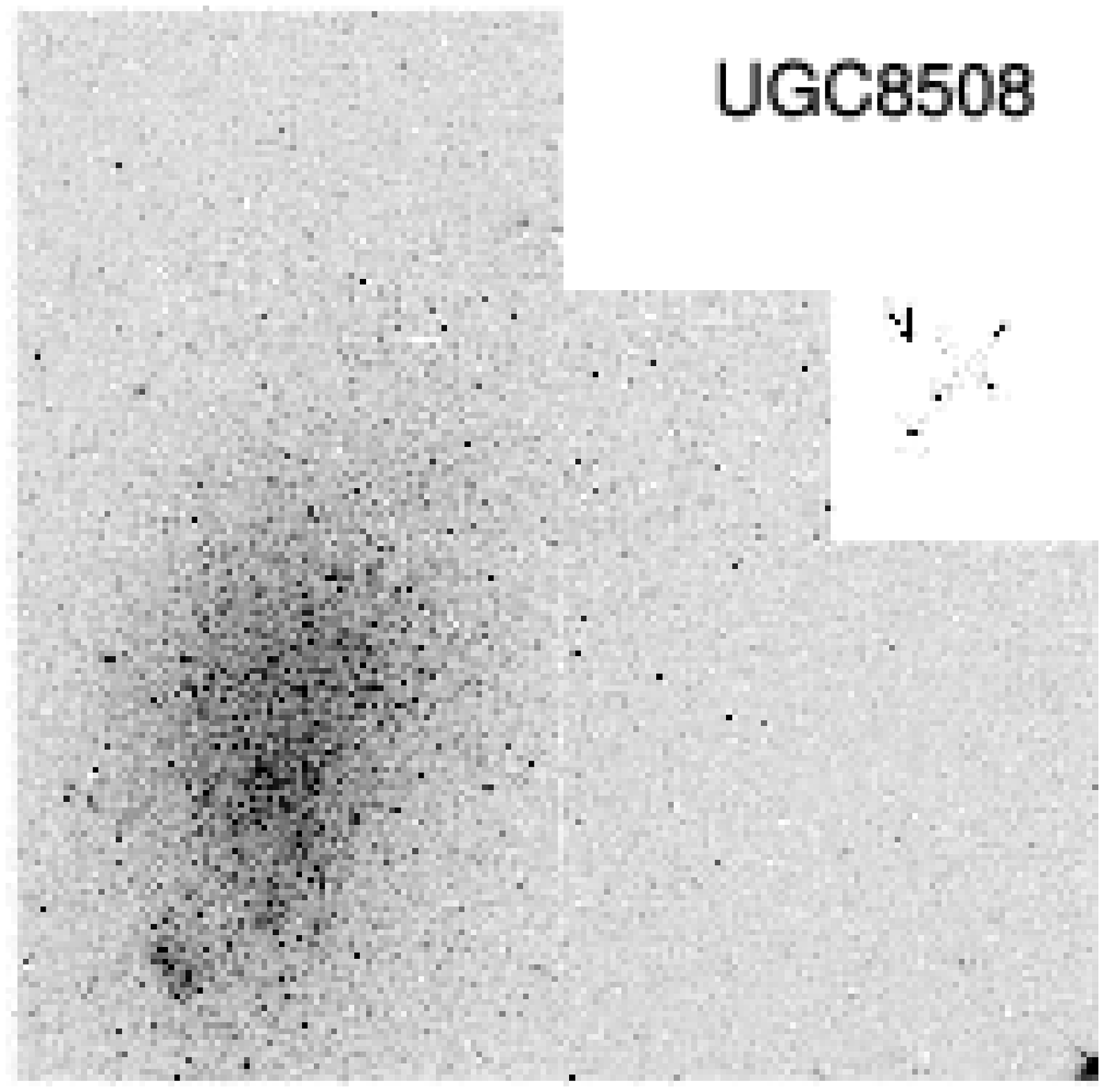}
\includegraphics[width=0.26\textwidth]{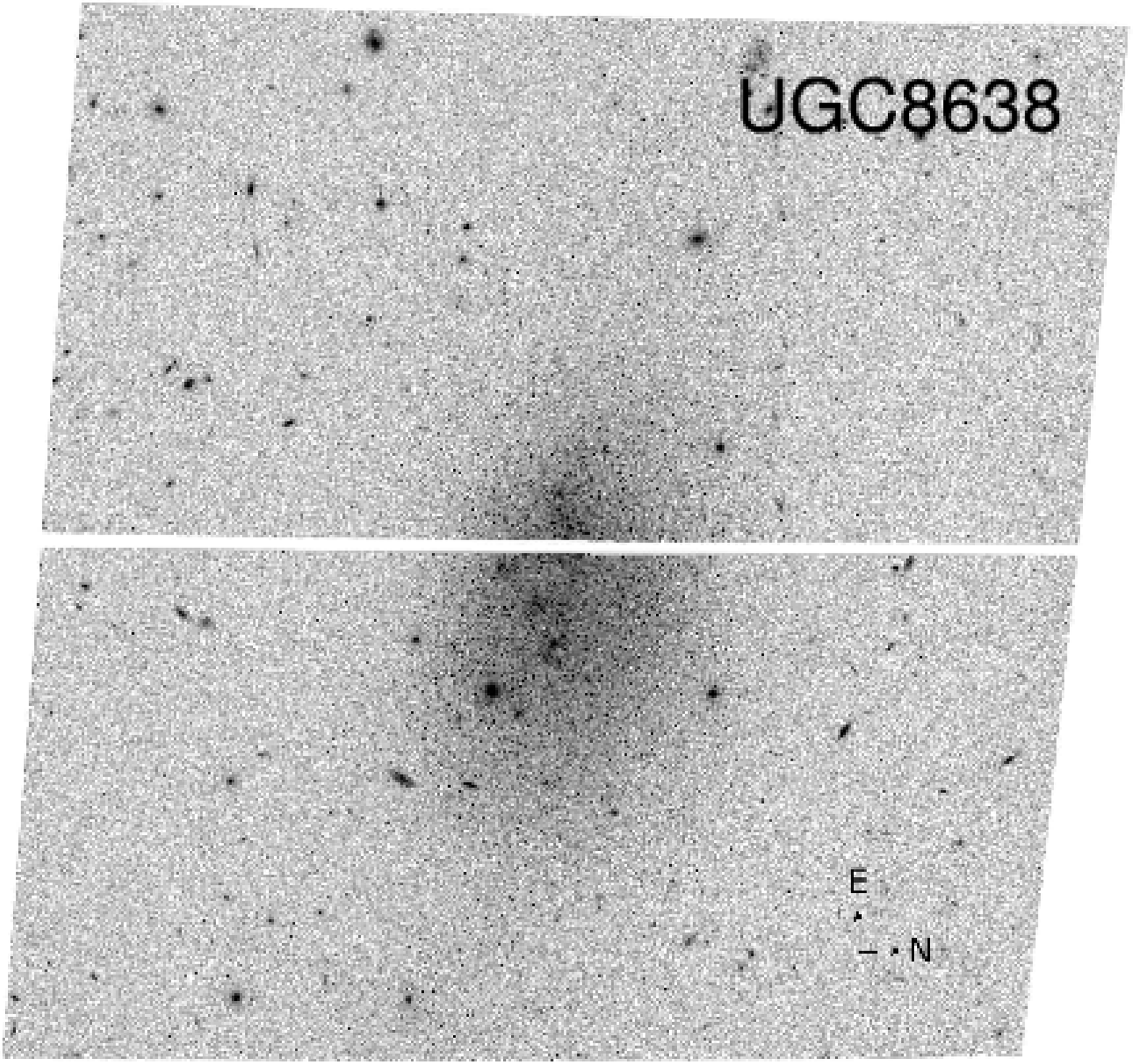}
\includegraphics[width=0.26\textwidth]{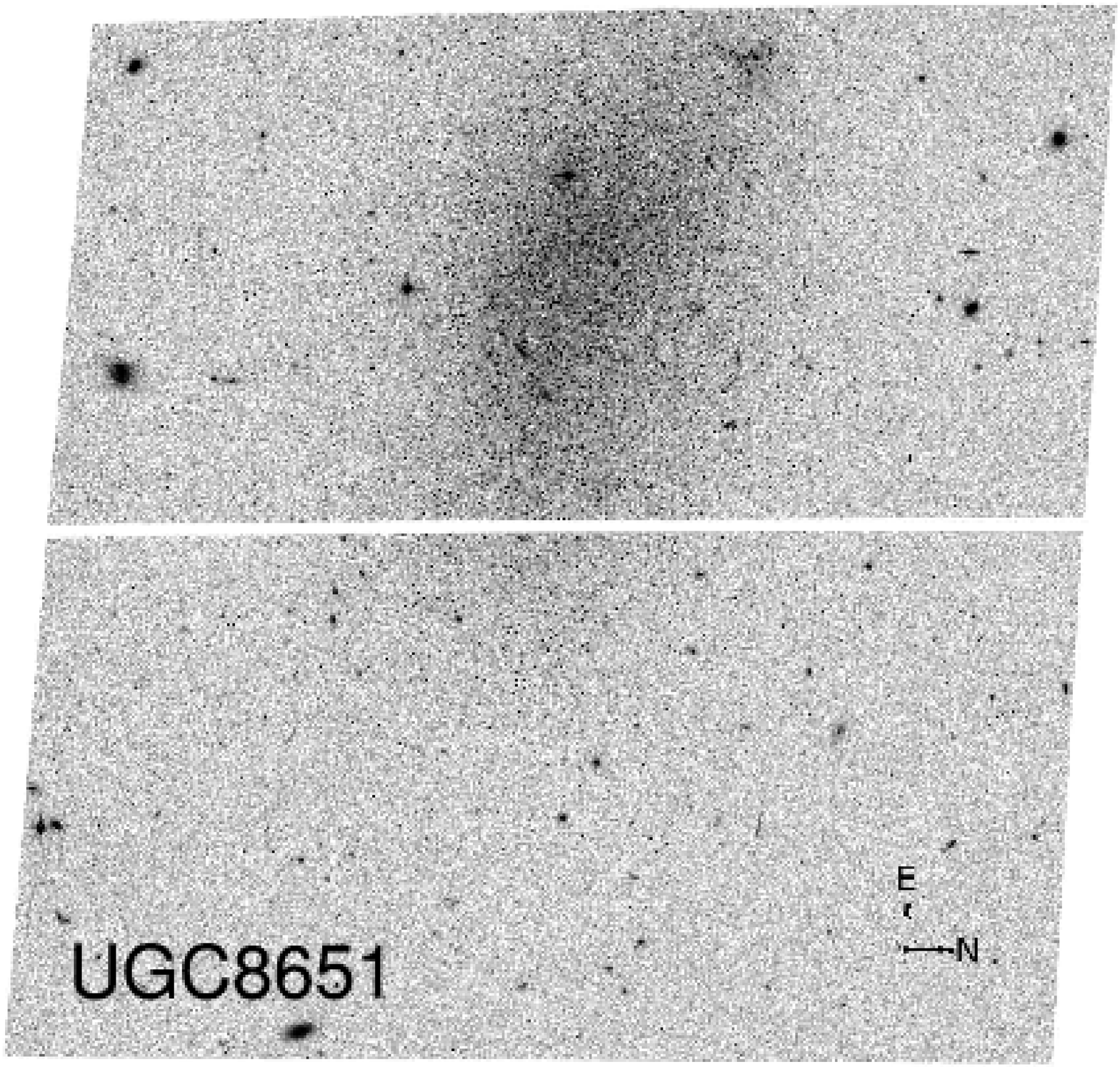}
\includegraphics[width=0.26\textwidth]{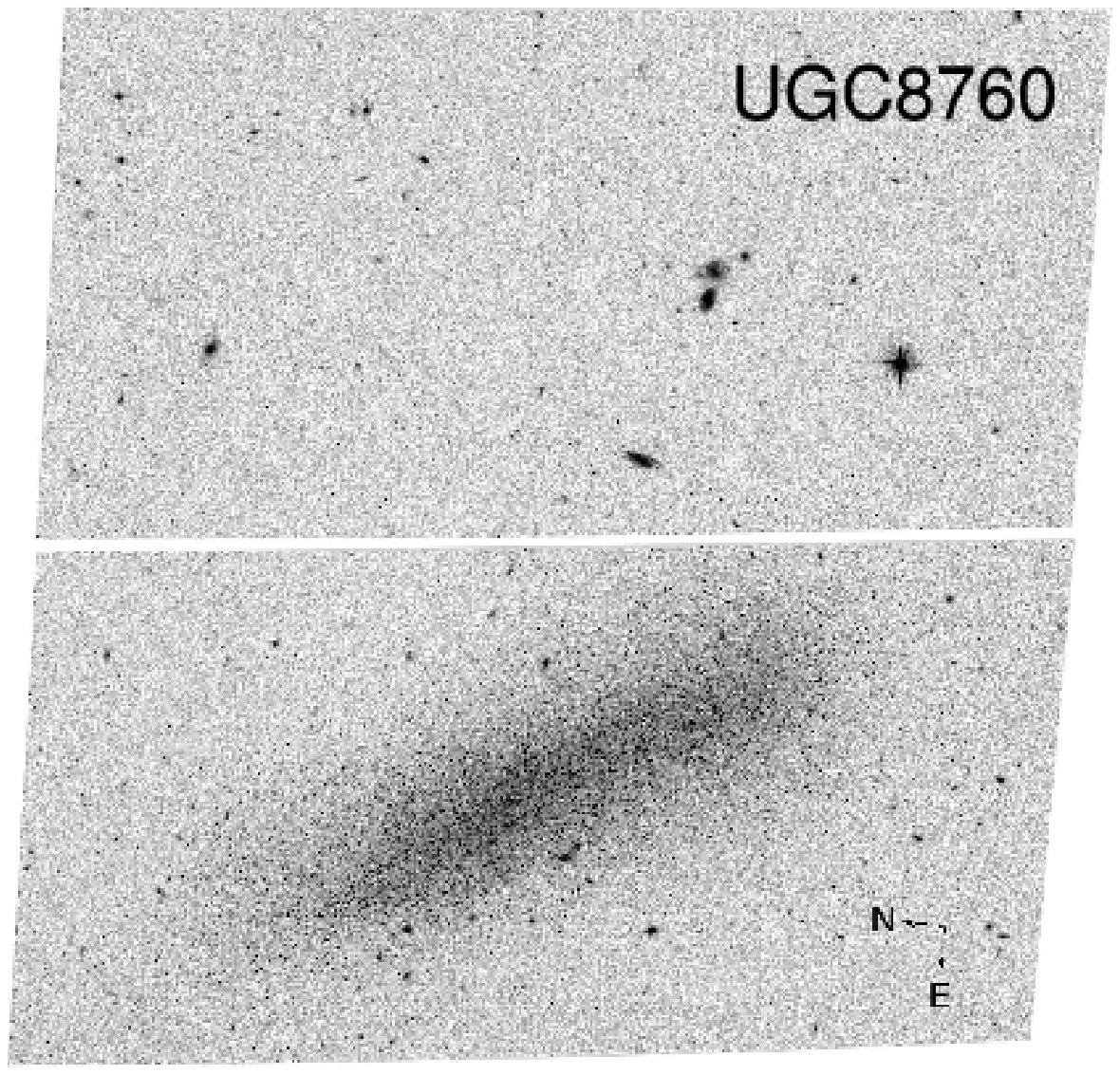}
\includegraphics[width=0.26\textwidth]{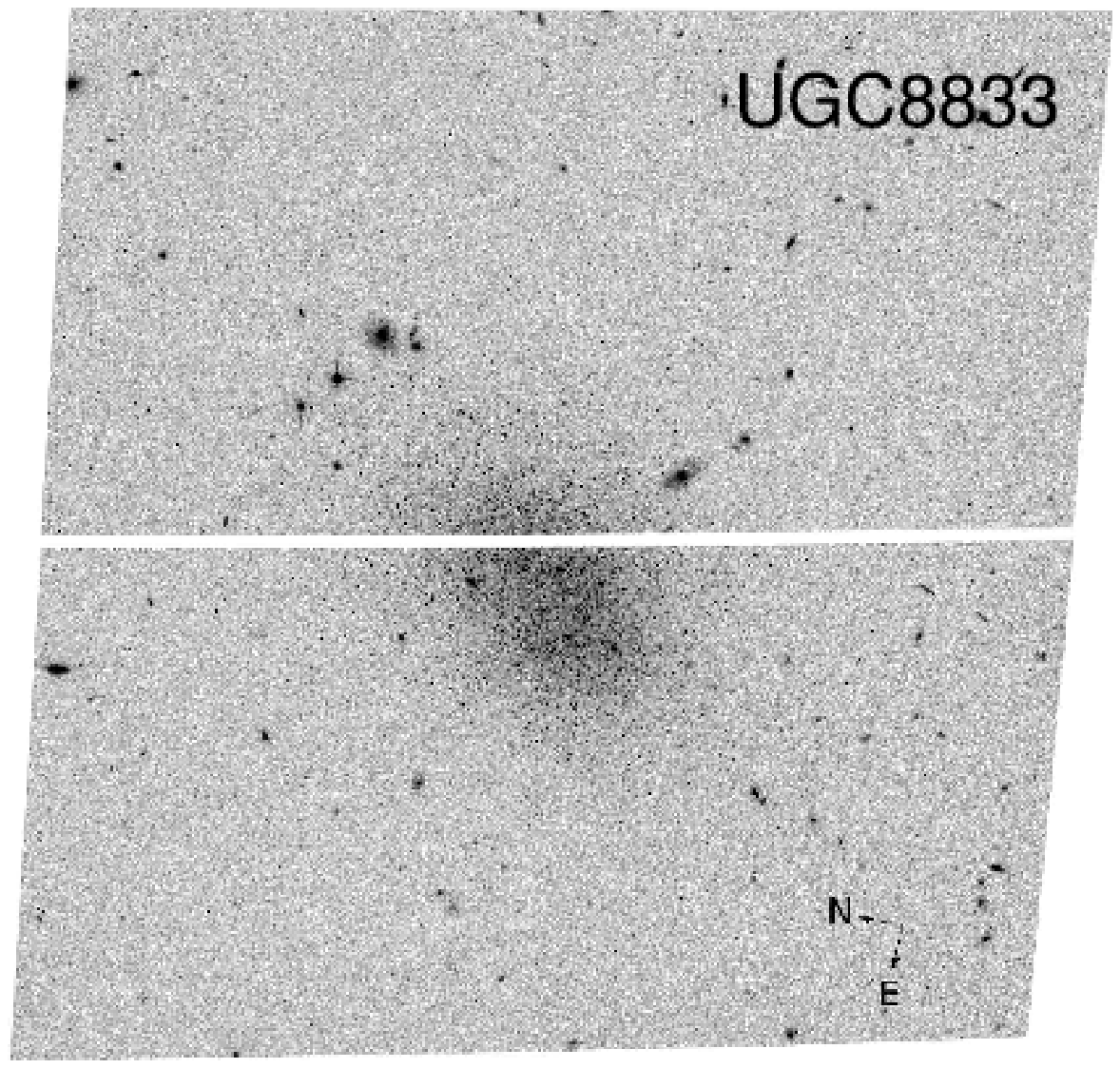}
\includegraphics[width=0.26\textwidth]{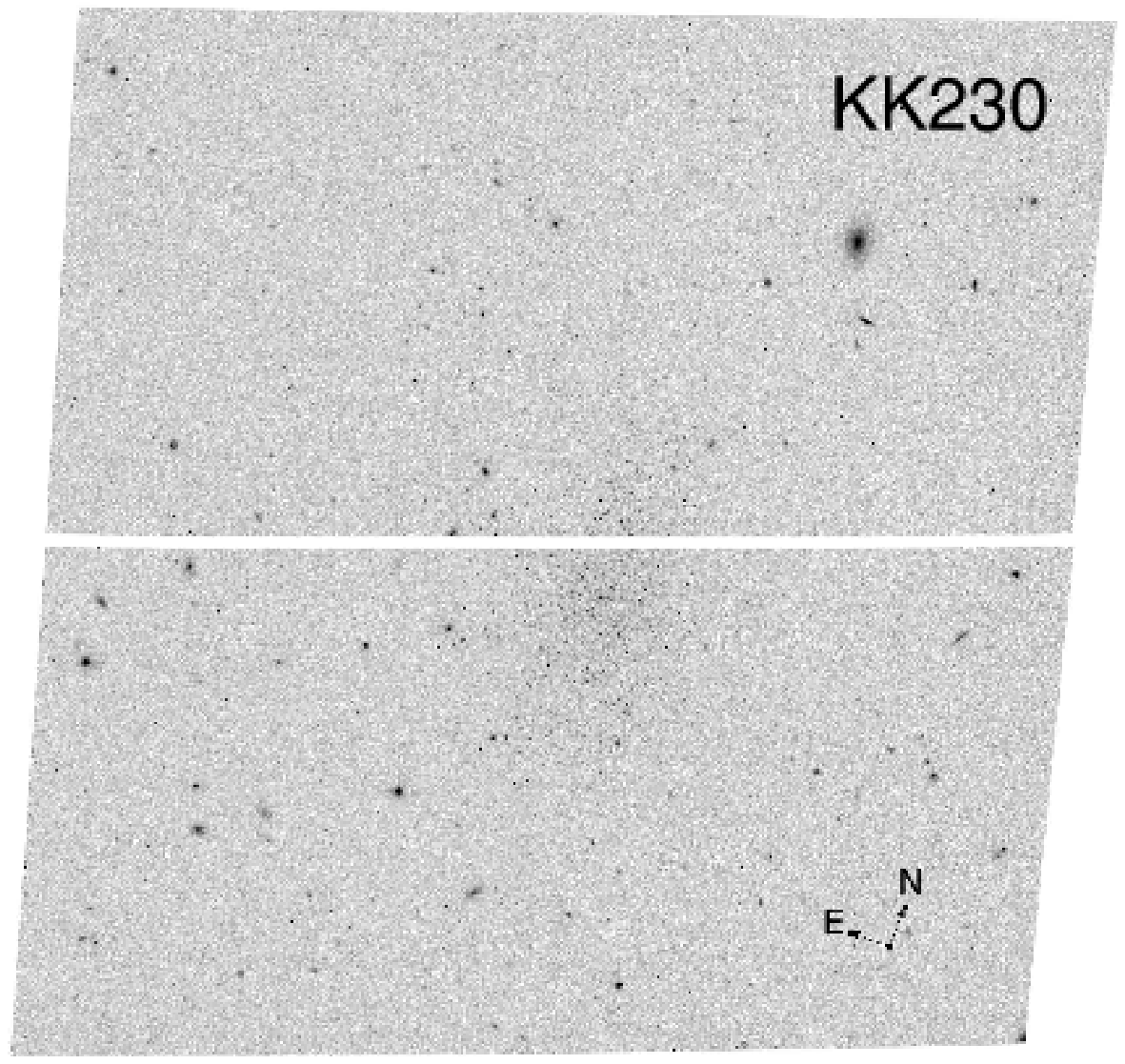}
\includegraphics[width=0.26\textwidth]{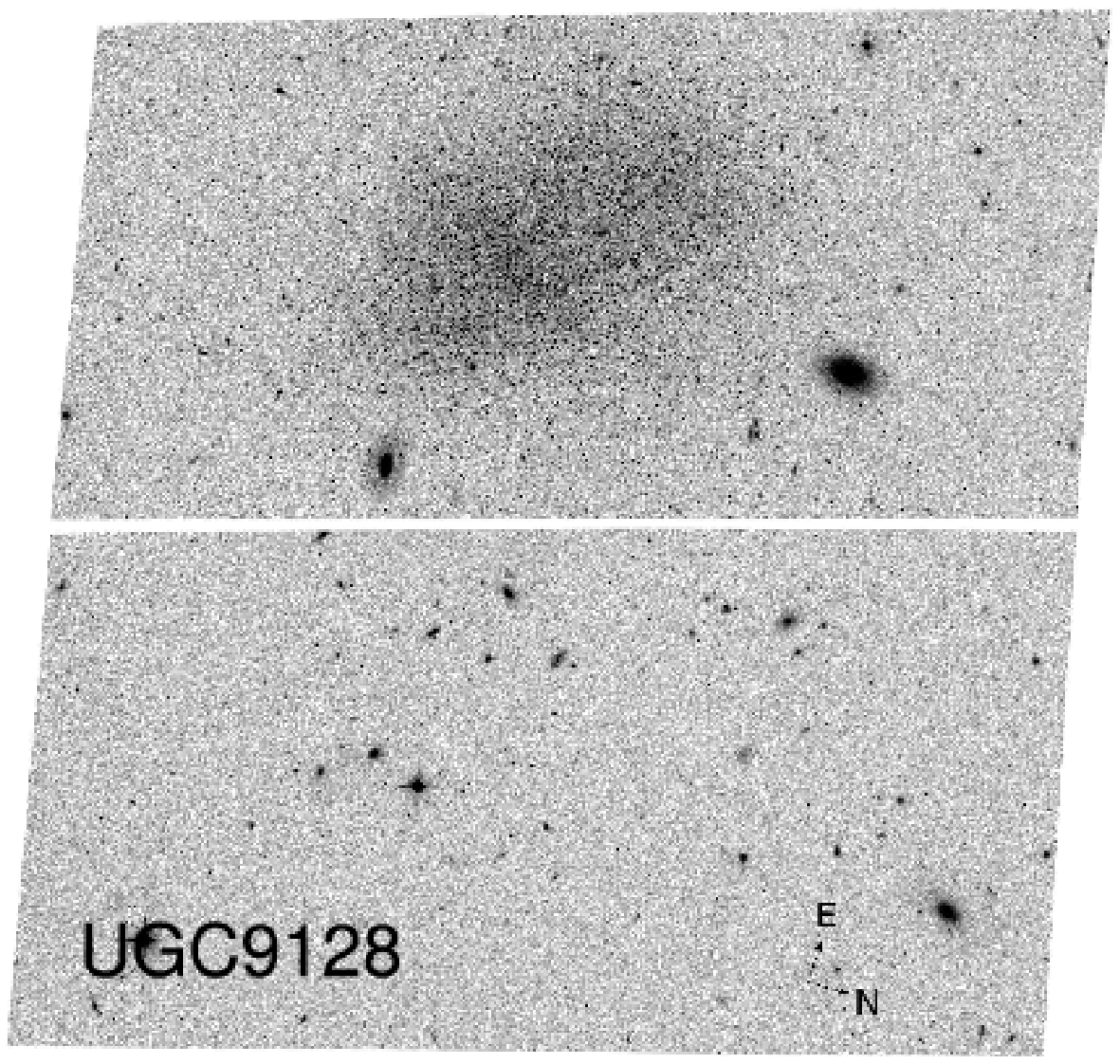}
\caption{(Contd.).} 
\end{figure*}

\newcommand{\refmag}{
$^a$Makarova, 1999~\cite{makarova1999:Makarov_n};
$^b$Taylor et al.,2005~\cite{taylor2005:Makarov_n};
$^c$Vaucouleurs et al., 1991~\cite{rc3:Makarov_n};
$^d$Makarova et al., 1998~\cite{makarova+1998:Makarov_n};
$^e$Makarova et al., 2009~\cite{makarova+2009:Makarov_n};
$^f$Karachentsev et al., 2004~\cite{cng:Makarov_n};
$^g$Jerjen et al., 2001~\cite{jerjen2001:Makarov_n};
$^h$Bremnes et al., 1999~\cite{bremnes1999:Makarov_n};
$^i$Karachentsev et al., 2013 (in press).}

\newcommand{\refvel}{
$^A$Begum et al., 2008~\cite{begum2008:Makarov_n};
$^B$Kova{\v c} et al., 2009~\cite{kovac2009:Makarov_n};
$^C$Springob et al., 2005~\cite{springob2005:Makarov_n};
$^D$Stil and  Israel, 2002~\cite{stil2002:Makarov_n};
$^E$Huchtmeier and Richter, 1986~\cite{huchtmeier1986:Makarov_n};
$^F$Huchtmeier and Seiradakis, 1985~\cite{huchtmeier1985:Makarov_n};
$^G$Tifft and Cocke, 1988~\cite{tifft1988:Makarov_n};
$^H$Simpson and Gottesman, 2000~\cite{simpson2000:Makarov_n}. }

The photometry of resolved stars in the galaxies was performed
using the HSTphot~\cite{dolphin00:Makarov_n} and
DOLPHOT~\cite{dolphin02:Makarov_n} specialized software packages,
developed for the PSF photometry in a crowded stellar field
obtained with the WFPC2/HST and ACS/HST. The procedures of
photometric reduction included masking the ``bad'' columns and
pixels, removing the traces of cosmic ray particles from the
images, and the simultaneous PSF photometry of the detected stars
in two filters using the recommended
parameters~\cite{dolphin00:Makarov_n, dolphin02:Makarov_n}. Only
the stars with photometry satisfying a number of quality criteria
were used for the further measurements and analysis. Specifically,
we selected the stars with the signal-to-noise ratio of \mbox{$S/N
\ge5$}  in both filters, \mbox{$\chi^2\le2.5$} and \mbox{$\vert
\rm{sharp} \vert \le 0.3$.}

\begin{figure*}
\vbox{
\includegraphics[height=0.4\textwidth]{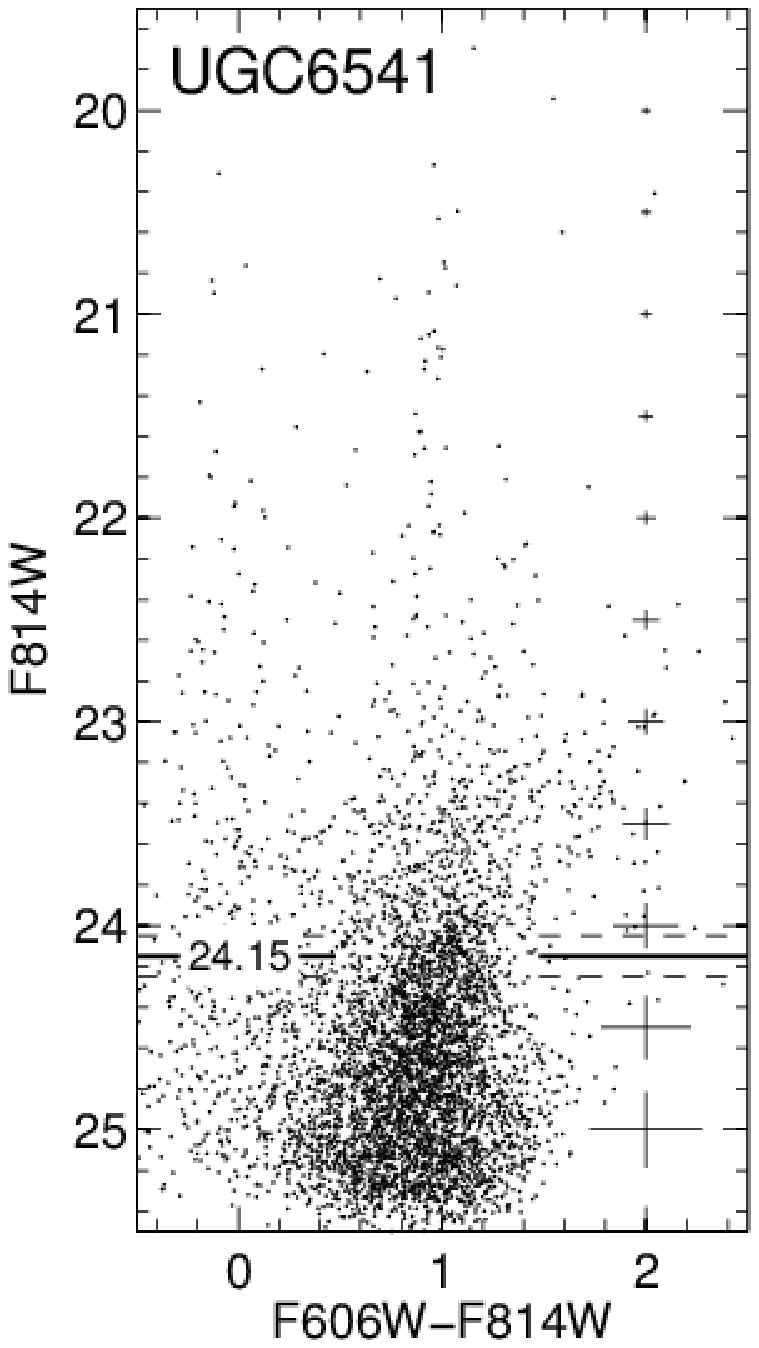}
\hspace{1mm}
\includegraphics[height=0.4\textwidth]{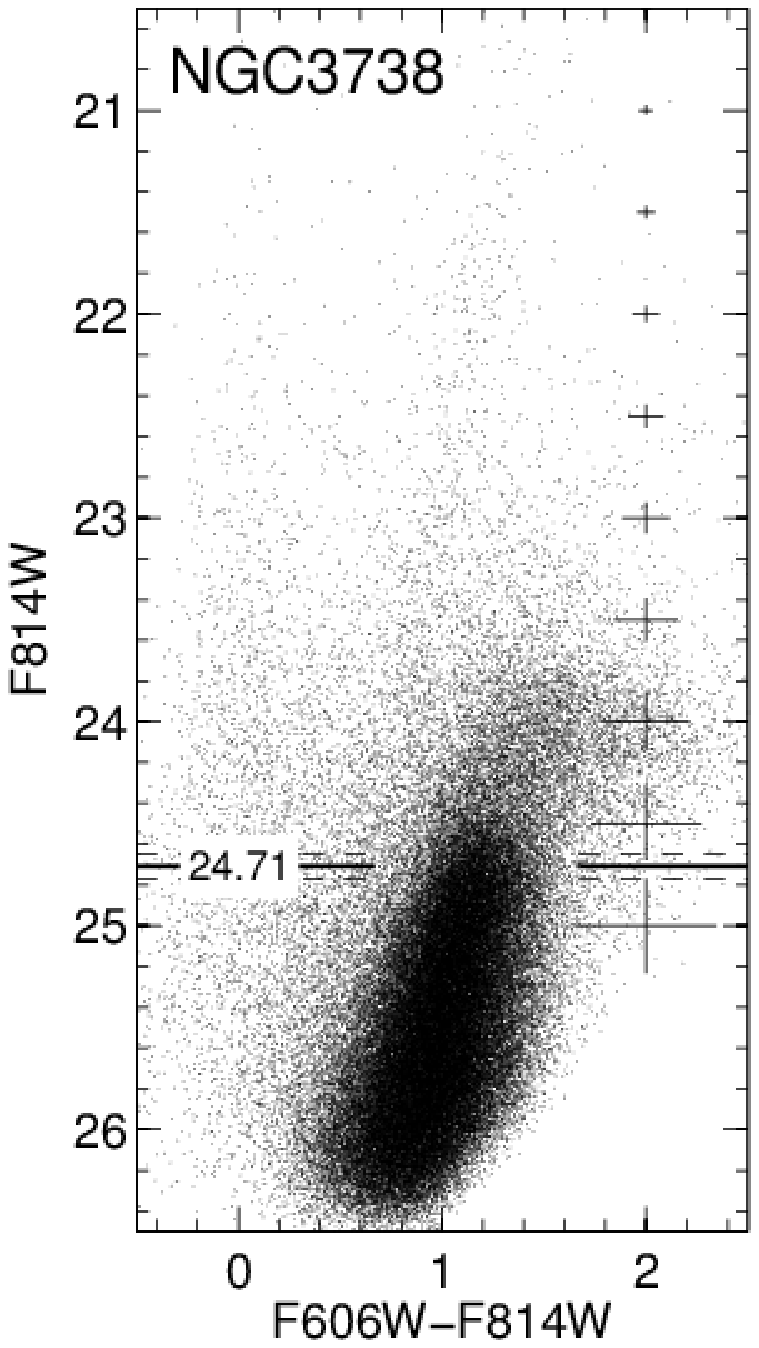}
\hspace{1mm}
\includegraphics[height=0.4\textwidth]{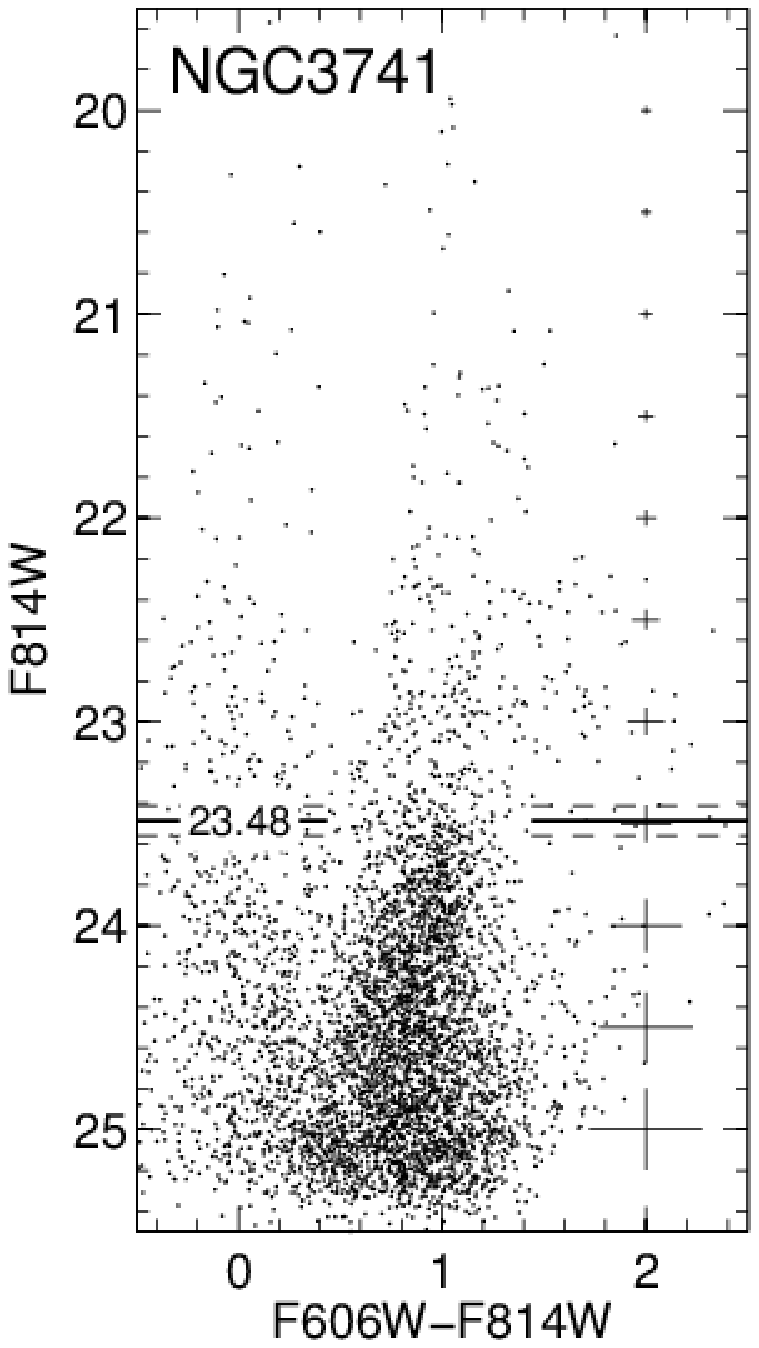}
\hspace{1mm}
\includegraphics[height=0.4\textwidth]{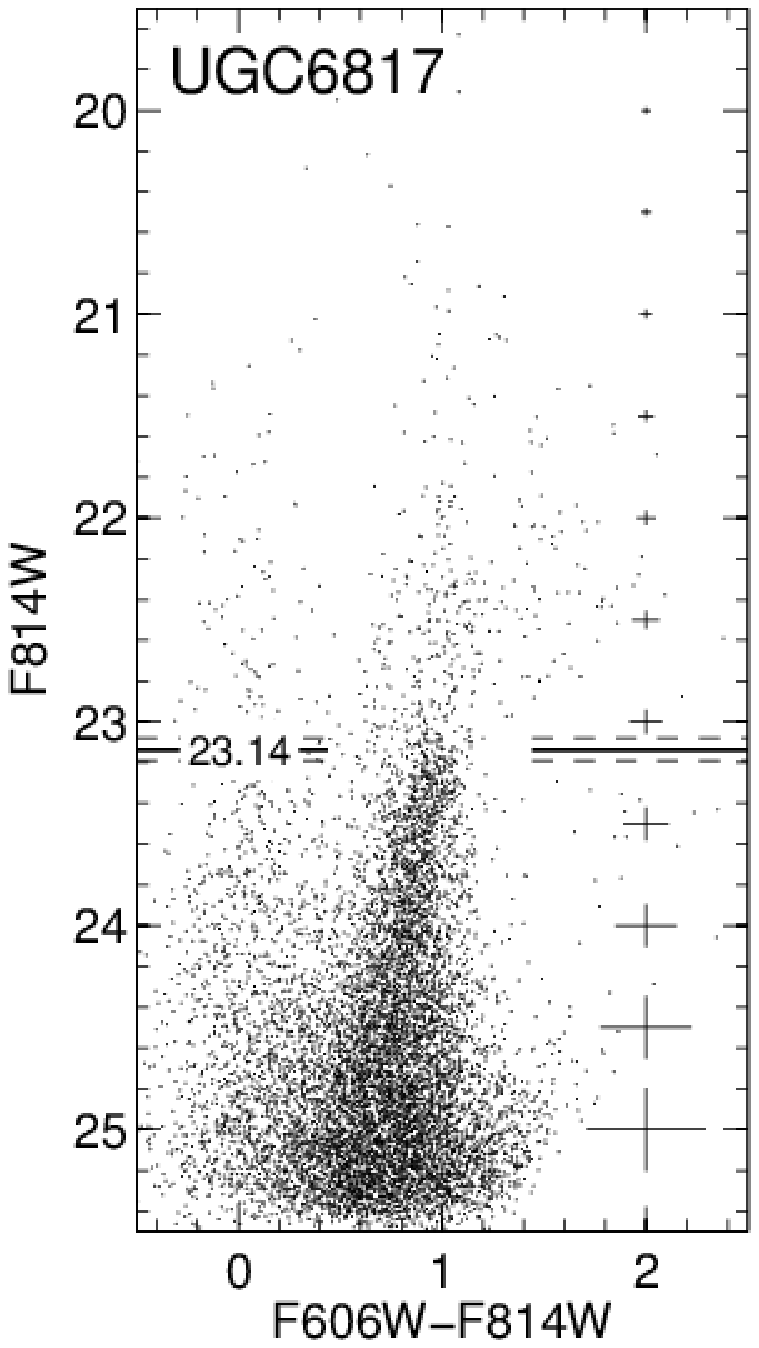}
} \vspace{2mm} \vbox{
\includegraphics[height=0.4\textwidth]{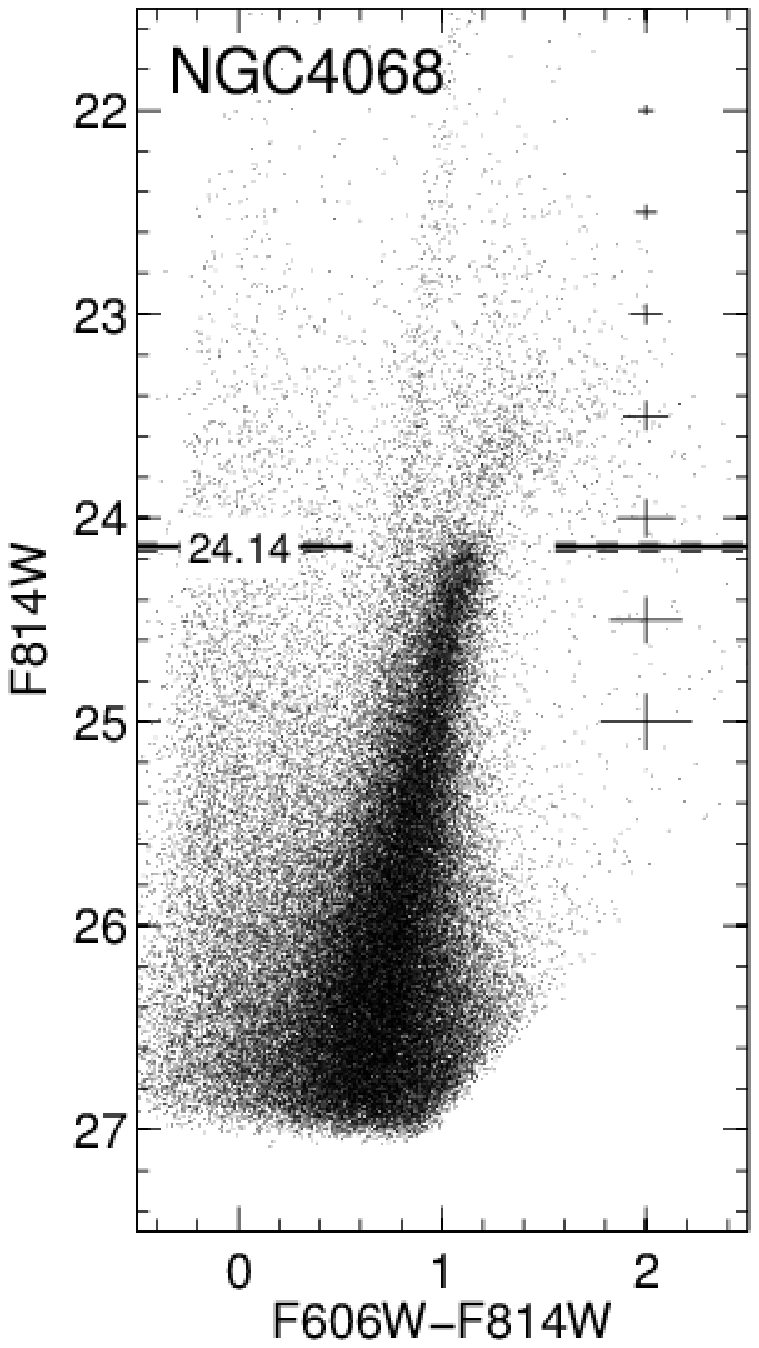}
\hspace{1mm}
\includegraphics[height=0.4\textwidth]{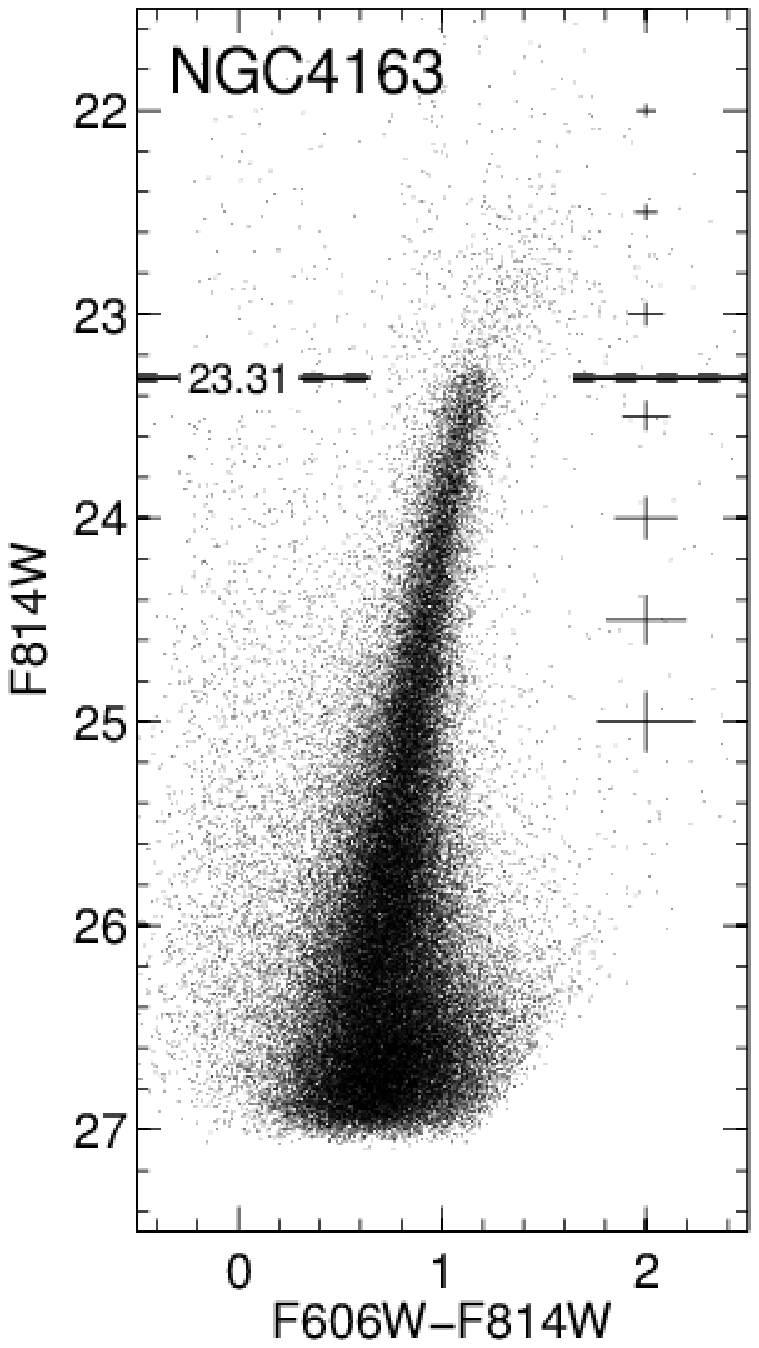}
\hspace{1mm}
\includegraphics[height=0.4\textwidth]{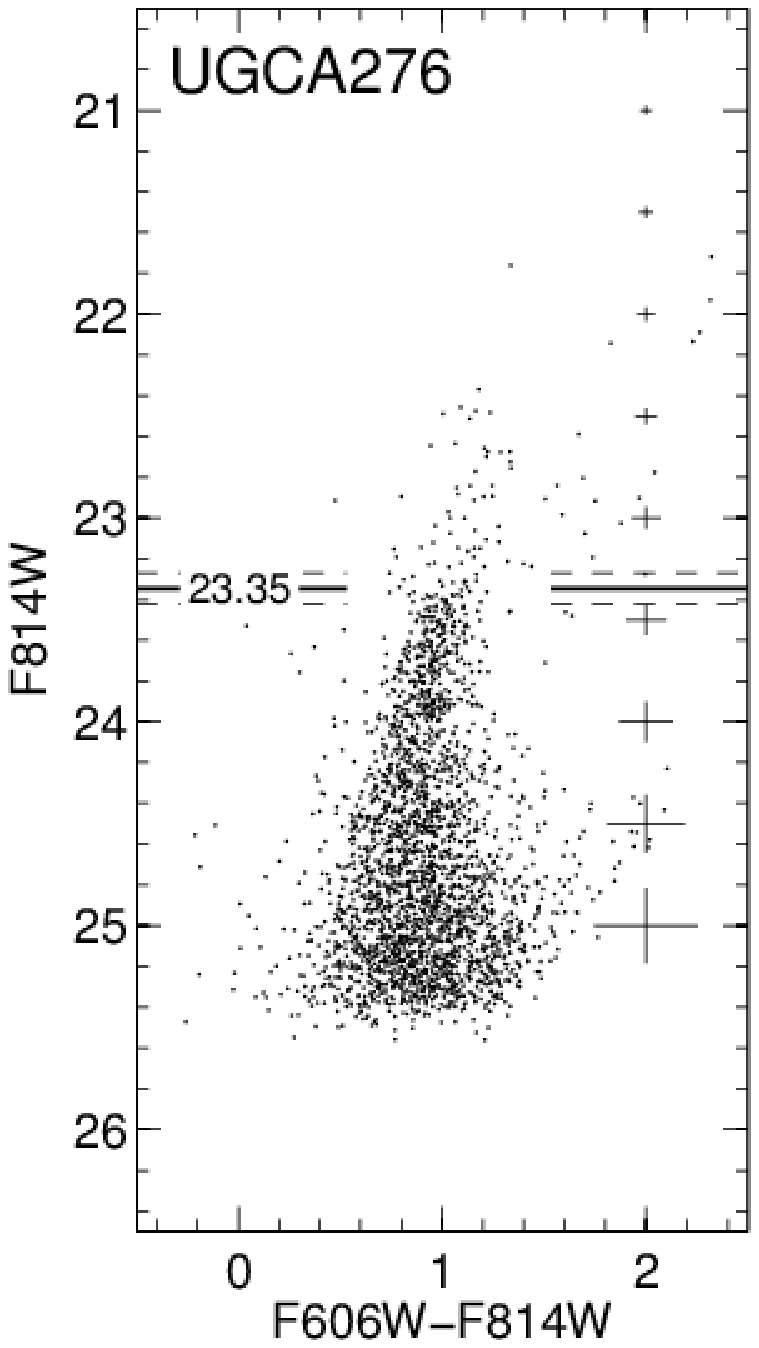}
\hspace{1mm}
\includegraphics[height=0.4\textwidth]{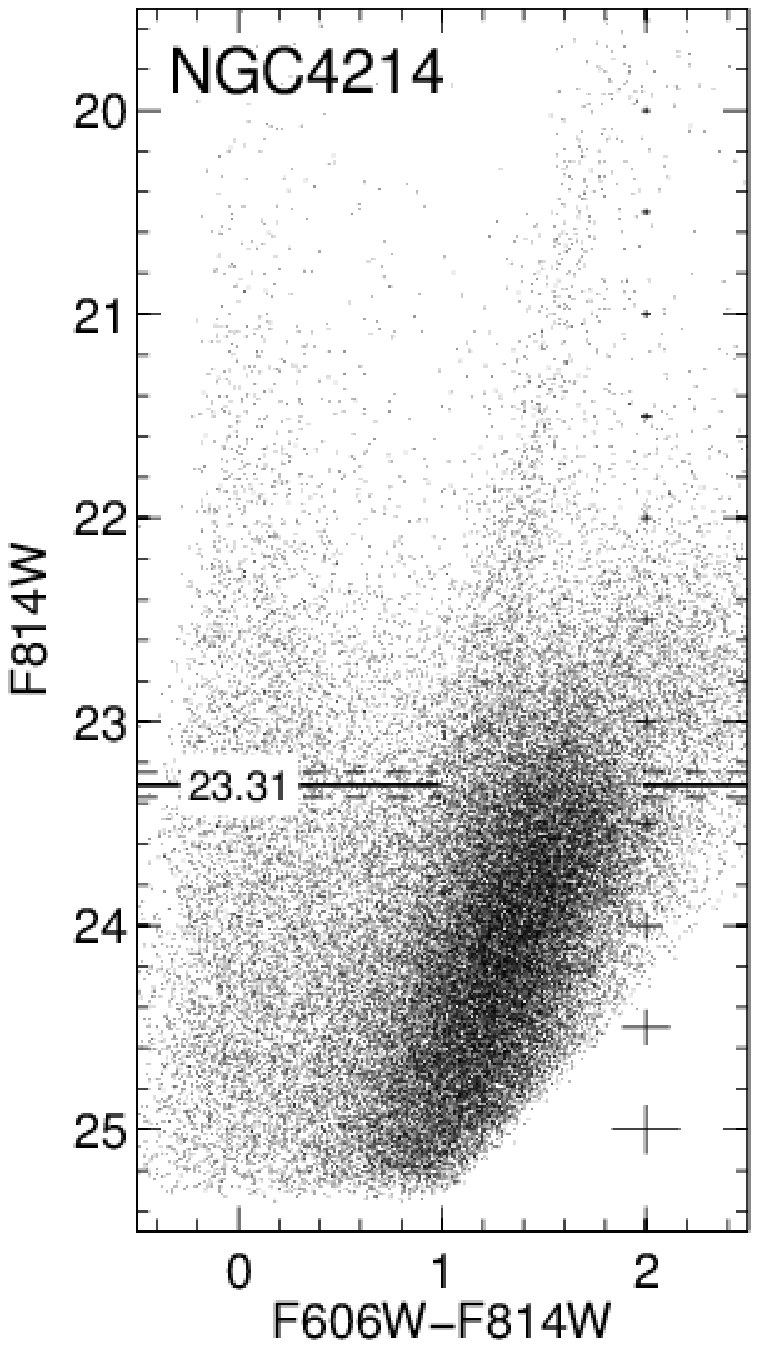}
} \vspace{2mm} \vbox{
\includegraphics[height=0.4\textwidth]{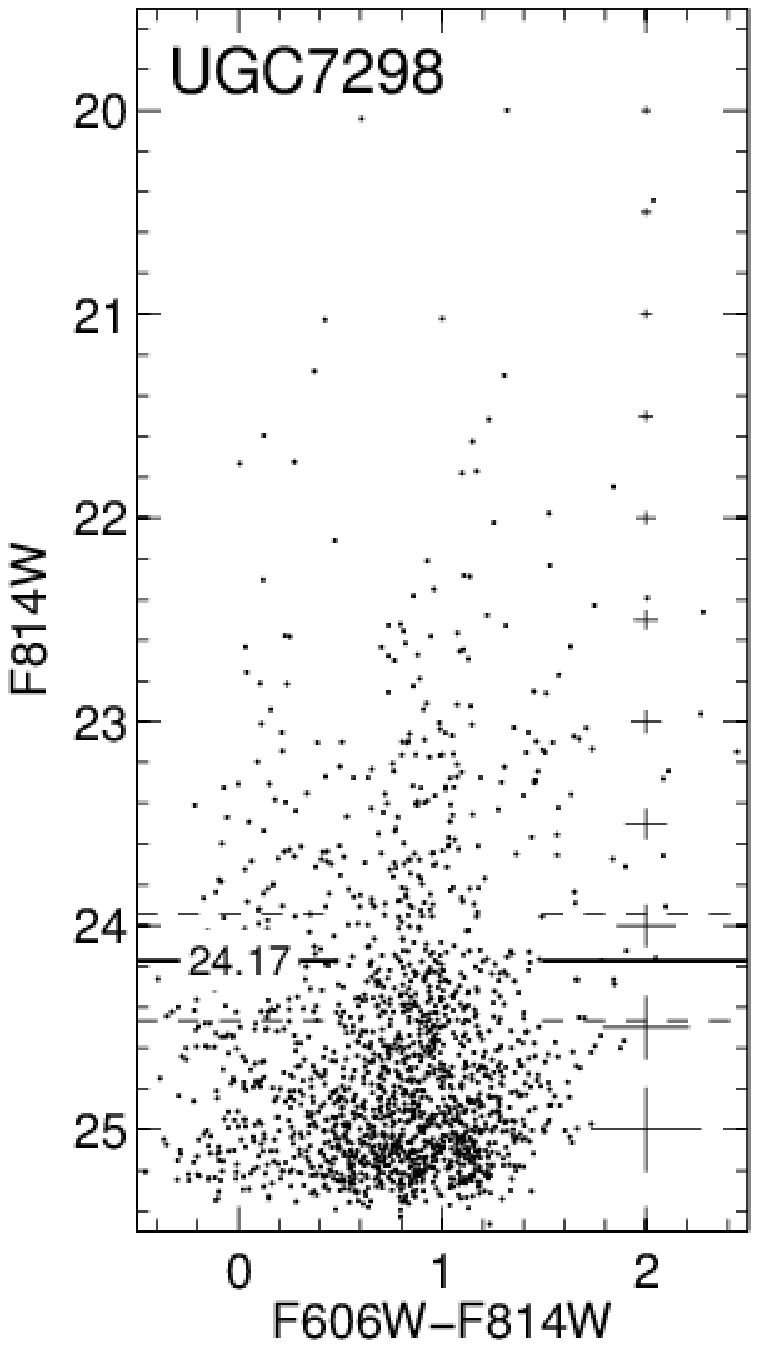}
\hspace{1mm}
\includegraphics[height=0.4\textwidth]{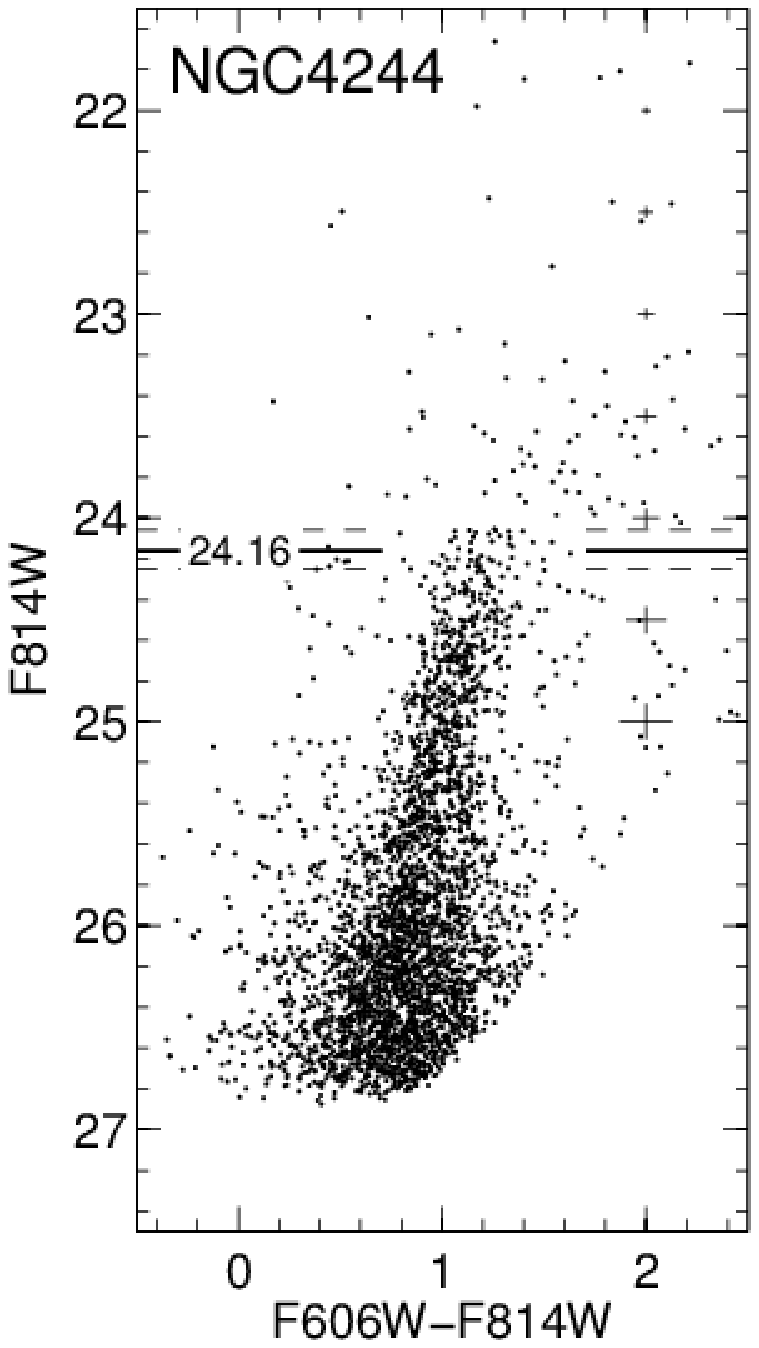}
\hspace{1mm}
\includegraphics[height=0.4\textwidth]{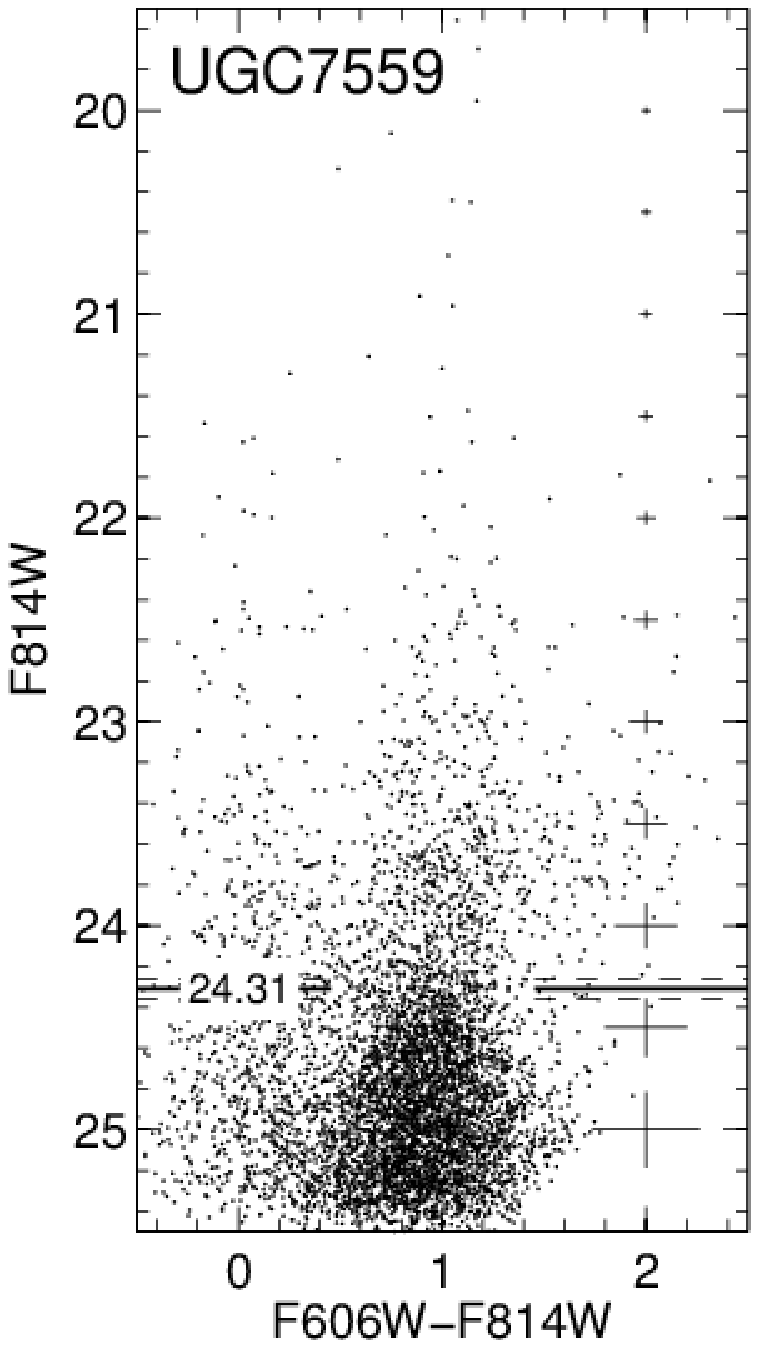}
\hspace{1mm}
\includegraphics[height=0.4\textwidth]{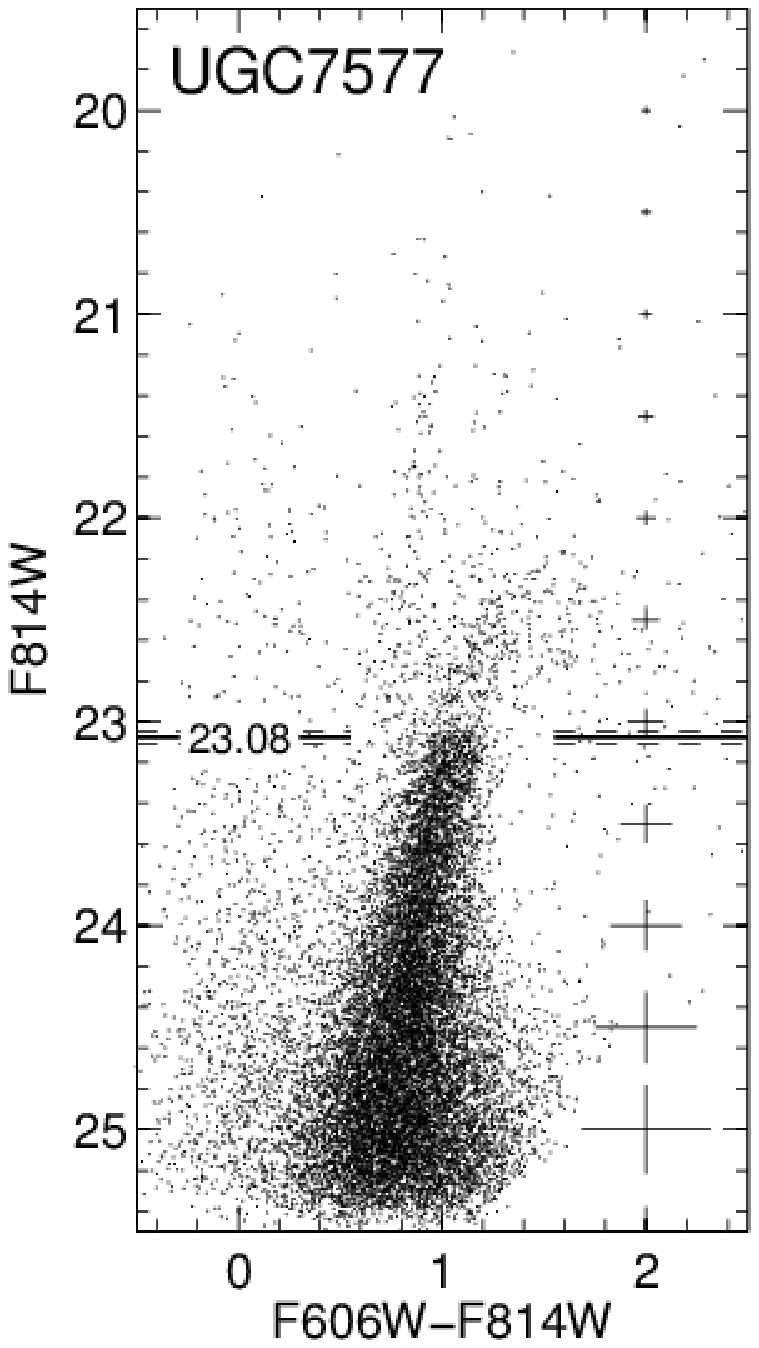}
} \caption{The color--magnitude diagrams  for the stars  of the
Canes Venatici I cloud galaxies. The crosses indicate the typical
errors of stellar photometry. The TRGB position is indicated by
the straight line, the corresponding measurement errors are
depicted by the dashed lines.} \label{fig:cmd1:Makarov_n}
\end{figure*}

\begin{figure*}
\addtocounter{figure}{-1} 
\vbox{
\includegraphics[height=0.4\textwidth]{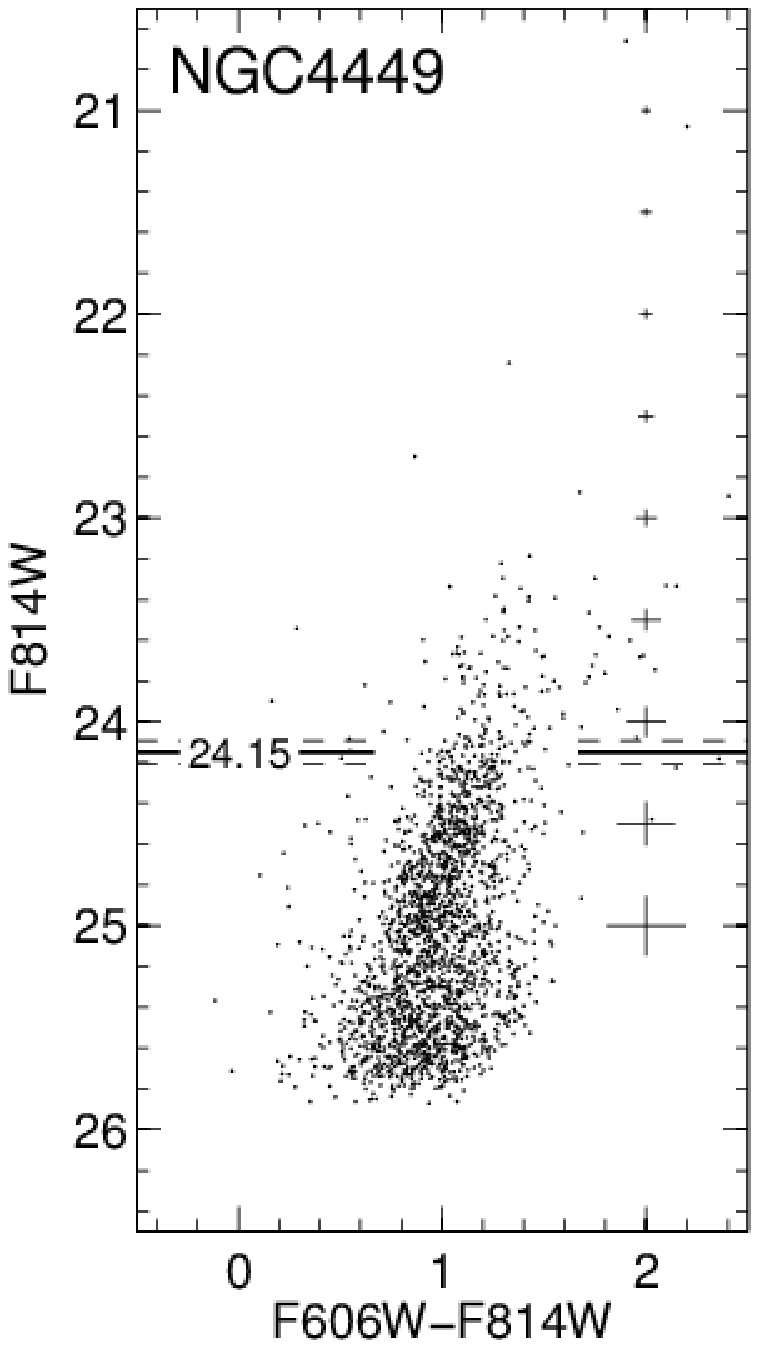}
\hspace{1mm}
\includegraphics[height=0.4\textwidth]{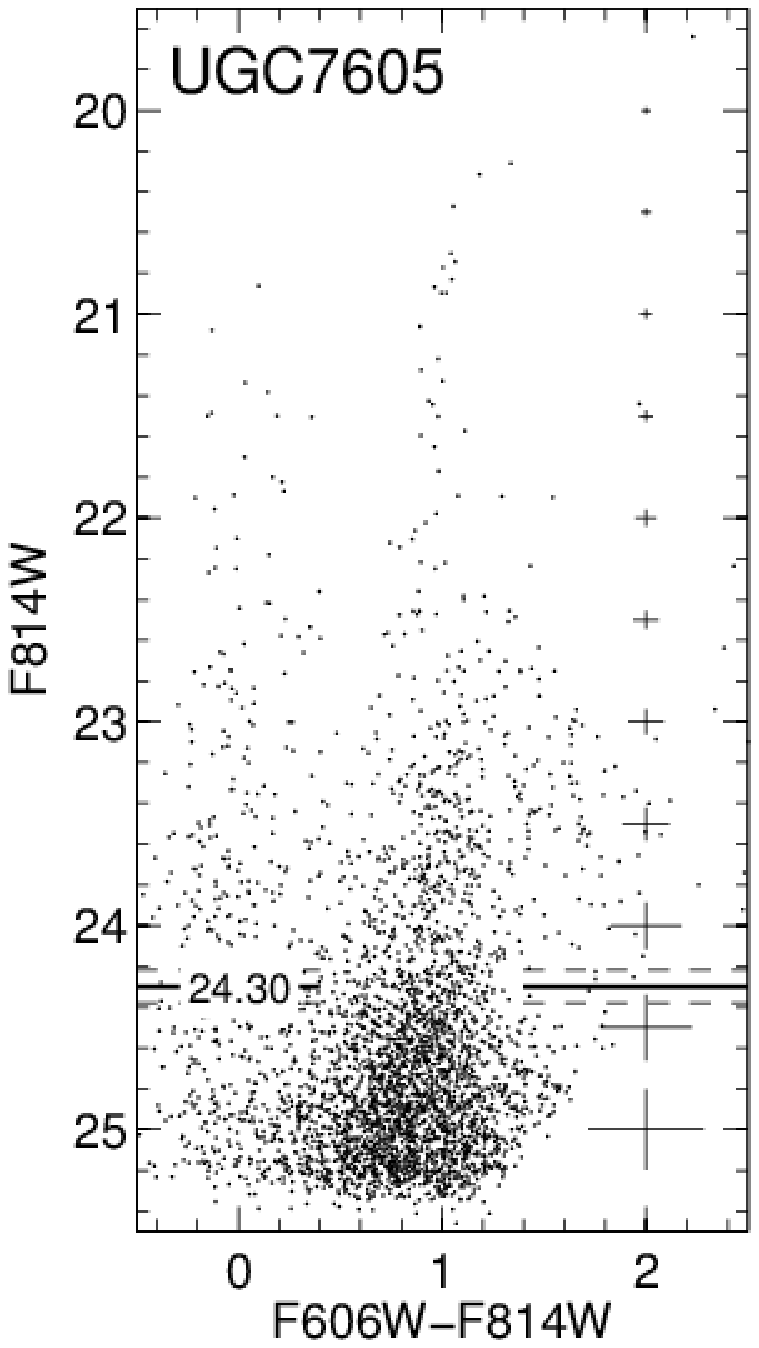}
\hspace{1mm}
\includegraphics[height=0.4\textwidth]{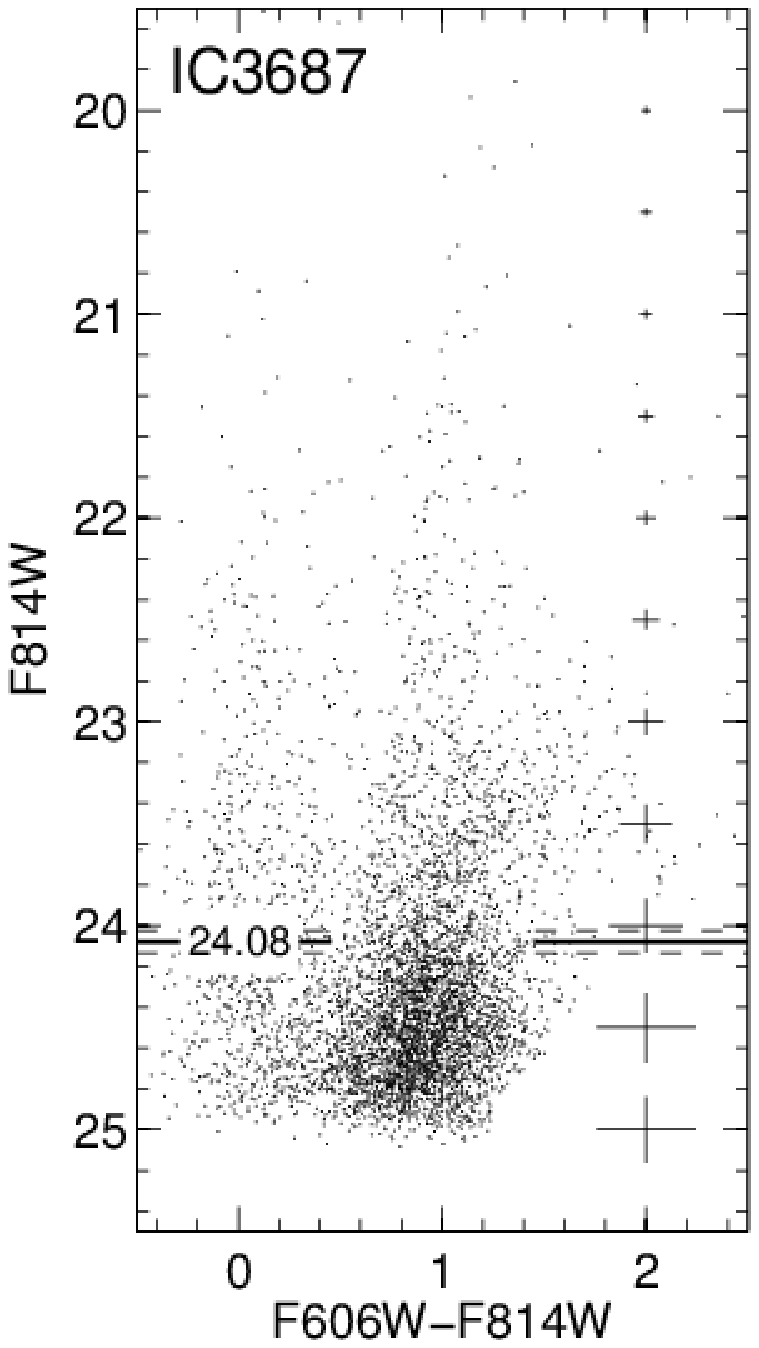}
\hspace{1mm}
\includegraphics[height=0.4\textwidth]{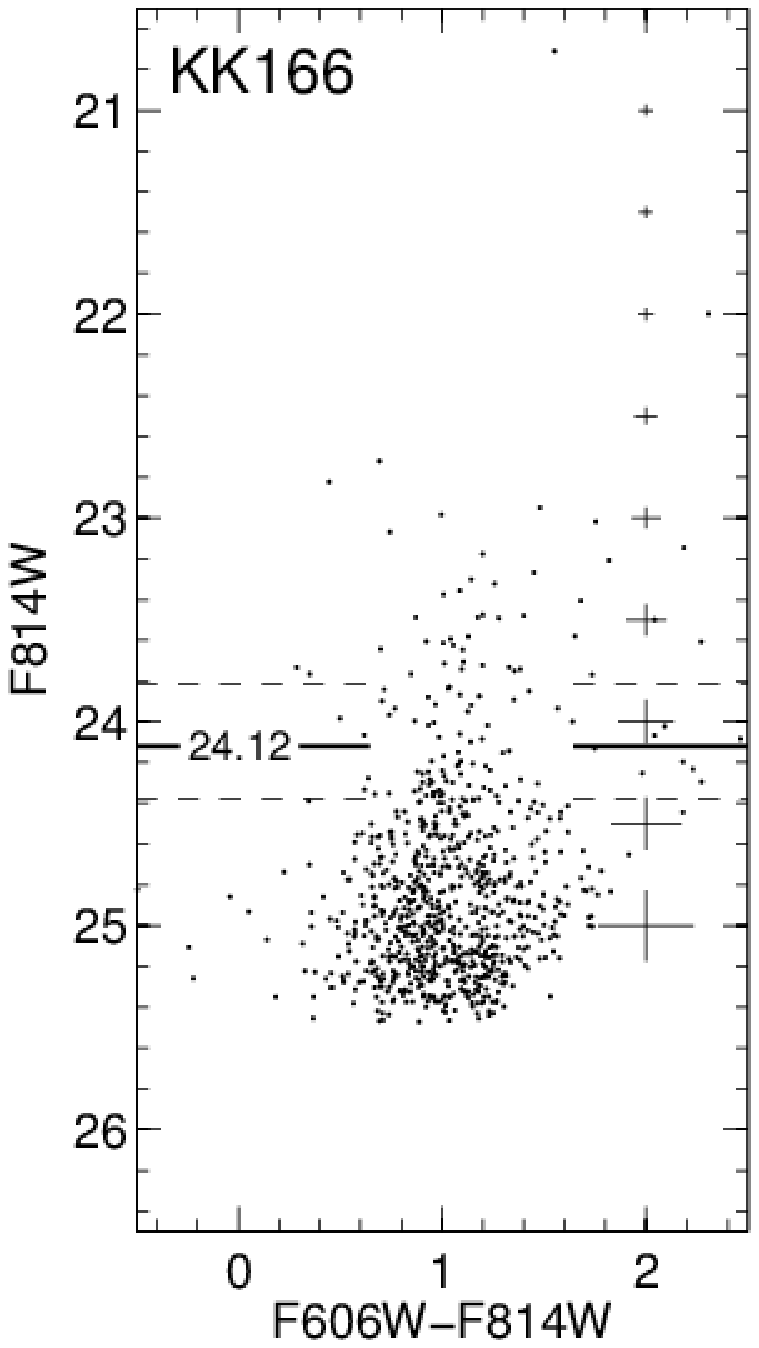}
} \vspace{2mm} \vbox{
\includegraphics[height=0.4\textwidth]{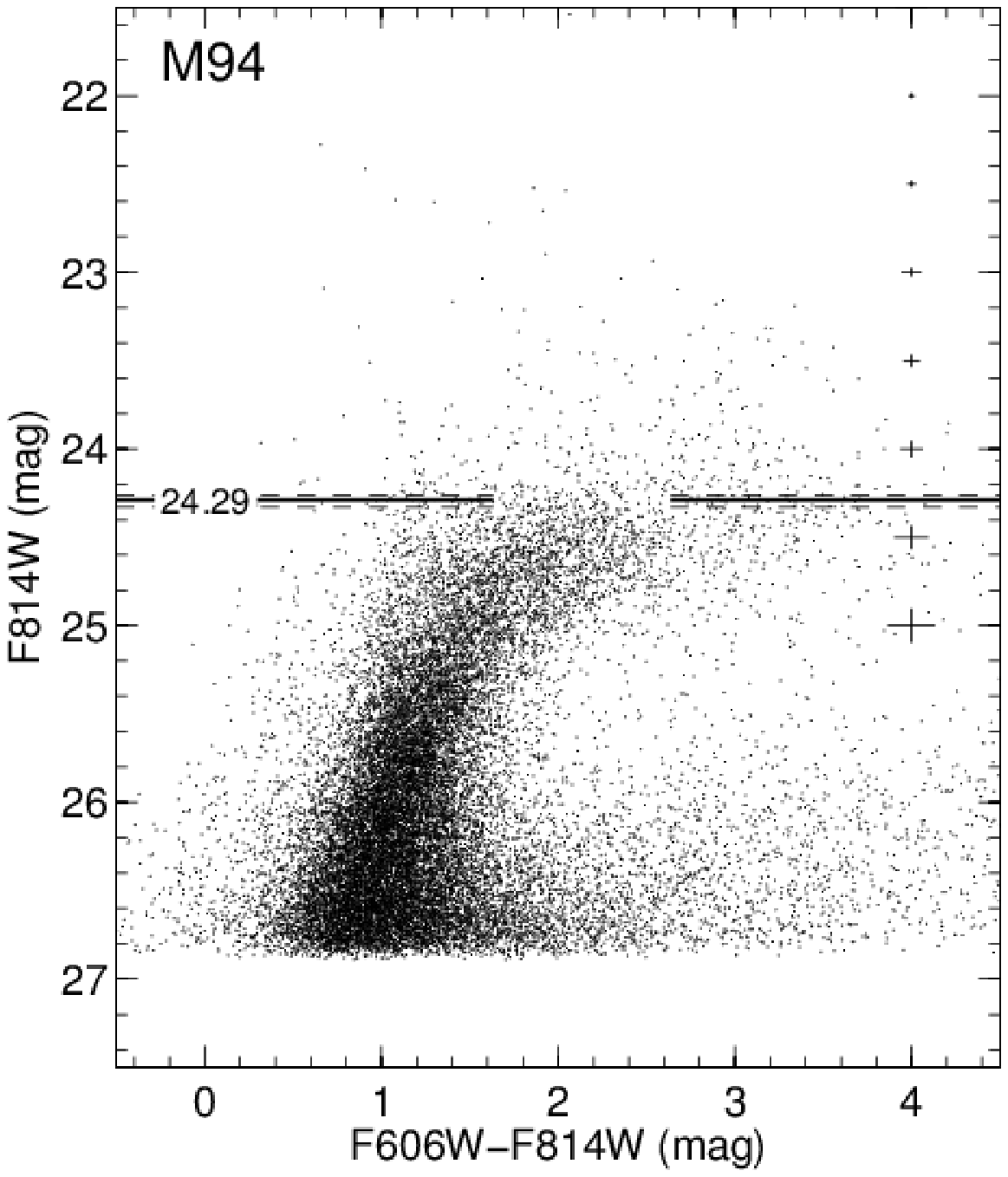}
\hspace{1mm}
\includegraphics[height=0.4\textwidth]{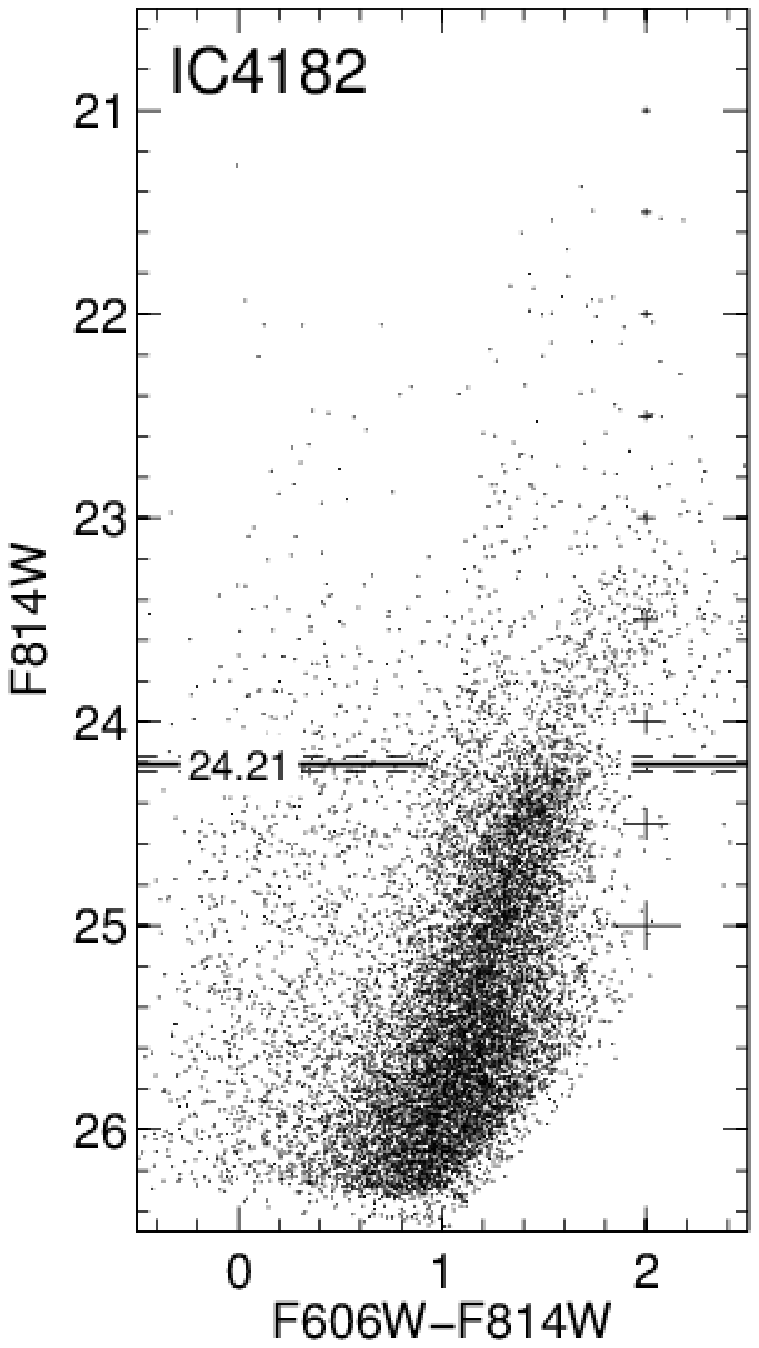}
\hspace{1mm}
\includegraphics[height=0.4\textwidth]{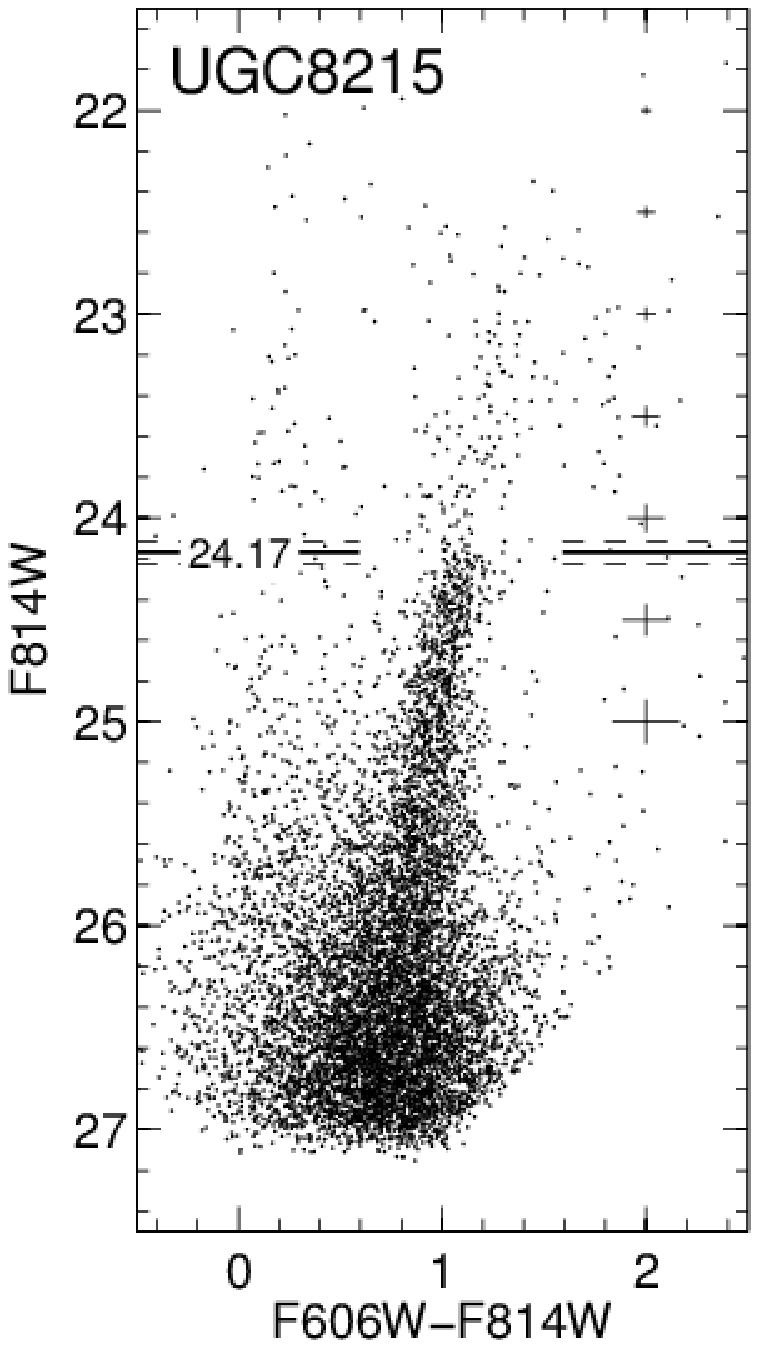}
} \vspace{2mm}\vbox{
\includegraphics[height=0.4\textwidth]{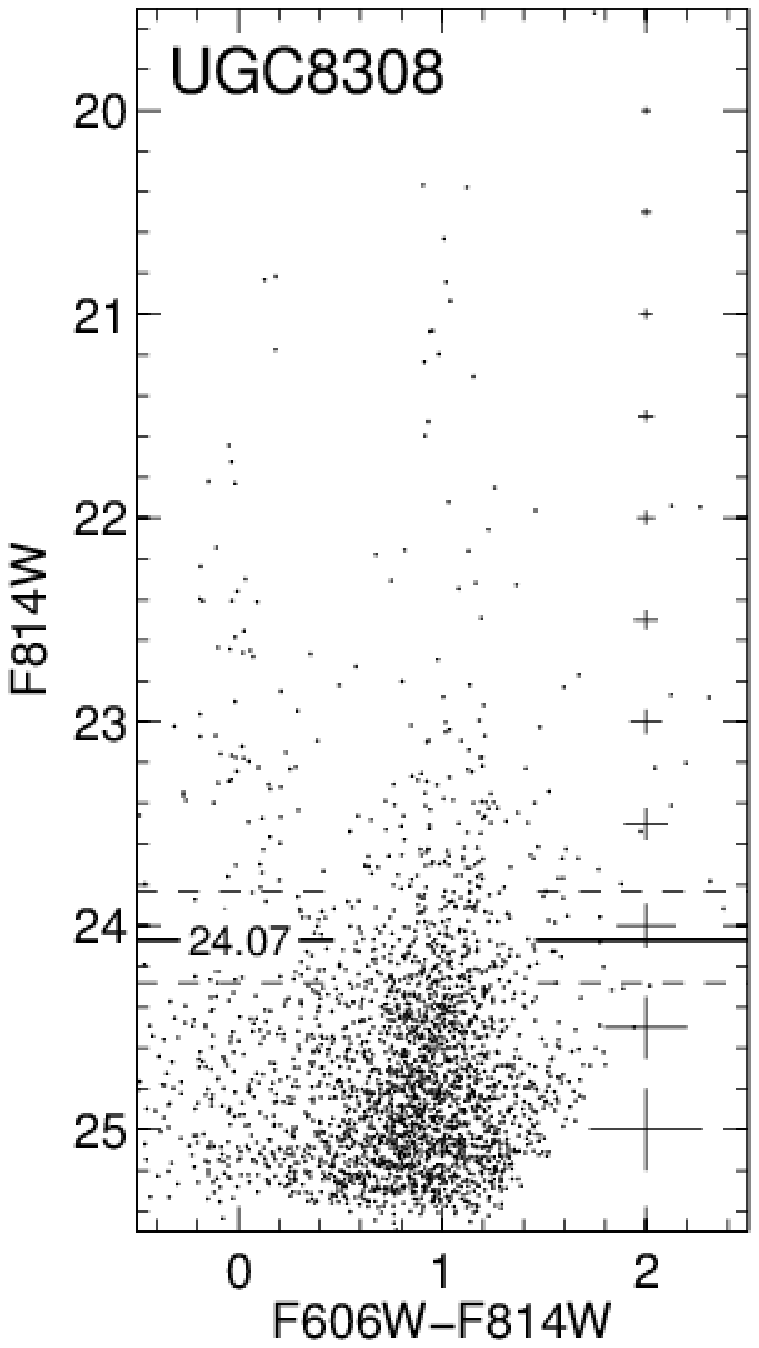}
\hspace{1mm}
\includegraphics[height=0.4\textwidth]{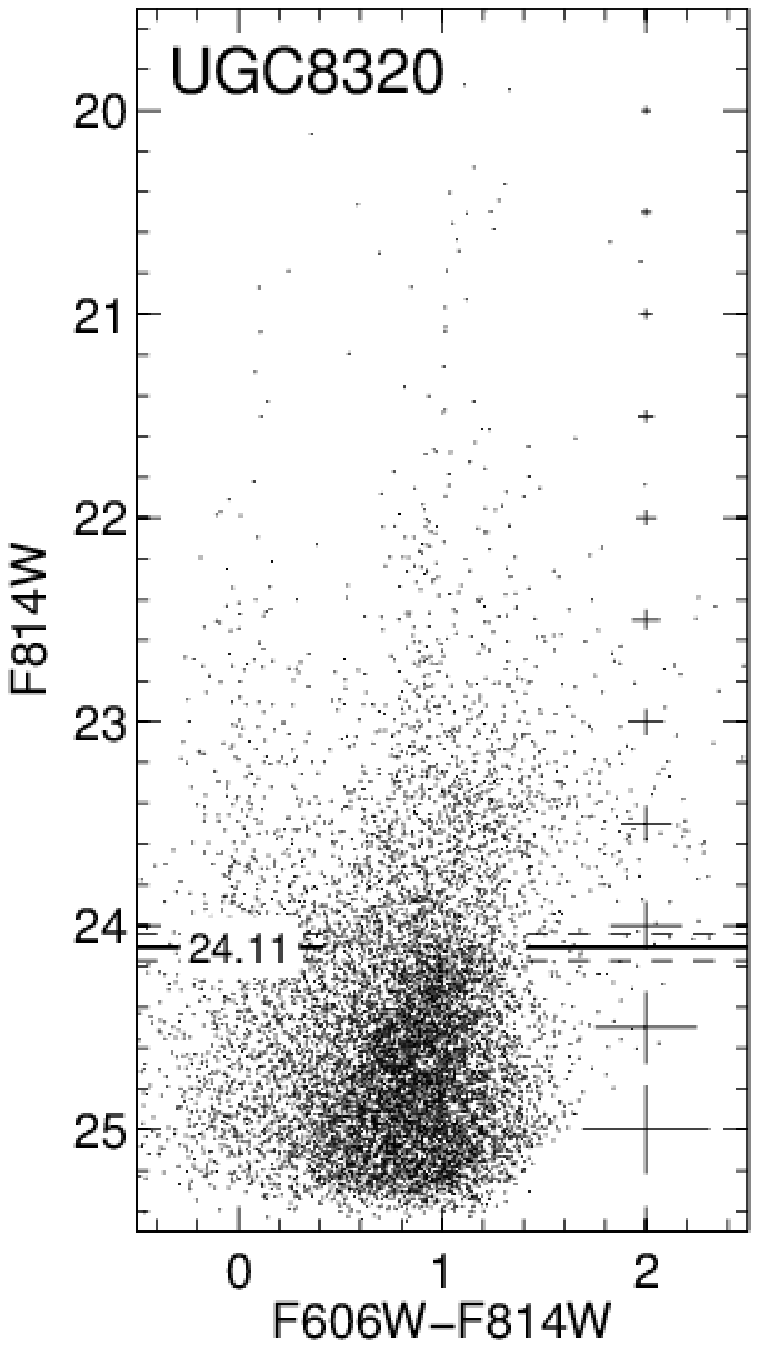}
\hspace{1mm}
\includegraphics[height=0.4\textwidth]{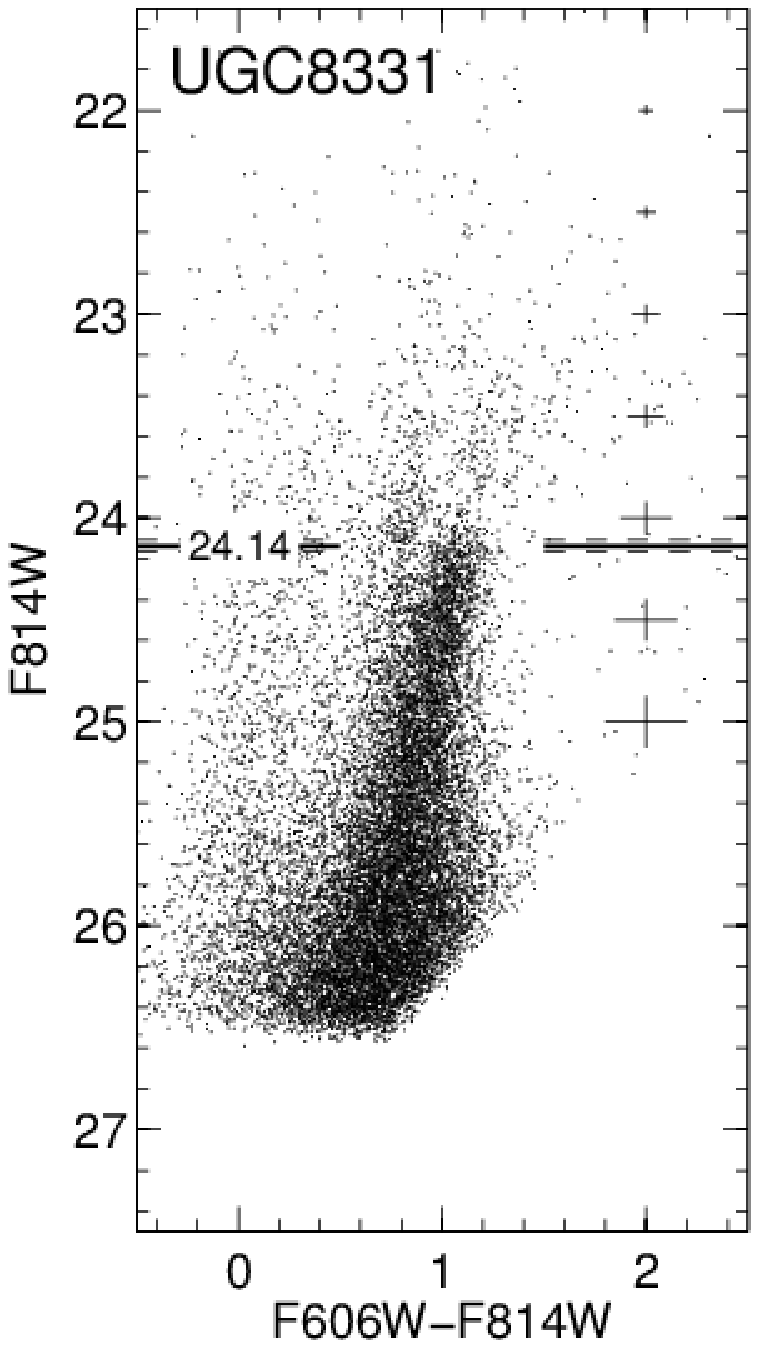}
\hspace{1mm}
\includegraphics[height=0.4\textwidth]{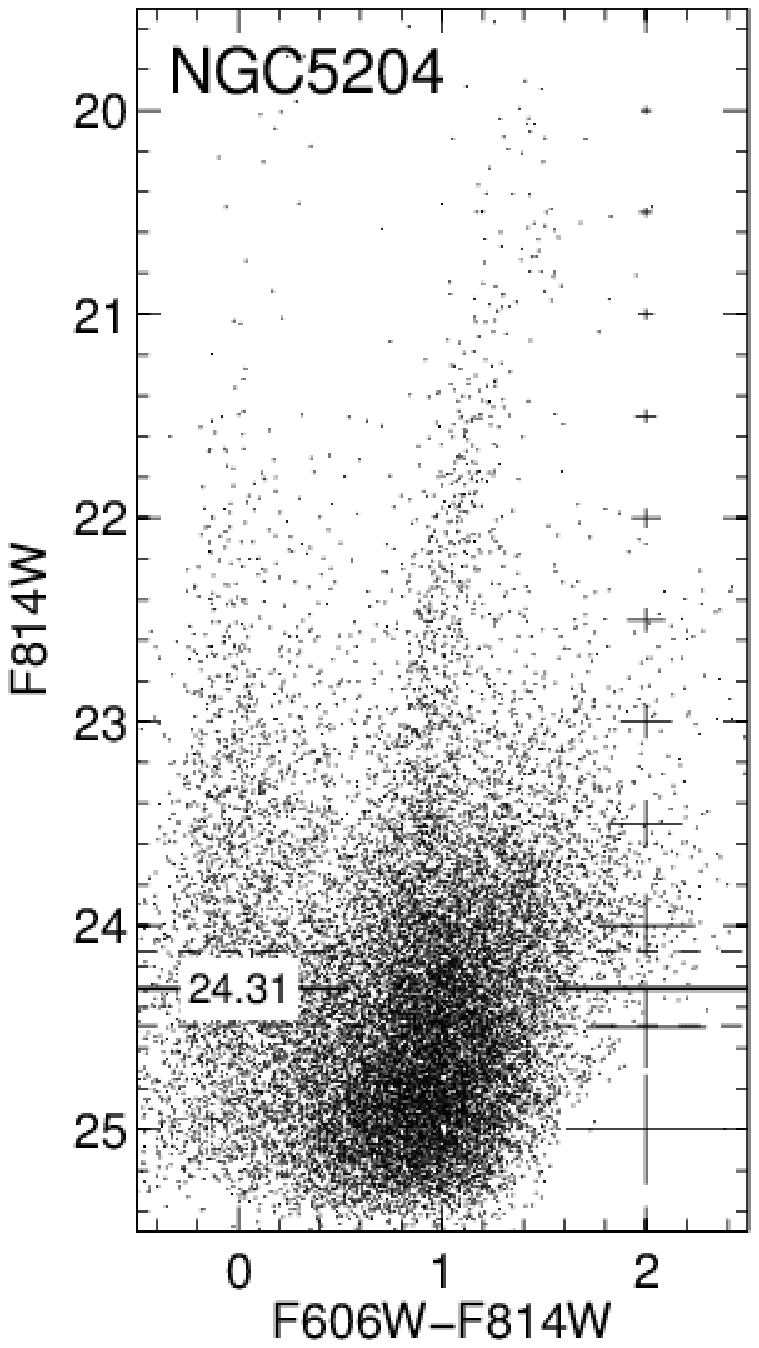}
} \caption{(Contd.).}
\end{figure*}

\begin{figure*}
\addtocounter{figure}{-1} 
\vbox{
\includegraphics[height=0.4\textwidth]{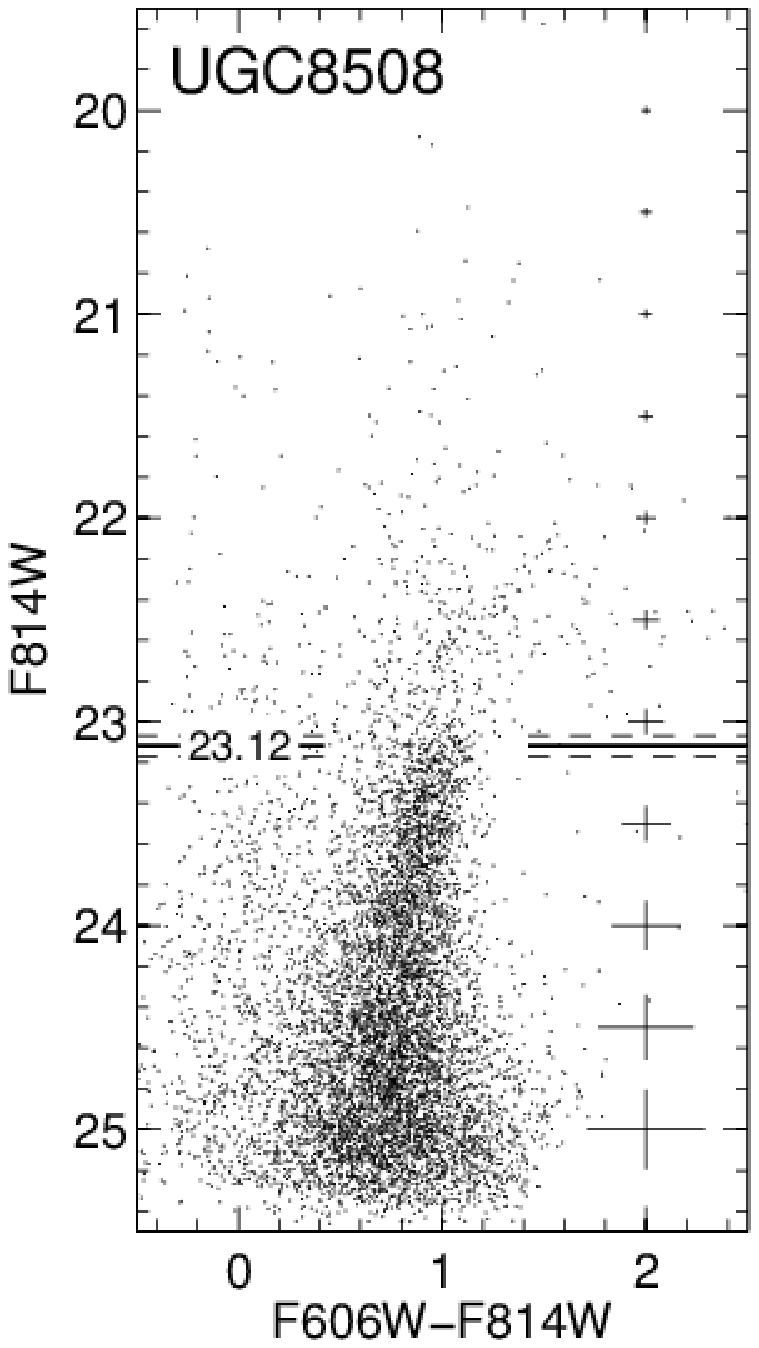}
\hspace{1mm}
\includegraphics[height=0.4\textwidth]{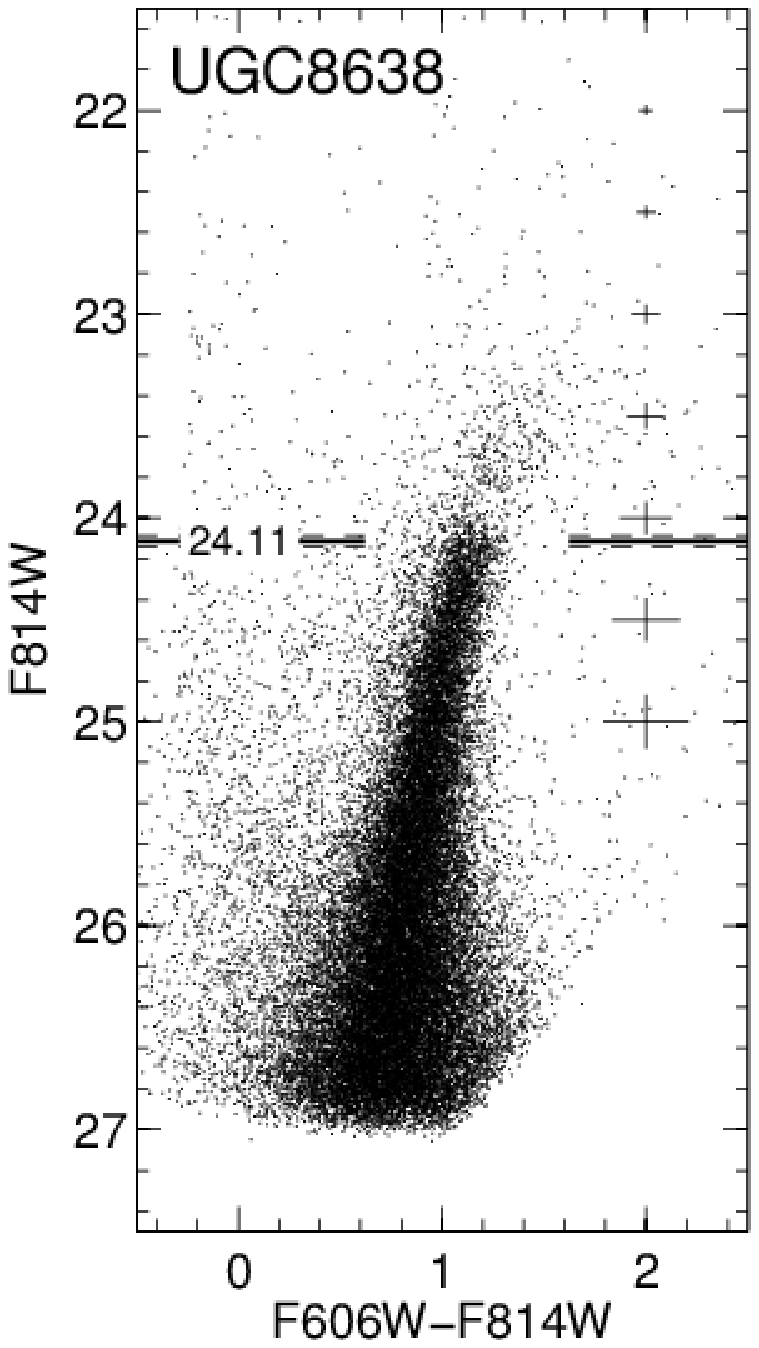}
\hspace{1mm}
\includegraphics[height=0.4\textwidth]{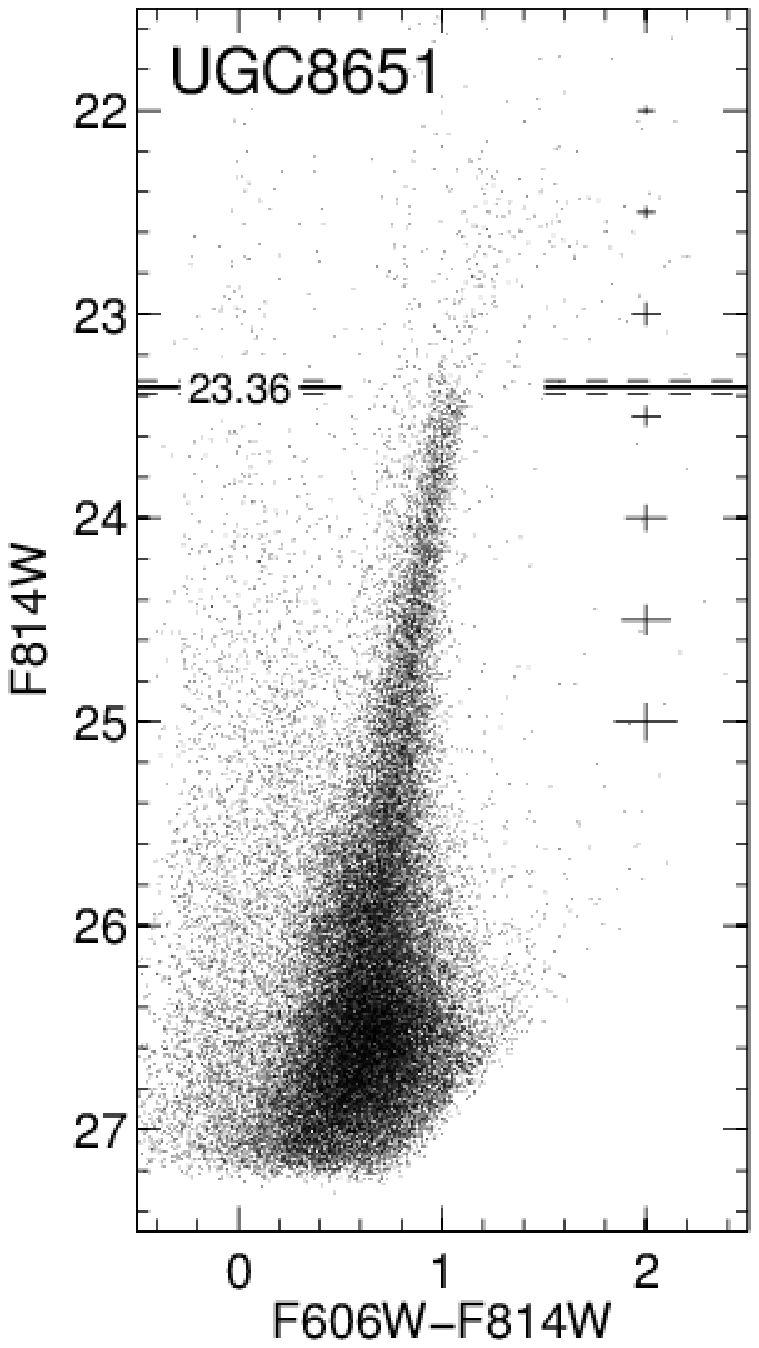}
\hspace{1mm}
\includegraphics[height=0.4\textwidth]{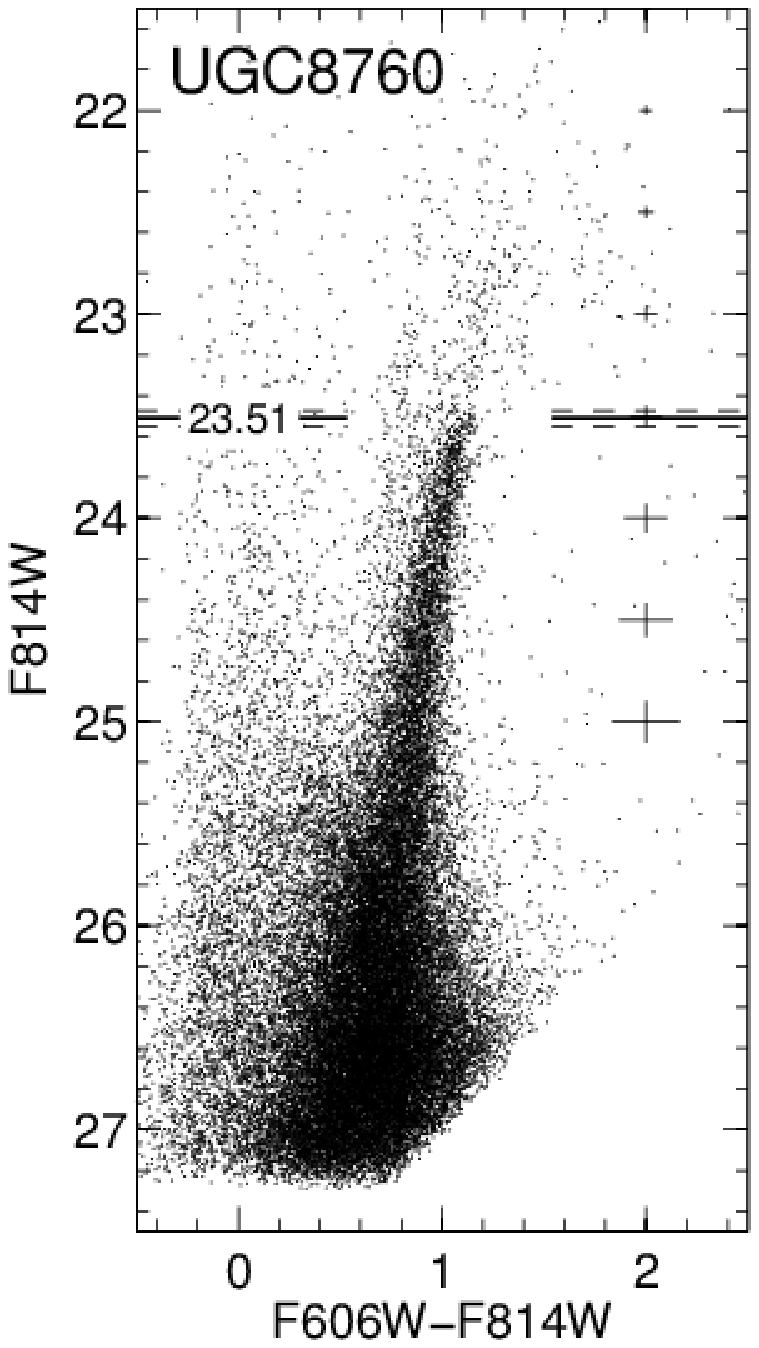}
} \vspace{2mm}\vbox{
\includegraphics[height=0.4\textwidth]{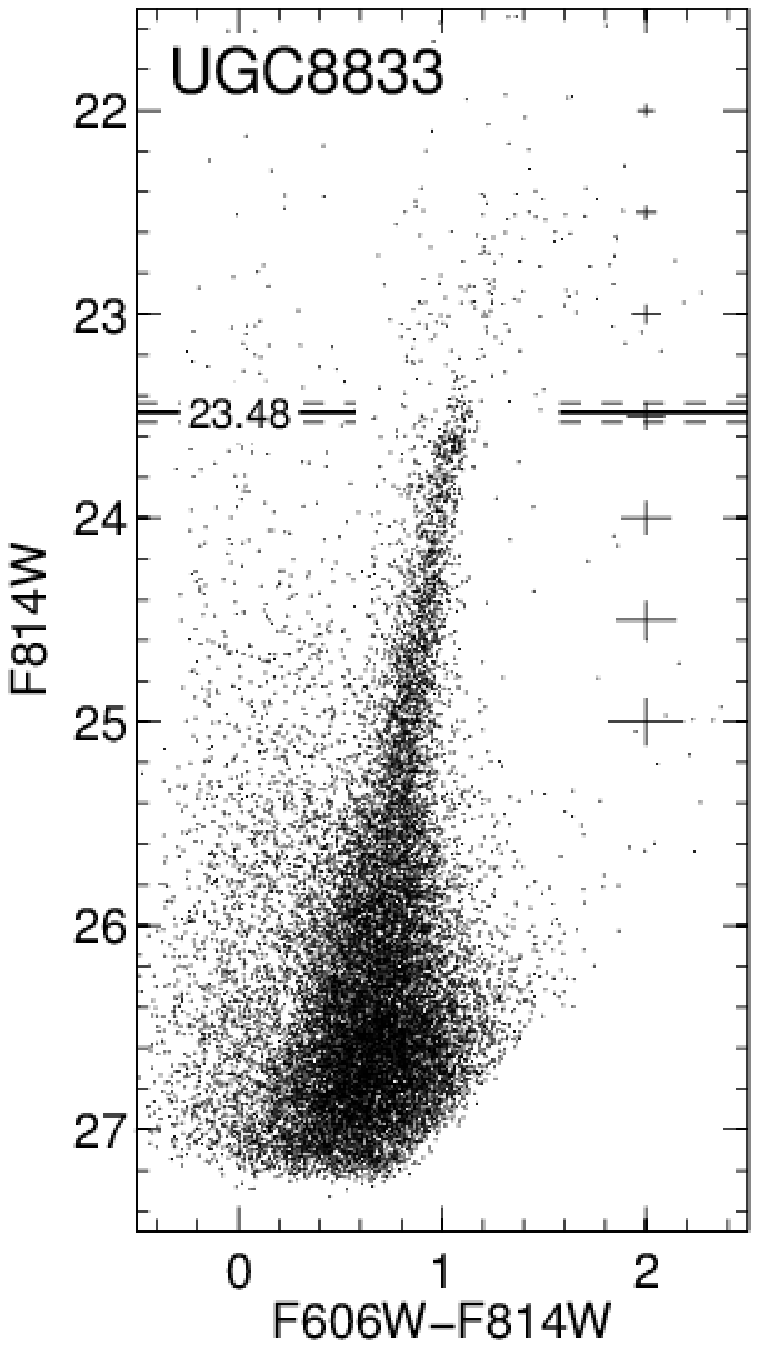}
\hspace{1mm}
\includegraphics[height=0.4\textwidth]{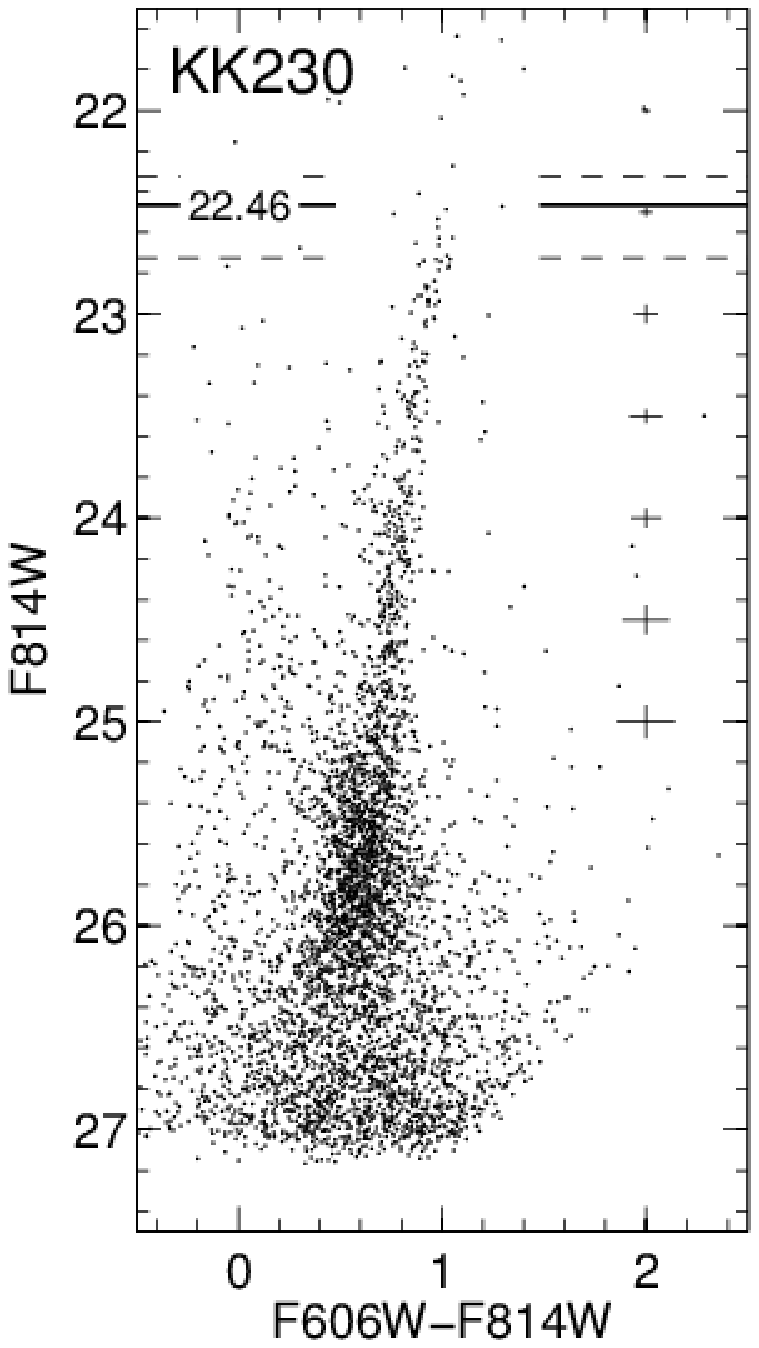}
\hspace{1mm}
\includegraphics[height=0.4\textwidth]{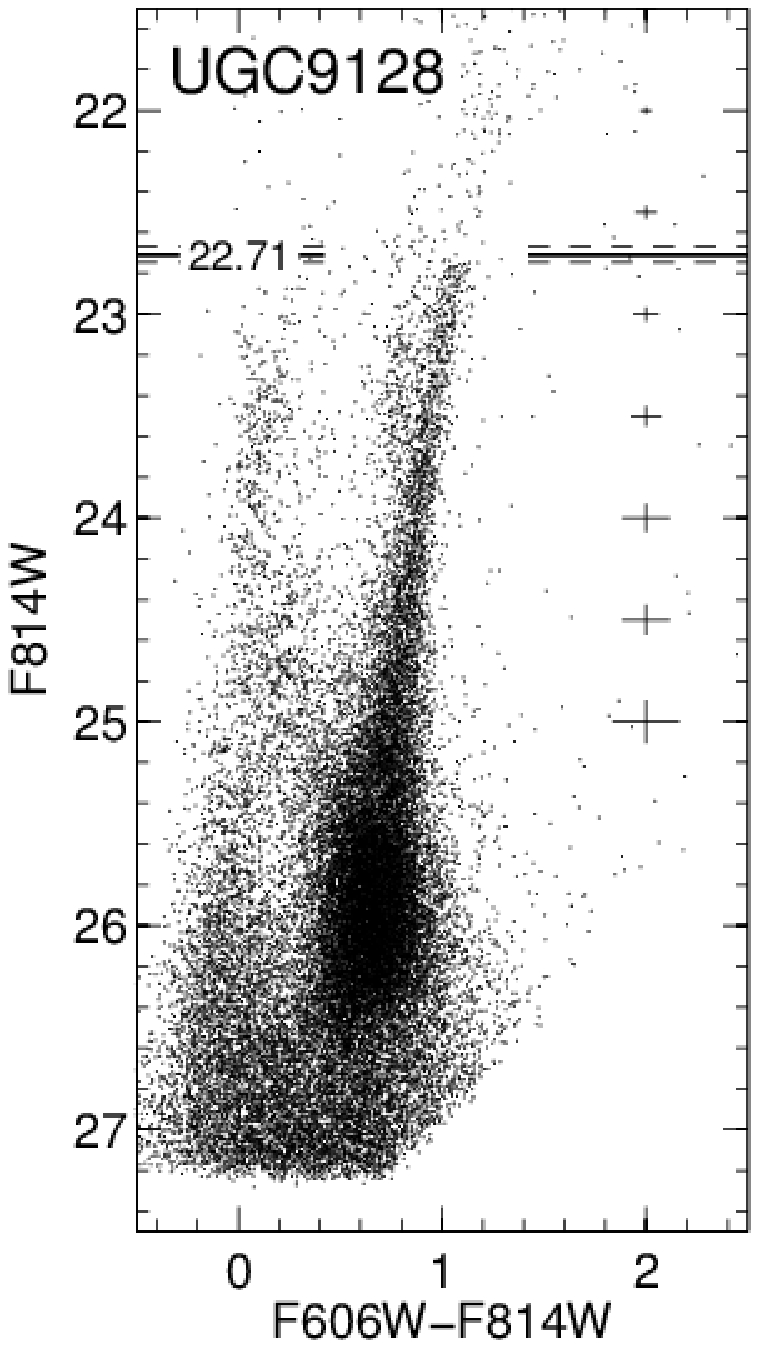}
} \caption{(Contd.).} 
\end{figure*}

During the photometry, we paid special attention to the
photometric errors. Since the quality of measurements in the
crowded star fields is severely affected by the presence of close
neighbors, we ran artificial star tests. We created a large
library of artificial stars for each galaxy, the
\mbox{color--magnitude} distribution of which would correspond to
the actual one. Exactly the same photometry procedures were
applied to these artificial stars in the galaxy images, as did the
real stars. This time consuming process is the best way to
estimate the real photometric errors, accounting for the
incompleteness of photometry, very close neighbors, and
superimposed stars.

{\renewcommand{\baselinestretch}{0.93}
\begin{table*}
 \caption{The distances of galaxies in the CVn\,I cloud} \label{t:distances:Makarov_n}
 \medskip
\begin{tabular}{l|r|r@{$\,\pm\,$}r|c|c|c|c|c|r|r|r|c}
\hline
 Name     &$B_t$\,,$^{*}$& \multicolumn{2}{c|}{$V_h\,$,$^{**}$}&$E(B-V)$,& TRGB,$\,^{***}$          & Color,$\,^{****}$              & $(m-M)_0$,               & $D$,                   &$V_{\rm LG}$,& $A_B^i$, & $M_B$, &  Ref \\ [-5pt]
          &  mag       & \multicolumn{2}{c|}{\kms}        & mag     & mag                   & mag                       & mag                      & Mpc                    & \kms        & mag      & mag      &           \\
\hline
UGC\,6541 & 14.44 & $249$&$ 2$ & 0.019 & $24.15^{+0.10}_{-0.10}$ & $0.977^{+0.034}_{-0.010}$ & $ 28.15^{+0.12}_{-0.12}$ & $4.26^{+0.23}_{-0.23}$ &  303 & 0.00 & $-13.79$ & $^b$, $^A$ \\
NGC\,3738 & 11.87 & $225$&$ 8$ & 0.010 & $24.71^{+0.06}_{-0.06}$ & $1.188^{+0.004}_{-0.004}$ & $ 28.76^{+0.08}_{-0.08}$ & $5.65^{+0.22}_{-0.22}$ &  306 & 0.02 & $-16.95$ & $^b$, $^D$ \\
NGC\,3741 & 14.40 & $229$&$ 2$ & 0.025 & $23.48^{+0.07}_{-0.07}$ & $0.937^{+0.023}_{-0.021}$ & $ 27.48^{+0.09}_{-0.09}$ & $3.13^{+0.14}_{-0.13}$ &  263 & 0.00 & $-13.18$ & $^b$, $^A$ \\
UGC\,6817 & 13.70 & $251$&$ 2$ & 0.026 & $23.14^{+0.05}_{-0.06}$ & $0.942^{+0.010}_{-0.024}$ & $ 27.13^{+0.08}_{-0.08}$ & $2.66^{+0.10}_{-0.10}$ &  257 & 0.00 & $-13.54$ & $^a$, $^A$ \\
NGC\,4068 & 13.19 & $201$&$ 2$ & 0.022 & $24.14^{+0.02}_{-0.02}$ & $1.056^{+0.005}_{-0.007}$ & $ 28.20^{+0.06}_{-0.06}$ & $4.36^{+0.12}_{-0.12}$ &  281 & 0.00 & $-15.10$ & $^a$, $^C$ \\
NGC\,4163 & 13.63 & $162$&$ 5$ & 0.020 & $23.31^{+0.02}_{-0.02}$ & $1.144^{+0.004}_{-0.005}$ & $ 27.35^{+0.06}_{-0.06}$ & $2.96^{+0.08}_{-0.08}$ &  163 & 0.00 & $-13.81$ & $^a$, $^H$ \\
UGC\,A276 & 15.86 & $285$&$ 5$ & 0.020 & $23.35^{+0.07}_{-0.08}$ & $1.033^{+0.030}_{-0.057}$ & $ 27.34^{+0.10}_{-0.09}$ & $2.93^{+0.13}_{-0.12}$ &  288 & 0.00 & $-11.56$ & $^g$, $^E$ \\
NGC\,4214 & 10.24 & $291$&$ 5$ & 0.022 & $23.31^{+0.06}_{-0.06}$ & $1.652^{+0.014}_{-0.045}$ & $ 27.26^{+0.05}_{-0.04}$ & $2.84^{+0.06}_{-0.06}$ &  295 & 0.01 & $-17.13$ & $^c$, $^F$ \\
UGC\,7298 & 15.95 & $174$&$ 2$ & 0.023 & $24.17^{+0.29}_{-0.23}$ & $0.992^{+0.043}_{-0.138}$ & $ 28.16^{+0.23}_{-0.30}$ & $4.28^{+0.46}_{-0.59}$ &  256 & 0.00 & $-12.31$ & $^a$, $^A$ \\
NGC\,4244 & 10.88 & $248$&$17$ & 0.021 & $24.16^{+0.09}_{-0.10}$ & $1.206^{+0.051}_{-0.074}$ & $ 28.19^{+0.12}_{-0.11}$ & $4.34^{+0.24}_{-0.21}$ &  260 & 0.68 & $-18.08$ & $^c$, $^B$ \\
UGC\,7559 & 14.12 & $217$&$17$ & 0.014 & $24.31^{+0.05}_{-0.05}$ & $0.956^{+0.019}_{-0.018}$ & $ 28.32^{+0.08}_{-0.08}$ & $4.61^{+0.16}_{-0.16}$ &  230 & 0.00 & $-14.26$ & $^d$, $^B$ \\
UGC\,7577 & 12.95 & $206$&$ 2$ & 0.020 & $23.08^{+0.03}_{-0.03}$ & $1.049^{+0.007}_{-0.008}$ & $ 27.06^{+0.06}_{-0.06}$ & $2.58^{+0.08}_{-0.08}$ &  251 & 0.00 & $-14.20$ & $^a$, $^A$ \\
NGC\,4449 & 10.06 & $202$&$33$ & 0.019 & $24.15^{+0.06}_{-0.06}$ & $1.170^{+0.026}_{-0.055}$ & $ 28.11^{+0.09}_{-0.09}$ & $4.19^{+0.17}_{-0.17}$ &  250 & 0.07 & $-18.21$ & $^a$, $^B$ \\
UGC\,7605 & 14.76 & $310$&$ 2$ & 0.014 & $24.30^{+0.08}_{-0.08}$ & $0.899^{+0.026}_{-0.032}$ & $ 28.32^{+0.10}_{-0.10}$ & $4.61^{+0.22}_{-0.21}$ &  317 & 0.00 & $-13.62$ & $^d$, $^A$ \\
IC\,3687  & 13.79 & $350$&$33$ & 0.020 & $24.08^{+0.06}_{-0.05}$ & $0.956^{+0.014}_{-0.012}$ & $ 28.08^{+0.08}_{-0.08}$ & $4.12^{+0.15}_{-0.15}$ &  377 & 0.00 & $-14.37$ & $^a$, $^B$ \\
KK\,166   & 17.62 & \multicolumn{2}{c|}{} & 0.015 & $24.12^{+0.26}_{-0.30}$ & $1.144^{+0.104}_{-0.055}$ & $ 28.10^{+0.31}_{-0.26}$ & $4.17^{+0.60}_{-0.50}$ &      & 0.00 & $-10.54$ & $^f$       \\
M\,94     &  8.74 & $308$&$ 8$ & 0.018 & $24.29^{+0.04}_{-0.03}$ & $2.106^{+0.033}_{-0.045}$ & $ 28.14^{+0.07}_{-0.08}$ & $4.25^{+0.15}_{-0.16}$ &  352 & 0.14 & $-19.62$ & $^f$, $^C$ \\
IC\,4182  & 12.02 & $321$&$ 2$ & 0.014 & $24.21^{+0.04}_{-0.04}$ & $1.431^{+0.015}_{-0.012}$ & $ 28.15^{+0.07}_{-0.07}$ & $4.26^{+0.14}_{-0.14}$ &  357 & 0.00 & $-16.19$ & $^a$, $^G$ \\
UGC\,8215 & 16.03 & $224$&$ 2$ & 0.011 & $24.17^{+0.05}_{-0.06}$ & $1.093^{+0.012}_{-0.042}$ & $ 28.24^{+0.08}_{-0.08}$ & $4.44^{+0.16}_{-0.16}$ &  303 & 0.00 & $-12.25$ & $^h$, $^A$ \\
UGC\,8308 & 15.45 & $150$&$ 2$ & 0.010 & $24.07^{+0.21}_{-0.24}$ & $0.969^{+0.022}_{-0.116}$ & $ 28.08^{+0.25}_{-0.22}$ & $4.14^{+0.47}_{-0.42}$ &  230 & 0.00 & $-12.68$ & $^a$, $^A$ \\
UGC\,8320 & 12.97 & $191$&$ 8$ & 0.015 & $24.11^{+0.07}_{-0.07}$ & $0.914^{+0.031}_{-0.037}$ & $ 28.12^{+0.09}_{-0.09}$ & $4.20^{+0.17}_{-0.17}$ &  270 & 0.00 & $-15.21$ & $^a$, $^D$ \\
UGC\,8331 & 14.46 & $262$&$ 5$ & 0.009 & $24.14^{+0.03}_{-0.03}$ & $1.072^{+0.008}_{-0.009}$ & $ 28.22^{+0.06}_{-0.06}$ & $4.40^{+0.13}_{-0.13}$ &  348 & 0.00 & $-13.80$ & $^a$, $^H$ \\
NGC\,5204 & 11.73 & $201$&$ 2$ & 0.013 & $24.31^{+0.18}_{-0.18}$ & $1.117^{+0.024}_{-0.024}$ & $ 28.30^{+0.19}_{-0.19}$ & $4.57^{+0.40}_{-0.40}$ &  339 & 0.11 & $-16.73$ & $^c$, $^C$ \\
UGC\,8508 & 14.12 & $ 57$&$ 2$ & 0.015 & $23.12^{+0.05}_{-0.05}$ & $0.923^{+0.016}_{-0.026}$ & $ 27.13^{+0.08}_{-0.08}$ & $2.67^{+0.09}_{-0.09}$ &  181 & 0.00 & $-13.07$ & $^a$, $^A$ \\
UGC\,8638 & 14.44 & $276$&$ 2$ & 0.013 & $24.11^{+0.03}_{-0.03}$ & $1.122^{+0.005}_{-0.005}$ & $ 28.17^{+0.06}_{-0.06}$ & $4.31^{+0.13}_{-0.13}$ &  275 & 0.00 & $-13.79$ & $^d$, $^A$ \\
UGC\,8651 & 14.22 & $214$&$ 2$ & 0.006 & $23.36^{+0.03}_{-0.03}$ & $1.004^{+0.014}_{-0.012}$ & $ 27.45^{+0.07}_{-0.07}$ & $3.10^{+0.10}_{-0.09}$ &  284 & 0.00 & $-13.26$ & $^e$, $^A$ \\
UGC\,8760 & 14.47 & $188$&$ 2$ & 0.016 & $23.51^{+0.04}_{-0.04}$ & $1.037^{+0.024}_{-0.029}$ & $ 27.58^{+0.07}_{-0.07}$ & $3.28^{+0.10}_{-0.11}$ &  254 & 0.00 & $-13.18$ & $^e$, $^A$ \\
UGC\,8833 & 15.30 & $221$&$ 2$ & 0.012 & $23.48^{+0.05}_{-0.05}$ & $1.077^{+0.011}_{-0.011}$ & $ 27.55^{+0.08}_{-0.07}$ & $3.24^{+0.11}_{-0.11}$ &  280 & 0.00 & $-12.30$ & $^e$, $^A$ \\
KK\,230   & 17.50 & $ 63$&$ 2$ & 0.014 & $22.46^{+0.26}_{-0.14}$ & $0.979^{+0.022}_{-0.025}$ & $ 26.55^{+0.15}_{-0.27}$ & $2.04^{+0.14}_{-0.25}$ &  127 & 0.00 & $-9.11$ & $^i$, $^A$ \\
UGC\,9128 & 14.38 & $160$&$ 2$ & 0.023 & $22.71^{+0.04}_{-0.04}$ & $0.922^{+0.067}_{-0.100}$ & $ 26.79^{+0.07}_{-0.07}$ & $2.28^{+0.08}_{-0.07}$ &  180 & 0.00 & $-12.51$ & $^d$, $^A$ \\
\hline

\multicolumn{13}{p{0.96\textwidth}}{ \footnotesize \vspace{-4.5mm}
\begin{tabular}{lrp{0.85\textwidth}}
Notes: &  *   & Links for photometry: \refmag\\[-5pt]
       & **   & Links for the line-of-sight velocity measurements: \refvel\\[-5pt]
       & ***  & TRGB location, measured in the F814W$_{\rm TRGB}$ filter.\\[-5pt]
       & **** & For virtually all cases we mean the value of
(F814W--F606W)$_{\rm TRGB}$,  with the exception of the
NGC4214 galaxy, for which the color (F814W--F555W)$_{\rm TRGB}$ was used.\\
\end{tabular}
}\\

\end{tabular}
\end{table*}
}

\section{COLOR--MAGNITUDE DIAGRAMS}

The color--magnitude diagrams  (CMD) of stars from the studied
galaxies are shown in Fig.~\ref{fig:cmd1:Makarov_n}. Irregular
dwarf galaxies constitute the vast majority of objects in our
sample. The upper part of the Main Sequence is clearly discernible
in all the diagrams, it contains blue stars with the mean color
index close to zero. The top right part of the CMD is occupied by
red supergiants and the stars of the asymptotic giant
branch~(AGB), both these branches are populated differently for
various galaxies. The largest population in each diagram is the
RGB stars. We have selected from the Hubble archive only those
exposures which are deep enough to safely identify the RGB in the
galaxy, and therefore, determine its distance with a good accuracy
by the TRGB method.

\section{FINDING THE DISTANCES}

\begin{figure*}
\includegraphics[width=1\textwidth]{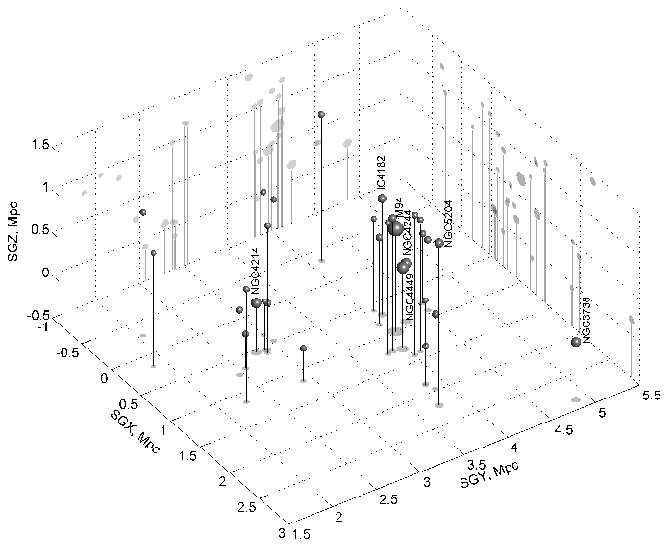}
\caption{A three-dimensional structure of the Canes Venatici I
cloud. The sphere size is inversely proportional to the absolute
magnitude of the galaxies. The names are denoted for the brightest
galaxies with $M_B<-16{\rm^m}$.} \label{fig:3d:Makarov_n}
\end{figure*}

To accurately determine the structure of the cloud of galaxies in
the Canes Venatici constellation, we need to know the precise
photometric distances to these galaxies, uniformly estimated. The
distance estimates to many galaxies of the cloud were performed
earlier, using the tip of the red giant branch method as one of
the most accurate distance indicators. However, a uniform
estimation of such distances is fulfilled  for the first time.
Moreover, an important factor for refining the distances is to
improve the method itself, which was done by authors. In order to
determine the TRGB values, we used our {\tt TRGBTOOL} code, based
on the maximum likelihood algorithm for the luminosity function of
stars in a galaxy, and taking into account accurate photometric
errors, determined from artificial star
tests~\cite{TRGB1:Makarov_n}. The calibration of the TRGB method
has been recently improved~\cite{TRGB2:Makarov_n}. In that study,
the color index--absolute TRGB magnitude dependence, as well as
the zero-points of ACS and WFPS2 photometric systems were
determined.
The position of the luminosity function cut-off and the
corresponding 1$\sigma$ error are shown in the CMDs of the studied
galaxies by horizontal lines (Fig.~\ref{fig:cmd1:Makarov_n}).

New photometric distances to the galaxies of the Canes Venatici~I
cloud are summarized in Table~\ref{t:distances:Makarov_n}. It
lists: (1)~the name of the galaxy in the known catalogs; (2)~total
apparent magnitude $B_t$ in the \mbox{$B$-band;} (3)~heliocentric
velocity $V_{h}$; (4)~color excess \mbox{$E(B-V)$} in the
direction of the galaxy according to~\cite{SFD98:Makarov_n};
(5)~TRGB position, determined using the {\tt
TRGBTOOL}; (6)~the averaged color {(${\rm
F814W}-{\rm F606W}$)$_{\rm TRGB}$} of stars in the TRGB region
(for the NGC\,4214 galaxy, we used the F555W filter instead of
F606W); (7) \mbox{$(m-M)_0$} is the measured distance modulus of
the galaxy in mag; (8)~$D$ is the corresponding photometric
distance in~Mpc; (9)~the line-of-sight velocity relative to the
centroid of the Local Group  $V_{\rm LG}$,~according
to~\cite{apex1996:Makarov_n}; (10)~$A_B^i$~ is the internal
absorption in the galaxy~in~ the $B$-filter, according
to~\cite{UNGC:Makarov_n}; (11)~$M_B$ is the absolute stellar
B-magnitude of the galaxy; (12)~references to the photometry and
line-of-sight velocities of galaxies. The total apparent magnitude
of the galaxy, TRGB and the average color of RGB stars in the
cut-off region was not corrected for galactic extinction. Note
that these measurements are part of a much broader program of
uniform distance measurements to the nearby, mostly dwarf galaxies
via the TRGB method using {\tt
TRGBTOOL}~\cite{jacobs:Makarov_n}.\footnote{\tt
http://edd.ifa.hawaii.edu/}

\subsection{Distance to M\,94}

M\,94 is a giant spiral galaxy, which is located within the CVn\,I
cloud and could be claimed as its gravitating center. Thus, it is
very important to estimate the exact photometric distance to this
object. M\,94 was resolved into individual stars for the first
time in the course of our SNAPshot observations with the HST/WFPC2
(Prop.~8601). The distance modulus, estimated from the tip of the
red giant branch based on the results of photometry of the images
is \mbox{$28.34\pm0.29$~\cite{karachentsev+2003:Makarov_n}.}
However, the exposures with WFPC2 ($600$-s in the F606W filter and
$600$-s in the F814W  filter) yield a sufficiently dense stellar
field, while the tip of the red giant branch is located only
$1{\rm ^m}$ above the photometric limit. The estimate of the
photometric distance in giant galaxies encounters a number of
technical difficulties due to the presence of internal absorption
and high surface brightness. We have determined the distance to
M\,94, using deeper exposures, obtained with the HST/ACS
(Prop.~10,523). Three fields of M\,94, located far from the center
of the galaxy, were observed within this project, and hence the
effect of the above-mentioned difficulties is minimized.

The luminosity function of RGB stars in the M\,94 galaxy appears
to be much more complex than that for the normal dwarf galaxies.
It has a long, extended plateau near the cut-off. This is probably
related with the complex history of star formation and metal
enrichment in M\,94. This behavior of the luminosity function
differs from a simple power law, which usually describes the red
giant branch, therefore making it impossible to use the maximum
likelihood technique for determining the TRGB in the M\,94 galaxy.
We have thus used the classical edge detection technique,
described by Sakai et al.~\cite{SMF1996:Makarov_n} and the same
up-to-date TRGB method calibrations~\cite{TRGB2:Makarov_n} as for
the other galaxies.

Our measurements give the distance modulus of
{$28.14\pm0.08$} and the distance to the giant spiral
M\,94 \mbox{$D = 4.25\pm0.15$~Mpc.} This value has a better
accuracy and is in a good agreement both with the earlier estimate
and with the estimates made by Radburn-Smith et
al.~\cite{radburn-smith:Makarov_n} from the same images. The
latest work gives a distance modulus amounting to
\mbox{$28.17\pm0.13$.}

\subsection{Distance to NGC\,5204}

Active star formation is underway in the center of the NGC\,5204
galaxy. It is extremely difficult to determine the distance to the
galaxy because of a dense stellar field, a large number of young
stars and the proximity of the TRGB to the photometric limit
(about $1{\rm^m}$). To get rid of ``contamination'' of the diagram
by a large number of young stars, and to avoid the excessive
influence of the nearby stellar fields on the photometry results,
we have selected only the stars, located far from the star forming
regions. This approach has allowed us to increase the contrast of
the red giant branch and determine the position of its cut-off. As
in the case M\,94, we used the edge detection method. The measured
distance modulus to the NGC\,5204 galaxy is
{$(m-M)_0=28.30\pm0.19$,} which is in excellent agreement
with the estimate \mbox{$(m-M)_0=28.34\pm0.27$,} obtained
in~\cite{karachentsev+2003:Makarov_n} based on the same data.

\section{DISCUSSION AND CONCLUSIONS}

Figure~\ref{fig:3d:Makarov_n} demonstrates  the spatial
distribution of galaxies, obtained from our distance measurements
in supergalactic coordinates. The sphere size is proportional to
the logarithm of the galaxy luminosity. The concentration of
galaxies around M\,94, the brightest member of the CVn\,I,  is
clearly visible here.

\begin{figure}
\includegraphics[width=0.5\textwidth]{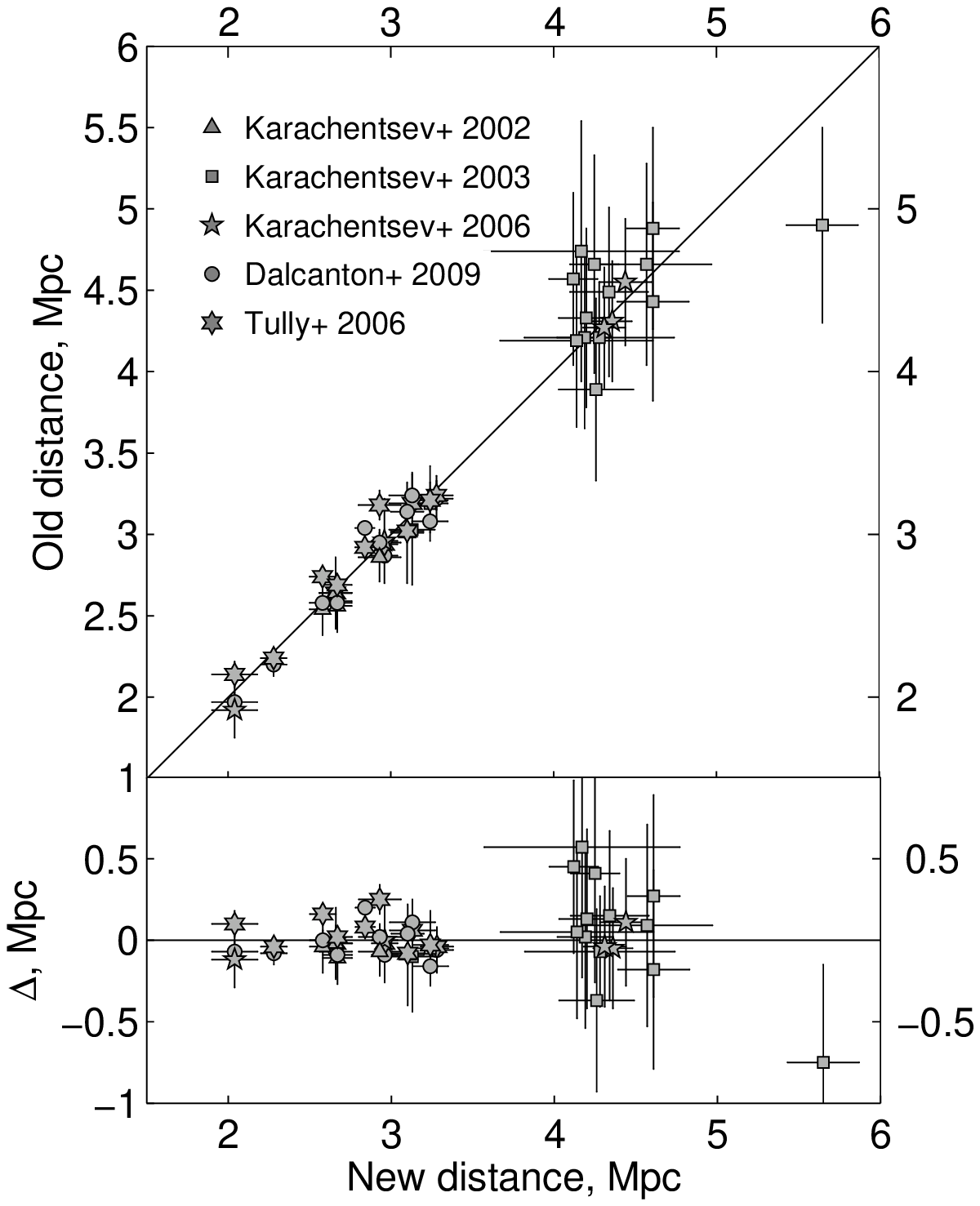}
\caption{A comparison of the distances obtained in this work, with
earlier measurements. The data are taken
from~\cite{karachentsev+2002:Makarov_n,karachentsev+2003:Makarov_n,karachentsev+2006:Makarov_n,dalcanton+2009:Makarov_n,tully+2006:Makarov_n}.
}\label{fig:dd:Makarov_n}
\end{figure}

\begin{figure*}
\includegraphics[width=0.8\textwidth]{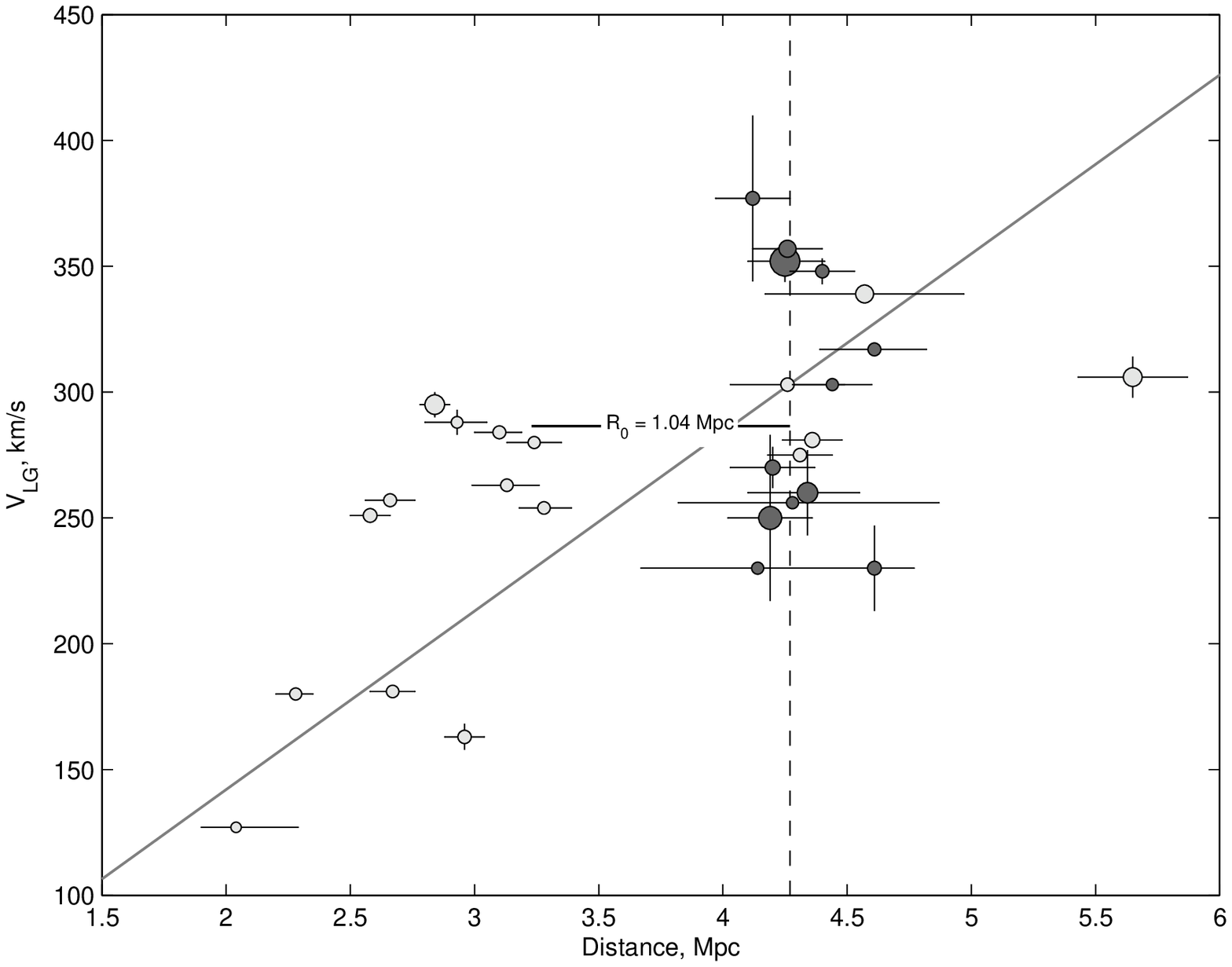}
\caption{The velocity--distance diagram for the galaxies in the
Canes Venatici region. The linear Hubble law   with
\mbox{$H_0=71$~km/(s$\times$Mpc)} is depicted by the solid line.
The galaxies within~1~Mpc from the center of the system are marked
in dark gray. The median average distance of the central
concentration is shown by the dashed line.}
\label{fig:hubble:Makarov_n}
\end{figure*}

A comparison of our distance estimates with the measurements by
other
authors~\cite{karachentsev+2002:Makarov_n,karachentsev+2003:Makarov_n,karachentsev+2006:Makarov_n,tully+2006:Makarov_n,dalcanton+2009:Makarov_n}~
shows a very good agreement with a generally better accuracy
(Fig.~\ref{fig:dd:Makarov_n}). The increased measurement accuracy
is due both to the deeper observations of the same galaxies,
carried out at a later time, and the use of a more refined
technique of distance estimation.

The Hubble diagram of the recession of galaxies in the direction
of the Canes Venatici is shown in Fig.~\ref{fig:hubble:Makarov_n}.
Note that in the previous study of the structure of this cloud of
galaxies, this region looks much more ``fuzzy'' (Fig.~6
in~\cite{karachentsev+2003:Makarov_n}), which did not allow to
make a conclusion on the virialized state of this group of
galaxies. The higher quality of  observations allows us to
identify an area of chaotic motions around the center of the
system. The group of galaxies around M\,94 is characterized by a
median velocity of \CVnIVmed, median distance of \CVnIDist, the
line-of-sight-velocity dispersion of \CVnIVsigma, corrected for
the measurement errors, the mean projected distance from the
galactic center of the system of \CVnIRmean, and the total
luminosity of \CVnILBtot{}. The mass of the system, estimated by
the virial theorem amounts to \CVnIMvir, which corresponds to the
\mbox{mass--luminosity} ratio of \CVnIMLvir{}. The projection mass
estimate~\cite{HTB1985:Makarov_n} of this  system is \CVnIMproj{}
and the corresponding mass--luminosity ratio amounts to
 \CVnIMLproj{}. Note, however, that the crossing time of
the CVn\,I cloud of galaxies,   \CVnITcr makes up about half the
age of the Universe \mbox{$T=13.7$ Gyr.} Therefore, the issue of
the system's proximity to the steady state requires further
consideration, and one should use the virial theorem to estimate
the mass of the system with  caution.


As noted in \cite{karachentsev+2003:Makarov_n}, almost all the
galaxies, located closer than the central concentration CVn\,I
have positive peculiar velocities and form the characteristic
``wave,'' caused by the infall of matter onto the massive clusters
of galaxies (see, e.g.,~\cite{KKMT2009:Makarov_n}). Unfortunately,
the current data on the distances of galaxies, located behind the
studied CVn\,I cloud, does not allow to unambiguously claim that
there is a similar infall of matter on the opposite side of the
group, although some hints imply this. According to our data, only
the  NGC\,3738 galaxy has a fairly deep CMD to measure the
distance of 5.65~Mpc. Apparently, this galaxy ``falls'' onto the
CVn\,I cloud from the opposite side and has a large negative
peculiar velocity of   $V_{\rm pec}=-95$~\kms. If we suppose that
the observed distribution of galaxies on the Hubble diagram at the
distances of less than 3.5~Mpc is due to the gravitational effect
of the group of galaxies around M\,94, we can estimate the radius
of the zero velocity sphere  \CVnIRta{} as the mean between the
forward and reverse regressions of the velocity and distance of
galaxies. It corresponds to the mass \CVnIMta{} (formula
from~\cite{KKMT2009:Makarov_n}). This value is in a good agreement
with the projection mass estimate. The analysis of peculiar
velocities of field galaxies is independent of the virial theorem
by the system mass measurement method. A more accurate mass
estimate of the CVn\,I cloud should include a simulation of the
distribution of galaxies by peculiar velocities and their spatial
distribution.

Our mass--luminosity estimate, $(M/L)_{p}=$
$159~(M/L)_{\odot}$
for the CVn\,I cloud of galaxies greatly exceeds the typical ratio
\mbox{$M/L_B\sim30$} for the nearby groups of galaxies, such as
the Local Group \mbox{($M/L_B=15\textrm{--}20$)} and
M\,81 group
($M/L_B=19\textrm{--}32$)~\cite{karachentsev2005:Makarov_n}. Note
that compared with the well-known nearby groups, such as the Local
Group \mbox{($L_B=10.1\times10^{10}L_{\odot}$),} M\,81
\mbox{($L_B=6.1\times10^{10}$ $L_{\odot}$)} and Centaurus\,A
\mbox{($L_B=5.5\times10^{10}$ $L_{\odot}$),} the CVn\,I cloud of
galaxies (\CVnILBtot) contains about \mbox{4--5}~times less
luminous matter, and M\,94 is at least $1{\rm ^m}$  fainter than
any other central galaxy of these
groups~\cite{karachentsev2005:Makarov_n}. However, the
concentration of galaxies in the Canes Venatici may have a
comparable total mass.

The catalog of groups of galaxies in the Local Supercluster
\cite{MK2011:Makarov_n} has demonstrated that the mean density of
gravitating matter on the scale of 80~Mpc is about 2.5 times
smaller than the cosmological constant $\Omega_m=0.27$. One
possible explanation for this striking difference between the
global and local estimates of density of the Universe can be the
presence of a significant proportion of dark matter outside the
virialized regions associated with luminous matter. Such ``dark
aggregates'' can be quite numerous. For example, Tully et
al.~\cite{tully+2006:Makarov_n} have identified the associations
of nearby dwarf galaxies by the high-precision photometric
distances obtained by the Hubble Space Telescope. Moreover, it was
noted in this paper  that on the scale of up to 3~Mpc, with the
exception of the KKR\,25 galaxy, all the known galaxies are either
combined in groups, or associations. Such rarefied   structures
may possess the \mbox{mass--luminosity} ratios in the range from
100 to 1000 $(M/L)_\odot$.  It was shown
in~\cite{MU2012:Makarov_n} that the groups consisting of dwarf
galaxies only  may be numerous, and they have higher
\mbox{mass--luminosity} ratios than the typical galaxy groups in
the Local Supercluster. It is possible that the cloud of galaxies
in the Canes Venatici is in fact one of these concentrations of
dark matter, where the ratio of dark to luminous matter
significantly exceeds a similar proportion in the typical groups
of galaxies.

\begin{acknowledgments}
The authors thank prof.~I.~D.~Karachentsev for constructive
discussions. This work was supported by the RFBR grant
no.~11-02-00639 and the grant of the Ministry of Education and
Science of the Russian Federation \mbox{no.~8523}. The study was
also supported by the program of the Physical Sciences Division
RAS PSD-17 ``Active Processes in Galactic and Extragalactic
Objects.'' We made use of the HyperLEDA database ({\tt
http://leda.univ-lyon1.fr}).

\end{acknowledgments}

\end{document}